\PassOptionsToPackage{hyphens}{url}
\documentclass[format=acmsmall, review=false, screen=true, dvipsnames]{acmart_arXiv}

\usepackage{fancyhdr}
\usepackage{etoolbox}
\usepackage[normalem]{ulem}
\usepackage{microtype}
\usepackage{array}
\usepackage{enumitem}
\usepackage{hhline}
\usepackage{dblfloatfix}
\usepackage{booktabs}
\usepackage[binary-units=true]{siunitx}
\usepackage{gensymb}
\usepackage{multirow}
\usepackage{xspace}
\usepackage{setspace}
\usepackage[ruled]{algorithm2e} 
\usepackage[caption=false]{subfig}
\usepackage[us,12hr]{datetime}
\usepackage{xstring}
\usepackage{pifont}

\renewcommand\footnotetextcopyrightpermission[1]{}
\setlist{leftmargin=15pt, itemsep=2pt, topsep=2pt}
\DeclareSIUnit\transfer{T}

\SetAlFnt{\small}
\SetAlCapFnt{\small}
\SetAlCapNameFnt{\small}
\SetAlCapHSkip{0pt}
\IncMargin{-\parindent}

\acmJournal{POMACS}
\acmYear{2018}
\acmMonth{0}
\copyrightyear{2018}
\setcopyright{acmcopyright}
\acmPrice{15.00}
\acmDOI{}

\received{February 2018}
\received[accepted]{April 2018}

\newcommand{\todo}[1][]{}
\newcommand{\ch}[0]{}
\newcommand{\chI}[0]{}
\newcommand{\chII}[0]{}
\newcommand{\chIII}[0]{}
\newcommand{\chIV}[0]{}
\newcommand{\rev}[0]{}

\newlength{\basewidth}
\newlength{\basewidthwide}
\newcommand{\figspace}[0]{\qquad\qquad}

\newcommand{\ignore}[1]{}
\newcommand{\paratitle}[1]{\vspace{6pt}\noindent\textbf{#1.}}
\newcommand{\incircle}[1]{%
    \IfEqCase{#1}{%
        {1}{\ding{182}}%
        {2}{\ding{183}}%
        {3}{\ding{184}}%
        {4}{\ding{185}}%
        {5}{\ding{186}}%
        {6}{\ding{187}}%
        {7}{\ding{188}}%
        {8}{\ding{189}}%
    }[\PackageError{mycirc}{Undefined option to mycirc: #1}{}]%
}

\newcommand{\idd}[1][]{IDD{#1}\xspace}
\newcommand{\chipcnt}[1][]{200\xspace}
\newcommand{\dimmcnt}[1][]{50\xspace}

\newcommand{\affilCMU}[0]{Carnegie Mellon University\xspace}
\newcommand{\affilETH}[0]{ETH Z\"{u}rich\xspace}
\newcommand{\affilNVIDIA}[0]{NVIDIA\xspace}
\newcommand{\affilUT}[0]{The University of Texas at Austin\xspace}

\title[What Your DRAM Power Models Are Not Telling You]{What Your DRAM Power Models Are Not Telling You: \\Lessons from a Detailed Experimental Study}
\author{Saugata Ghose}
\affiliation{\affilCMU}

\author{Abdullah Giray Ya\u{g}l{\i}k\c{c}{\i}}
\affiliation{\institution{\affilETH \& \affilCMU}}

\author{Raghav Gupta}
\affiliation{\institution{\affilCMU}}

\author{Donghyuk Lee}
\affiliation{\institution{\affilNVIDIA}}

\author{Kais Kudrolli}
\affiliation{\institution{\affilCMU}}

\author{William X. Liu}
\affiliation{\institution{\affilCMU}}

\author{Hasan Hassan}
\affiliation{\institution{\affilETH}}

\author{Kevin K. Chang}
\affiliation{\institution{\affilCMU}}

\author{Niladrish Chatterjee}
\affiliation{\institution{\affilNVIDIA}}

\author{Aditya Agrawal}
\affiliation{\institution{\affilNVIDIA}}

\author{Mike O'Connor}
\affiliation{\institution{\affilNVIDIA \& \affilUT}}

\author{Onur Mutlu}
\affiliation{\institution{\affilETH \& \affilCMU}}

\authorsaddresses{}

\renewcommand{\shortauthors}{S. Ghose et al.}

\ccsdesc[500]{Hardware~Dynamic memory}
\ccsdesc[500]{Hardware~Power estimation and optimization}
\ccsdesc[500]{Hardware~Board- and system-level test}
\ccsdesc[300]{Computing methodologies~Model development and analysis}
\ccsdesc[300]{Computer systems organization~Architectures}

\keywords{DRAM; memory systems; energy; power consumption; power modeling; experimental characterization;
data encoding; low-power design}

\begin{abstract}

Main memory (DRAM) consumes as much as half of the total system power in a
computer today, due to the increasing demand for memory capacity and
bandwidth.
There is a growing need to understand and analyze DRAM power consumption, 
which can be used to research new DRAM architectures and systems that 
consume less power.
A major obstacle against such research is the 
lack of detailed and accurate information on the power consumption
behavior of modern DRAM devices.
Researchers have long relied on DRAM power models that are predominantly
based off of a set of standardized current measurements
provided by DRAM vendors, called IDD values.
Unfortunately, we find that state-of-the-art DRAM power models are often highly inaccurate, 
as these models do \emph{not} reflect the actual
power consumed by \emph{real} DRAM devices.

To build an \emph{accurate} model and provide insights into DRAM power consumption,
we perform the first comprehensive experimental characterization of the power 
consumed by modern real-world DRAM modules.  Our extensive characterization
of \dimmcnt~DDR3L DRAM modules from three major vendors
yields four key new observations about DRAM power consumption that
prior models cannot capture:
(1)~\rev{across all IDD values that we measure,}
the current consumed by real DRAM modules
\rev{\emph{varies significantly}} from the current specified by the vendors;
(2)~DRAM power consumption strongly \emph{depends on the data value} 
that is read or written;
(3)~there is significant \emph{structural variation},
where the same banks and rows across multiple DRAM modules from the 
same model consume more power than other banks or rows; and
(4)~over successive process technology generations, DRAM power consumption
has not decreased by as much as vendor specifications have indicated.
Because state-of-the-art DRAM power models do \emph{not} account for any 
of these four key characteristics, they are highly \ch{inaccurate
compared} to the actual,
measured power consumption of \dimmcnt real DDR3L modules.

Based on our detailed analysis and characterization data, we develop
the \emph{Variation-Aware model of Memory Power Informed by Real Experiments}
(VAMPIRE).
VAMPIRE is a new, accurate power consumption model for DRAM
that takes into account
(1)~module-to-module and intra-module variations, and
(2)~power consumption variation due to data value dependency.
\ch{We show that VAMPIRE has a mean absolute percentage error of only 6.8\%
compared to actual measured DRAM power.}
VAMPIRE enables a wide range of studies that were \emph{not} possible using prior DRAM power 
models.
As an example, we use VAMPIRE to evaluate the energy efficiency of three different 
encodings that can be used to store data in DRAM.
We find that a new power-aware data encoding mechanism can reduce total DRAM 
energy consumption by an average of 12.2\%, across a wide range of applications.
We plan to open-source both VAMPIRE and our extensive raw data collected during our
experimental characterization.

\end{abstract}

\begin{document}
\maketitle



\section{Introduction}
\label{sec:intro}

\ch{As processor power consumption has been reduced via many techniques designed
over multiple decades, main memory, \chI{which is built using the dynamic random access memory
(DRAM) technology,} has now become a significant source of power consumption in modern
computer systems.}
This is because the amount of DRAM in a computer has been increasing over the
years, to keep up with the growing demand for memory capacity and 
bandwidth in modern applications~\cite{wang2014bigdatabench, 
ferdman.sigplan2012, malladi.isca12, mutlu.imw2013}.
In a contemporary system, DRAM power consumption accounts for as much as \chI{46\%} 
of the \emph{total system power}\ch{~\cite{Lefurgy, ware.power7, david.icac11, 
holzle.book09, malladi.isca12, yoon.isca12, elmore.tr16, paul.isca15}}.
In response, vendors have developed several low-power and low-voltage variants 
of DRAM (e.g., DDR3L~\cite{ddr3l.jedec13}, LPDDR3~\cite{lpddr3.jedec15} 
and LPDDR4~\cite{lpddr4.jedec17}), and there has been some
research on reducing the power consumed by modern DRAM architectures
(e.g., \chI{\cite{liu.isca2012, liu.sigplan2012, seshadri.micro13, chang.pomacs2017, 
Cooper-Balis, Bhati, bhati.islped13, ozturk.isqed06, delaluz.dac02b, 
delaluz.hpca01, Diniz, Deng1, malladi.micro12, Fan, chang.thesis17, Patel, PARBOR,
Khan, khan.micro17, lee.taco16}}).
However, as DRAM consumes a growing fraction of the total system power, a much
greater effort is necessary to \chI{invent} \ch{new} low-power solutions.

One major hindrance towards further research is the relative lack 
of information available on the low-level power consumption behavior of 
modern DRAM devices.
It has historically been difficult to collect \emph{accurate} power consumption
data from real DRAM devices, as computer systems
(1)~do not offer fine-grained control \ch{over commands}
being issued to DRAM, instead exposing only high-level operations such as \ch{loads
and stores}; and
(2)~often lack dedicated monitors that track the \ch{power
consumed} by DRAM.
As a result, for many years, researchers have instead relied on current 
specifications that vendors provide for each DRAM \ch{part}, which are known as
\emph{IDD values}~\cite{micron.2015}.
A vendor determines IDD values by using a standardized set of benchmarks that are
meant to represent common high-level DRAM operations, such as reading 
one cache line of data from DRAM, or \emph{refreshing} data (i.e., restoring 
the charge lost from DRAM cells due to charge leakage).
State-of-the-art DRAM power models (e.g., \cite{Jung, ChandrasekarRunTime, drampower, micron.2015}),
which researchers currently use to perform DRAM power studies, and 
which are used by many popular simulators (e.g., \cite{gem5, rosenfeld-cal2011, 
bakhoda.ispass09, patel.dac11, kim.cal15}),
are predominantly based on these IDD values.

We find that state-of-the-art DRAM power models are often highly inaccurate
when compared with the \ch{power consumed by real DRAM chips}.
This is because existing DRAM power models
(1)~are based off of the \emph{worst-case} power consumption of devices, as vendor
specifications list the current consumed by the most power-hungry device sold;
(2)~do \emph{not} capture variations in DRAM power consumption due to 
different \emph{data value} patterns; and
(3)~do \emph{not} account for any \emph{variation} across different devices or within
a device.
Because existing DRAM power models do \ch{\emph{not}} capture these characteristics,
it is often difficult for researchers to accurately 
(1)~identify sources of inefficiency within DRAM; and
(2)~evaluate the effectiveness of memory energy saving techniques,
including new hardware designs and \ch{new} software mechanisms.
\emph{Our goal} in this work is to \emph{rigorously} measure and analyze
the power consumption of \emph{real} DRAM devices, and to use our analysis to 
develop an accurate and detailed DRAM power model, which can be useful
for a wide variety of purposes.

To this end, we perform the first extensive characterization of 
the power consumed by real DRAM devices.  To overcome prior obstacles to
collecting real power measurements from DRAM, 
\rev{we significantly extend the SoftMC FPGA-based DRAM testing
infrastructure~\cite{hassan.hpca17, softmc.repo} to work with
high-precision current measurement equipment}, which we describe in detail in Section~\ref{sec:char}.
Our testing infrastructure allows us to execute precise test procedures to characterize \chII{the effects of}
\ch{(1)~intra-chip and inter-chip} variation and
\ch{(2)}~data dependency \chII{on} the power consumed by DRAM.  We collect detailed power 
measurement data \ch{from} \dimmcnt~DDR3L DRAM modules, comprising of \chipcnt~chips, 
which were manufactured by three major vendors (A, B, and C).
Our testing infrastructure allows us to make four key new observations about 
DRAM power consumption that prior models \chI{\emph{cannot}} capture:
\begin{enumerate}
  \item
  \rev{Across all IDD values that we measure,}
  the current consumed by real DRAM modules \rev{varies significantly} from the
  current specified by the vendors (Section~\ref{sec:idd}).
  For example, to
  read one cache line of data from DRAM, the measured current
  of modules from Vendor~A is lower than the current specified 
  in the datasheet by an average of 54.1\% (up to 61.6\%).

  \item
  DRAM power consumption strongly depends on the data value
  that is read or written (Section~\ref{sec:datadep}).
  Reading a cache line where all bits are \emph{ones} uses 
  an average of 39.2\% (up to 91.6\%)
  more power than reading a cache line where all bits are \emph{zeroes}.

  \item
  There is significant \emph{structural variation}, where the current varies
  based on which bank or row is selected in a DRAM module
  (Section~\ref{sec:var}).
  For example, in modules from Vendor~C, the idle current consumed
  when one of the eight banks is active
  (i.e., open) can vary by an average of 15.4\% (up to 23.6\%) depending on the bank.

  \item
  Across successive process technology generations, the \rev{actual
  power \chI{reduction}} of DRAM is \emph{much lower} than \rev{the savings indicated by the
  vendor-specified IDD values in the datasheets} \ch{(Section~\ref{sec:gen})}. 
  Across five key IDD values, the measured savings 
  of modules from Vendor~A are 
  lower than indicated by an average of 48.0\% (up to 66.7\%).
\end{enumerate}
Because state-of-the-art DRAM \ch{power} models\ch{~\cite{ChandrasekarRunTime, drampower, micron.2015, Jung}} 
do \emph{not} \rev{adequately} capture these four key characteristics, their predicted DRAM power is highly inaccurate 
compared to the actual measured DRAM power of \dimmcnt real DDR3L modules.
\rev{\ch{We perform a validation of two state-of-the-art models, DRAMPower~\cite{ChandrasekarRunTime, 
drampower} and the Micron power model~\cite{micron.2015}, using our FPGA-based 
current measurement platform, and find that the mean absolute percent error
compared to real DRAM power measurements is 32.4\% for DRAMPower and
160.6\% for the Micron power model.}}

Building upon our \chI{new} insights and characterization data, we develop the
\emph{Variation-Aware model of Memory Power Informed by Real Experiments} (VAMPIRE).
VAMPIRE is a new power model for
DRAM, which captures important characteristics such as 
module-to-module and intra-module variations,
and power consumption variation due to data value dependency 
(Section~\ref{sec:model}).
\rev{We show that VAMPIRE is highly \ch{accurate:} it has \ch{a mean absolute
percentage} error of
only \ch{6.8\%} compared to actual measured DRAM power.}

VAMPIRE enables a wide range of studies that were \chI{\emph{not}} possible using prior DRAM power 
models.
For example, we use VAMPIRE to evaluate the impact of different data
encoding mechanisms on DRAM power consumption (Section~\ref{sec:studies}).
We find that a new power-aware data encoding \ch{technique} can reduce DRAM energy by an average
of 12.2\% (up to 28.6\%) across a wide range of applications.

We plan to open-source both VAMPIRE and our extensive raw data
collected during our experimental characterization~\cite{vampire.github}.
We hope that \ch{our findings} and our new power model model will inspire new
research directions, new ideas, and rigorous evaluations
in power- and energy-aware DRAM design.

We make the following contributions in this work:
\begin{itemize}
  \item
    \rev{We conduct a} detailed and accurate experimental characterization of the
    power consumed by \dimmcnt~real, state-of-the-art DDR3L DRAM modules
    \chI{from three major DRAM vendors,} comprising \chipcnt~DRAM chips.
    \rev{To our knowledge, our characterization is the first to
    (1)~report real power \chII{consumption across} a wide variety of tests on
    a large number of DRAM modules from three major DRAM vendors,
    \chII{(2)~comprehensively demonstrate the inter-vendor and intra-vendor 
    module-to-module variation of DRAM power consumption,}
    (3)~study the impact of \chII{data dependency and structural variation} on DRAM power consumption, and
    (4)~examine power consumption \ch{trends} over multiple product generations.}

  \item
    We make four major new observations based on our measurements, showing that
    DRAM power consumption
    (1)~varies significantly from the values provided by DRAM vendors,
    (2)~depends on the data value that is read or written,
    (3)~varies based on which bank and which row are used,
    (4)~has not decreased as much as vendor specifications indicate 
    over successive generations.

  \item
    We build VAMPIRE, a new DRAM power consumption model based on our 
    \ch{new insights and} characterization
    data.  VAMPIRE provides significantly greater accuracy than existing power 
    models, and enables studies that were previously not easily possible.
    We plan to release our power model and our characterization data online\ch{~\cite{vampire.github}}.
\end{itemize}


\section{Background}
\label{sec:bkgd}

In this section, we first provide necessary DRAM background.
We discuss the hierarchical organization of a modern memory system in
Section~\ref{sec:bkgd:org}.
We discuss the fundamental operations performed on DRAM in
Section~\ref{sec:bkgd:operations}.
\ch{For a detailed overview of DRAM operation, we refer the reader
to our prior works\chIII{~\cite{Liu, chang.thesis17, lee.thesis16, kim.thesis15,
lee.hpca15, chang.sigmetrics2016, chang.pomacs2017, Tiered-Latency,
lee.sigmetrics17, hassan.hpca17, kim.isca14, SALP, lee.taco16,
Hassan, seshadri.micro13, seshadri.micro17, seshadri.bookchapter17,
seshadri.thesis16, kim.cal15, lee.pact15, Chang, LISA}}.}

\subsection{DRAM Organization}
\label{sec:bkgd:org}

\begin{figure}[h]
  \centering
  \subfloat[\ch{Memory system hierarchy}]{%
    \includegraphics[width=0.6\basewidthwide, trim=82 410 371 110, clip]{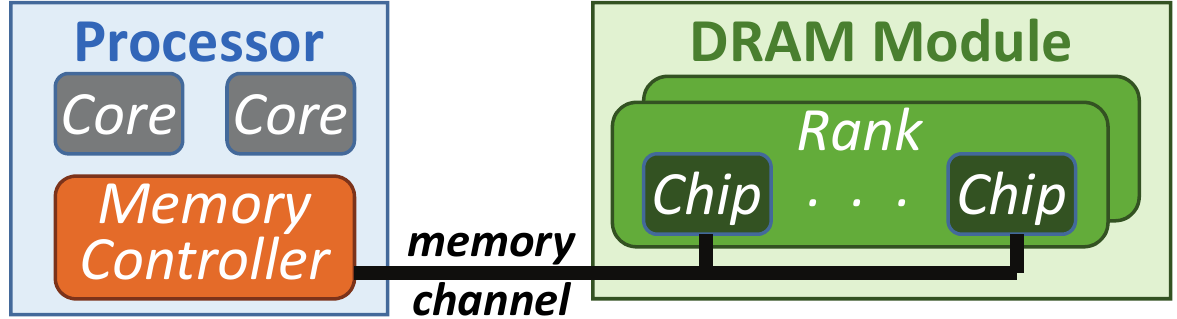}%
    \label{fig:hierarchy}
  }%
  \figspace%
  \subfloat[\ch{DRAM array structure}]{%
    \includegraphics[width=0.38\basewidthwide, trim=408 415 182 78, clip]{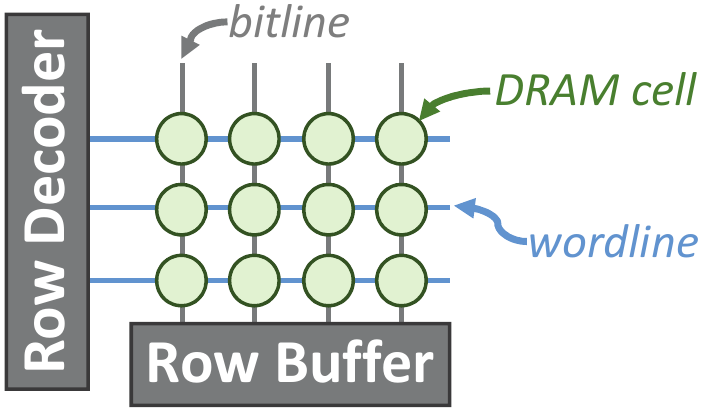}%
    \label{fig:array}
  }%
  \caption{\ch{High-level overview of DRAM organization.}}%
  \label{fig:memsys}
\end{figure}

\ch{Figure~\ref{fig:hierarchy}} shows the basic overview of a DRAM-based memory system.
The memory system is organized in a hierarchical manner.
The highest level in the hierarchy is a \emph{memory channel}.  Each channel consists
of its own bus to the host device, and has a dedicated \emph{memory controller} that interfaces
between the DRAM and the host.  A channel
can connect to one or more \emph{dual inline memory modules} (DIMMs).
Each DIMM contains multiple DRAM \emph{chips}.  A DRAM row typically spans
across several chips, which requires these chips to perform all 
operations in lockstep with each other.  Each group of chips operating in 
lockstep is known as a \emph{rank}.

Inside each rank, there are several \emph{banks}, where each bank can 
independently process DRAM \ch{commands sent} by the memory controller.
While each of the banks within a memory channel can operate concurrently, 
\chI{banks} share a single memory bus, and, thus, the controller must coordinate the
operations across multiple banks in order to avoid interference on the bus.
The ability to operate multiple banks concurrently is known as \emph{bank-level
parallelism}\ch{~\cite{mutlu.isca08, kim.micro10, lee.micro09}}.
DDR3 DRAM typically contains eight banks in each rank~\cite{ddr3.jedec12}.

Each bank contains a two-dimensional \emph{array} of DRAM \emph{cells}, 
\ch{as shown in Figure~\ref{fig:array},}
where each cell stores a single bit of data in a capacitor.
Within the array, cells can be selected one row at a time, and the access
transistors of the cells in one row are connected together using a \emph{wordline}.
Each bank contains a \emph{row buffer}, which consists of a row of sense
amplifiers that are used to temporarily buffer the data from a single row in
the array during read and write operations.  Cells within the array are
connected using vertical wires, known as \emph{bitlines}, to the sense
amplifiers.

A typical row in a DRAM module, which spans across all of the DRAM
chips within a rank, is \SI{8}{\kilo\byte} wide, and holds 128 64-byte cache lines of data.
\chI{For example, in a DDR3 DRAM module with four x16 chips per rank,
each chip contains a \SI{2}{\kilo\byte} portion of the 
\SI{8}{\kilo\byte} row.  Each chip holds a piece of each cache line within the row.}

\subsection{DRAM Operations}
\label{sec:bkgd:operations}

\begin{figure}[h]
  \centering
   \subfloat[\chI{Activate}]{%
    \includegraphics[width=0.29\basewidth, trim=408 393 250 73, clip]{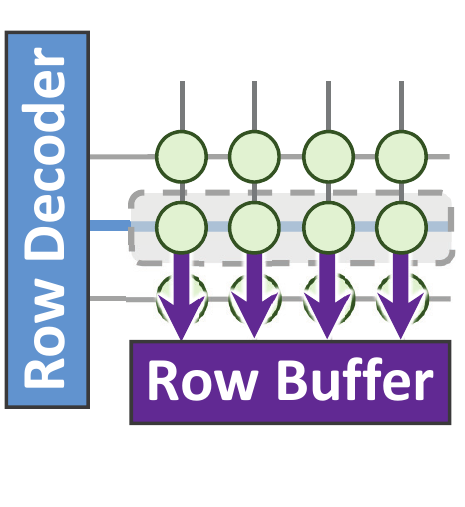}%
    \label{fig:ops:act}
  }%
  \figspace%
  \subfloat[\chI{Read/write}]{%
    \includegraphics[width=0.29\basewidth, trim=408 393 250 73, clip]{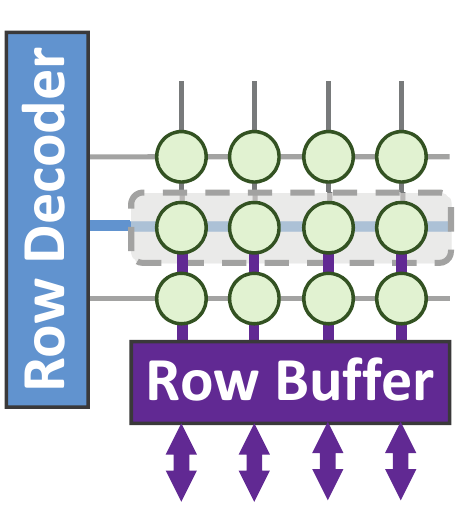}%
    \label{fig:ops:rd_wr}
  }%
  \figspace%
  \subfloat[\chI{Precharge}]{%
    \includegraphics[width=0.29\basewidth, trim=408 393 250 73, clip]{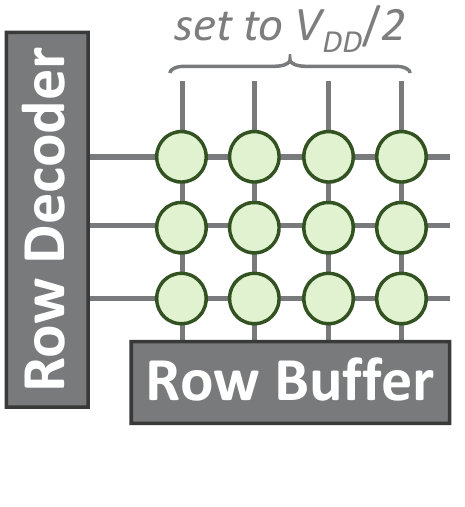}%
    \label{fig:ops:pre}
  }%
  \caption{\chI{Overview of fundamental DRAM commands.}}%
  \label{fig:ops}
\end{figure}

In order to access and update data stored within DRAM, the memory controller 
issues a series of commands across the memory channel to the DRAM chips.
\ch{\chI{Figure~\ref{fig:ops} shows the} four fundamental DRAM \chII{commands:}
\emph{activate}, \emph{read}, \emph{write}, and \emph{precharge}.  
We describe each of these commands below.}

\paratitle{\chI{Activate}}
To start processing a request, the controller issues a command to \emph{activate}
the row (i.e., open the row to perform reads and writes) 
\chI{within each DRAM chip} that contains \chI{a part of}
the desired cache line, as shown in \chI{Figure~\ref{fig:ops:act}}.
\ch{Initially, each bitline is set to half of $V_{DD}$, the full supply voltage
of the DRAM module.  When an activate command is issued,
a \emph{row decoder} turns on one of the wordlines on the array,
based on the row address of the request.  This activates the row of DRAM
cells that are connected to the wordline that is turned on.
Each DRAM cell in the activated row starts \emph{sharing its charge} with the bitline
that the cell is attached to, perturbing the bitline voltage by a small
amount.
Once the bitline voltage changes by more than a set threshold,
the sense amplifier connected to the bitline \chI{detects this \chII{charge},
and} amplifies the bitline voltage
to either $V_{DD}$ (if the DRAM cell connected to
the bitline holds a data value `1') or to \SI{0}{\volt} (if the DRAM
cell connected to the bitline holds a data value `0').
A latch \chI{in the row buffer} is enabled to hold the full voltage value.}

\chI{When a row is activated, the \chI{charge sharing process between
the cell and} the bitline drains charge from the cells in the selected row
\chII{(i.e., destroys the contents of the cells)}.  
This change in \chII{cell} charge can lead to data corruption if 
\chII{the cell charge is not restored to correspond to the cell's original
data value}.
To avoid \chII{such} data corruption, the DRAM chip automatically 
\emph{restores} the charge within the cell back to its starting voltage
once the sense amplifier detects the change in the bitline voltage.}

\begin{figure}[h]
  \centering
  \includegraphics[width=\linewidth, trim=90 171 77 159, clip]{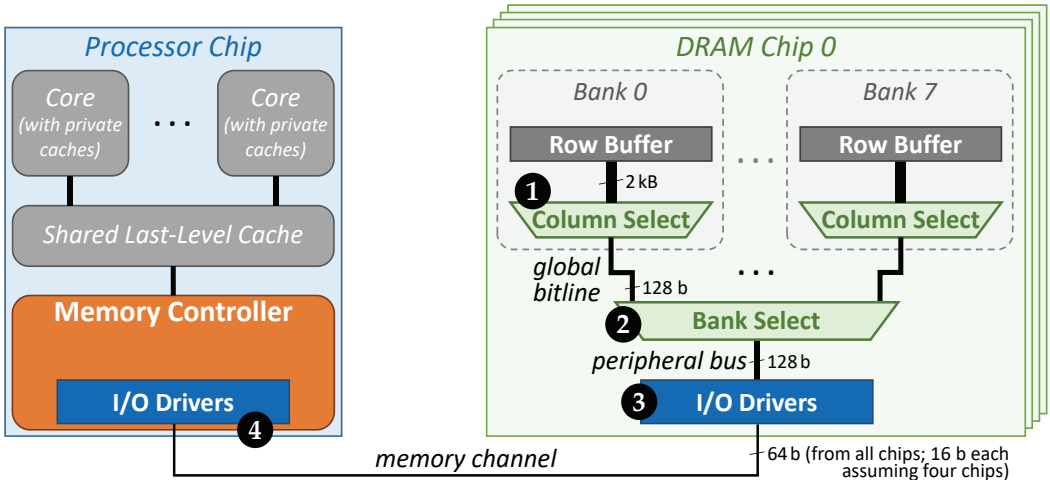}%
  \caption{\ch{Overview of peripheral circuitry and I/O in a four-chip DRAM module.}}%
  \label{fig:peripheral}
\end{figure}

\paratitle{\chI{Read/Write}}
\ch{Once an activated row is latched into the row buffer,}
the controller sends \emph{read and write commands} to 
the row \chI{buffer, as shown in Figure~\ref{fig:ops:rd_wr}}.  
Each read/write command
operates on one \emph{column} of data at a time 
\ch{in each chip of a single rank.  Across the entire rank,
the width of data operated on by a read/write command
(i.e., $column~width \times \#~chips$) is}
the same width as a processor cache line (\SI{64}{\byte}).
\ch{Figure~\ref{fig:peripheral} shows the \emph{peripheral circuitry} in a
DRAM module that is used by the read and write commands.
We walk through the four steps of an example read command as its requested
data moves through the peripheral circuitry \chI{of one x16 DRAM chip}.
First, the read command uses the
\emph{column select logic} (\incircle{1} in Figure~\ref{fig:peripheral})
to select the \chI{128-bit column (which is one part of the cache line)} 
that the request wants to read.
Second, the \chI{column} is sent over the \emph{global bitline} to the
\emph{bank select logic} (\incircle{2}), which is set by the read command
to select the bank that contains the requested cache line.
Third, the \chI{column} is then sent over the \emph{peripheral bus} to the
\emph{I/O drivers} (\incircle{3}).  The \chI{128-bit column} is split up into 
eight \chI{16-bit} \emph{data bursts}.  
\chI{Across all four x16 chips in our example module, \chI{64~bits} of data
are sent per data burst.}
The I/O drivers send the data bursts one at a
time across the memory channel, where each wire of the memory channel has
its own dedicated I/O driver.  In \emph{double data rate} (DDR) DRAM, a
burst can be sent on every positive \emph{or} negative DRAM clock edge,
allowing the entire cache line to be transmitted in four DRAM clock cycles.
Fourth, the bursts are received by the I/O drivers that sit inside the
memory controller at the processor (\incircle{4}).  The memory controller
reassembles the bursts into a \chI{64-byte} cache line, and sends the data to the
processor caches.
For a write operation, the process is similar, but \chII{in the reverse direction:}
the I/O drivers on the
memory controller side send the data across the memory channel.}

\paratitle{\chI{Precharge}}
Once the read and write
operations to the row are complete, the controller issues a \emph{precharge}
command, to prepare the array for commands to a different row,
\chI{as shown in Figure~\ref{fig:ops:pre}.}
\ch{During precharge, the latches in the row buffer are disabled, disconnecting
the DRAM cells from the peripheral circuitry, and the voltage of the bitlines
\chII{is set} to half of $V_{DD}$.
Note that a precharge command can be issued by the controller \chI{\emph{only after}}
the DRAM cells in the activated row are fully restored.}

\paratitle{\chI{DRAM Refresh}}
A key issue in DRAM is that charge \chI{leaks from a DRAM cell, as this
charge is stored in a capacitor.}
When a row has \ch{\emph{not} been accessed} for a long time, 
\chI{the amount of charge that leaks out of the cell can be large enough to
lead to data corruption.}
To avoid data loss due to excessive charge
leakage, the memory controller periodically issues \emph{refresh} commands,
which activate the row, restore the charge, and precharge the bank.
In DDR3 DRAM, refresh is typically performed on each row every 
\SI{64}{\milli\second}\ch{~\cite{ddr3.jedec12, liu.isca2012}}.
\chI{More detail about DRAM refresh can be found in our recent
works\chII{~\cite{Liu, liu.isca2012, Patel, kim.hpca18, Qureshi, Chang,
Khan, PARBOR, Wilkerson, khan.micro17}}.}


\section{Experimental Methodology}
\label{sec:char}

\ch{To develop a thorough understanding of the factors that affect \chII{DRAM} power consumption,
we} perform \rev{an} extensive experimental characterization and analysis of the power
consumption of real modern DRAM chips.
Each operation described in Section~\ref{sec:bkgd:operations} consumes
a different amount of current.  We can directly correlate current to power and
energy in DRAM, as (1)~DRAM operates at a constant voltage in modern systems;
and (2)~DRAM operations take a fixed amount of time to complete, which
is dictated by a series of \emph{timing parameters} provided by DRAM vendors
for each model.  Therefore, we provide current measurements in our 
characterization.

In this section, we \ch{describe} our methodology for measuring the current 
consumed by real DRAM modules.  In Section~\ref{sec:idd}, we show how real-world
current measurements differ significantly from the vendor-specified values that form the basis of
existing DRAM power models.  In Sections~\ref{sec:datadep} and \ref{sec:var}, we study 
several factors that existing power models neglect to account for, which
significantly affect DRAM current consumption.
\ch{In Section~\ref{sec:gen}, we show current consumption trends over several
generations of DRAM.}
We use our measurements to develop VAMPIRE, a new DRAM power model, in Section~\ref{sec:model}.
We plan to open-source our power model, along with all of our raw measurement 
data~\cite{vampire.github}.

\subsection{Current Measurement Infrastructure}
\label{sec:char:meth}

Collecting real power measurements from DRAM has historically been a 
challenging problem, \ch{because} in a real system, we do not have the ability to
determine or control the sequence of commands that are sent to DRAM,
making it difficult to correlate commands with measured power.
To work around these obstacles, we construct a custom FPGA-based infrastructure 
that allows us to 
(1)~precisely control the commands \chI{that are} issued to \chII{the DRAM chips}, and
(2)~accurately measure the \ch{current drawn} only by the module under test.

\rev{Our infrastructure makes use of a significantly-modified version of}
SoftMC~\cite{hassan.hpca17, softmc.repo}, an open-source
\chI{programmable memory control infrastructure}, 
and allows us to transparently send \chIII{customized sequences of} commands \chI{to DRAM \chII{chips}}
in order to \emph{reliably} measure current. 
One of our major modifications adds support to loop continuously
over a fixed set of DRAM commands, \rev{which the base SoftMC code does not
currently support}.  We do this because 
even high-end current measuring equipment can read the average current \ch{only} on
the order of \chI{every} hundreds of microseconds\ch{~\cite{keysight34134a.manual}}, 
whereas DRAM commands take on the order of tens of nanoseconds.
With our command loop support, we repeatedly perform the same microbenchmark
of DRAM commands back-to-back, providing us with enough time
to accurately measure the current.
Our looping functionality ensures that required periodic maintenance 
operations such as ZQ synchronization~\cite{xilinx_ar_36719} are issued
correctly to the DRAM \chII{chips}.
As these maintenance operations can alter the state of the DRAM row buffer,
we issue them only between loop iterations.
We guarantee that maintenance operations do not take more than 0.3\% of
the total microbenchmark execution time, and thus have a negligible impact on 
our current measurements.
Another of our major modifications adds support for power-down modes, which are 
an important technique \chI{employed in modern DRAM chips} to reduce idle power, but are not
supported by the base SoftMC code.
\rev{This requires us to develop new API calls and DRAM commands to
start and stop the power-down modes.
We plan to incorporate these modifications into the open-source release of
SoftMC~\cite{softmc.repo}.}

\begin{figure}[h]
    \centering
    \includegraphics[width=\basewidth, trim = 80 115 80 10, clip]{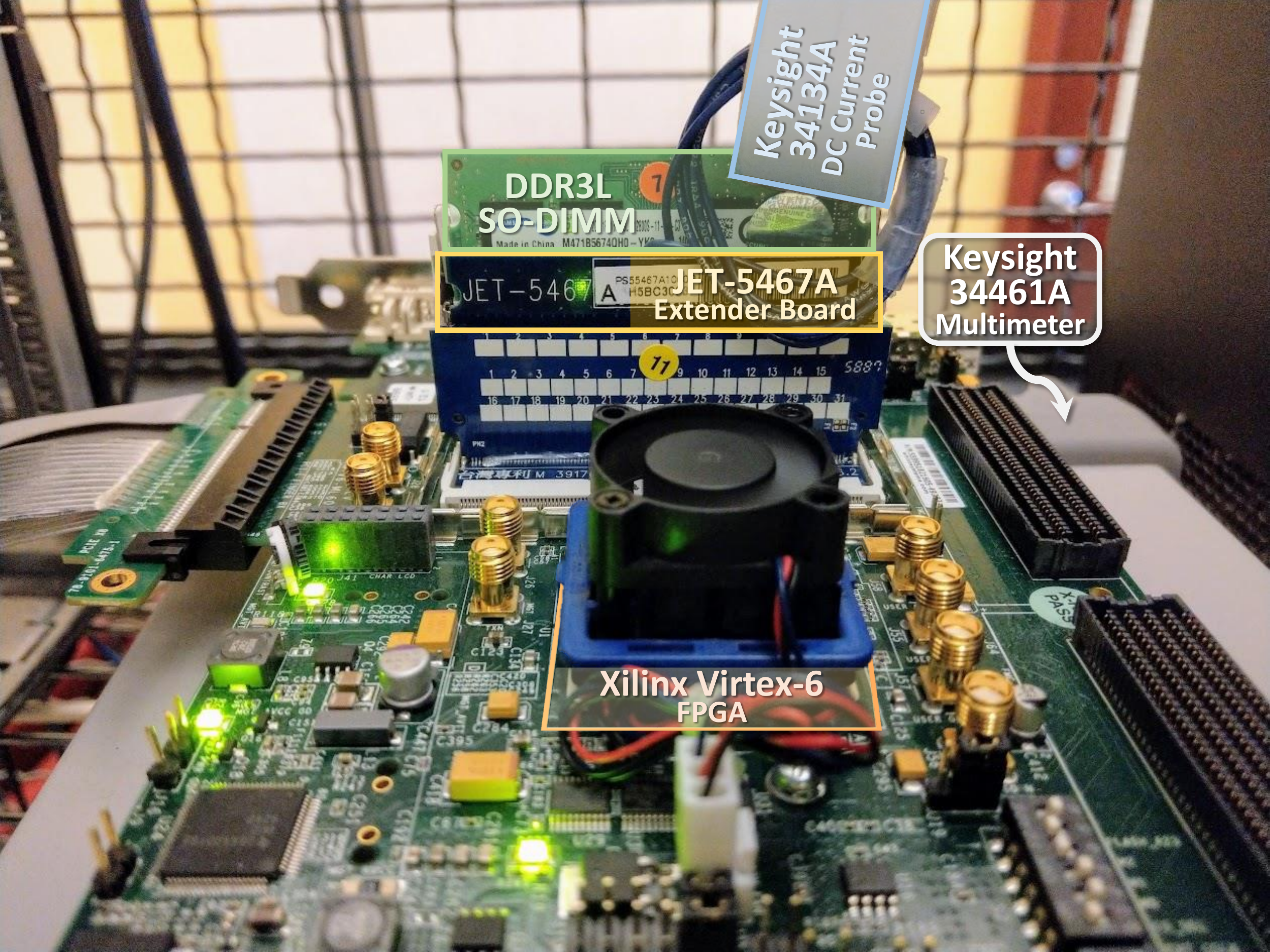}%
    \caption{\ch{Our experimental} infrastructure connected to an FPGA to measure DRAM current.}%
    \label{fig:infrastructure}
\end{figure}

Figure~\ref{fig:infrastructure} shows a photo of the current measurement
hardware used for one test setup in our infrastructure,
\rev{which extends upon the base infrastructure used for SoftMC~\cite{hassan.hpca17}.}
We program SoftMC on a Xilinx ML605~\cite{ml605.manual}, 
\ch{a Virtex-6\chI{~\cite{virtex6.website}} FPGA}
board, which is connected to a host PC and contains \chI{an}  
SO-DIMM (small outline dual in-line memory module)~\cite{sodimm.jedec14} socket.
To measure the current consumed by each DRAM module that we test,
we attach a module to a JET-5467A current-sensing extender 
board\ch{~\cite{jet5467a}}.  We remove the shunt resistor provided on the extender, and add in a 5-coil
wire.  We then insert the coil into a Keysight 34134A
high-precision DC current probe\ch{~\cite{keysight34134a.manual}}, 
which is coupled to a Keysight 34461A 
high-precision multimeter\ch{~\cite{keysight34461a.manual}}.
The current-sensing extender is then inserted into the 
SO-DIMM socket \ch{on the FPGA board}.
\rev{To validate the accuracy of our infrastructure, we
(1)~use independent power supplies to confirm the accuracy of the current
measurements \ch{that are} read from each DC current probe,
(2)~\chI{perform electrical connectivity tests to} verify against the 
DDR3 SO-DIMM standard~\cite{sodimm.jedec14} that
all power pins on our tested DRAM modules are connected through the
\ch{extender board's coiled wire,} and
(3)~read back data from the DRAM modules to verify that each FPGA
\chI{sends} the correct DRAM commands to the module that is attached to the
FPGA board.}

We write custom DRAM command microbenchmarks to perform each 
\ch{of our tests (see Sections~\ref{sec:idd} through \ref{sec:gen})}, 
controlling \chI{three factors:
(1)~}the command sequence issued to DRAM, 
\chI{(2)~}the data \ch{that is} read or written, and 
\chI{(3)~}the latency of each command.
We execute one microbenchmark at a time by launching the microbenchmark on the
host PC, which sends the DRAM command loop to the SoftMC controller \chI{on} the
FPGA, and we connect the multimeter to the host to sample current 
measurements while the microbenchmark \chI{iterates over the loop}.  
\rev{
\ch{For each test that we execute, we perform ten runs of the test per
DRAM module.}
During each run, we sample the current while the microbenchmark \chI{performs
the command loop}, ensuring that we capture at least ten \ch{current} samples for each run,
and determine the current reading for the overall run by averaging each
sample together.  \ch{We then average the current measured over the 
ten runs to represent the current consumed by the DRAM module under test}.
\chI{In other words, we collect at least 100~samples per test for each module.}}

Unless otherwise
stated, all DRAM modules are tested at an ambient temperature of \SI{20(1)}{\celsius}. 
We examine the effects of high ambient temperature (\SI{70(1)}{\celsius}) using a
custom-build heat chamber, where we use a controller connected to a heater to 
regulate the temperature\chI{~\cite{chang.sigmetrics2016, chang.pomacs2017, Patel, 
lee.hpca15, Liu, kim.hpca18}}.  We discuss high ambient temperature results in
Section~\ref{sec:var:temp}.

\subsection{DRAM Modules Tested}
\label{sec:char:dimms}

We characterize power \chI{consumption}
on \dimmcnt DDR3L~\cite{ddr3l.jedec13} DRAM modules, which 
(1)~are comprised of \chipcnt~DRAM chips, and 
(2)~use the SO-DIMM form factor~\cite{sodimm.jedec14}.  
Table~\ref{tbl:dimms} shows a summary of the modules that we test.
These modules are sourced from three major
DRAM vendors.  Each of our modules contains a single rank, and has a \SI{2}{\giga\byte}
capacity.  The modules support a channel frequency of up to 
\SI{1600}{\mega\transfer\per\second},\footnote{\ch{In double data rate (DDR) DRAM, the 
channel frequency is typically expressed as \emph{megatransfers per second} (MT/s),
where one transfer sends a single \chI{64-bit} burst of data across the channel.
DDR DRAM sends \emph{two} transfers per clock cycle (one on the positive clock
edge, and another on the negative clock edge).  This means that for a DRAM with
a channel frequency of \SI{1600}{\mega\transfer\per\second}, the channel uses
an \SI{800}{\mega\hertz} clock~\cite{ddr3.jedec12}.}} but all
of our tests are conducted at \SI{800}{\mega\transfer\per\second}, due to limitations on the maximum 
frequency at which our FPGA can operate.  In order to anonymize the vendors,
we simply refer to them as Vendors~A, B, and C in this paper.  
\rev{Many of these modules are the same ones used in our prior 
work~\cite{chang.pomacs2017, voltron.github}, where we characterize the 
latency and supply voltage behavior, but \ch{\emph{not}} the measured power consumption,
of each module.}
We supply the
modules with their nominal operating voltage of \SI{1.35}{\volt}~\cite{ddr3l.jedec13}.

\begin{table}[h]
  \centering
  \small
  \caption{Selected properties of the tested DDR3L DRAM modules.}%
  \label{tbl:dimms}%
  \vspace{-7pt}
    \setlength{\tabcolsep}{.45em}
    \begin{tabular}{cccccc}
        \toprule
        \multirow{2}{*}{\bf Vendor} & {\bf Total Number} & \textbf{Timing} (ns) &
        {\bf Assembly} & {\bf Supply} & {\bf Max. Channel} \\
                       & \bf{of Chips} & (t\textsubscript{RCD}/t\textsubscript{RP}/t\textsubscript{RAS}) & {\bf Year} & {\bf Voltage} & {\bf Frequency} (MT/s)  \\
        \midrule
        A (14 SO-DIMMs) & 56 & 13.75/13.75/35 & 2015-16 & \SI{1.35}{\volt} & 1600 \\
        B (13 SO-DIMMs) & 52 & 13.75/13.75/35 & 2014-15 & \SI{1.35}{\volt} & 1600 \\
        C (23 SO-DIMMs) & 112 & 13.75/13.75/35 & 2015 & \SI{1.35}{\volt} & 1600 \\
        \bottomrule
    \end{tabular}%
  \vspace{-5pt}%
\end{table}

We note that while DDR4 modules are available on the market, there is poor 
experimental infrastructure support available for such modules \ch{today;}
hence our use of DDR3L modules in our characterization.
In particular, at the time of writing, no \ch{tool equivalent} to SoftMC has support
\ch{for issuing} test routines to DDR4 DRAM at a command-level granularity, and it is
very difficult and time-consuming to develop a new current measurement
infrastructure for DDR4 modules (based on both our prior experience\chII{~\cite{Liu, 
lee.hpca15, chang.sigmetrics2016, chang.pomacs2017, Patel, PARBOR, Wilkerson,
lee.sigmetrics17, hassan.hpca17, khan.micro17, kim.isca14, luo.sigmetrics18,
cai.procieee17, cai.hpca17, cai.date12, cai.date13, cai.iccd12, cai.iccd13, 
cai.sigmetrics14, cai.hpca15, cai.dsn15, luo.hpca18, kim.hpca18, cai.book18,
Qureshi, cai.itj13}} and on
other prior work on building DRAM current measurement 
\chI{infrastructures}\ch{~\cite{Matthias, mathew.date18}}).
However, due to the large number of similarities between the design of
DDR3 memory~\cite{micron.ddr3.design} and DDR4 memory~\cite{micron.ddr4.design},
we believe that the general trends observed in our characterization should
apply to DDR4 DRAM as well.  We leave the exact \chI{adaptation} of the power models
that we develop to DDR4 modules \chI{\chII{and an investigation} of the differences
between DDR3 power consumption and the power consumption of other DDRx
DRAM architectures to} future work.


\section{Measuring Real IDD Current}
\label{sec:idd}

Most existing DRAM power models are based on \emph{\idd values}, which
are a series of \rev{current measurement tests~\cite{ddr3.jedec12} that are
standardized by the JEDEC Solid State Technology Association
(commonly referred to as JEDEC).
DRAM vendors conduct these current measurement tests for each \ch{DRAM part} that
they manufacture, and publish the \chII{measured values} in \ch{part}-specific datasheets.}
In order to perform these measurements,
a specific series of commands is \chI{executed continuously in a loop}, 
and average current measurements are taken while the loop executes.
We start our characterization by measuring the \ch{\emph{actual}} current consumed by the
modules listed in Table~\ref{tbl:dimms},
\rev{and present a summary of the \ch{\emph{actual}} measurements in this section}.

\rev{Recall from Section~\ref{sec:char} that due to limitations in the
maximum frequency attainable on an FPGA, our infrastructure can
operate the DRAM modules using a channel frequency of only
800~MT/s.}
While each vendor provides \idd values for multiple channel frequencies in
their datasheets, they do \ch{\emph{not}} provide \idd values for 
\ch{\SI{800}{\mega\transfer\per\second}, the} channel
frequency employed by our FPGA infrastructure.  
\rev{However, we can take advantage of the following relationship
to extrapolate the expected \idd values at \SI{800}{\mega\transfer\per\second}:
\begin{equation}
\label{eq:power}
P = IV \propto V^2 f
\end{equation}
where $P$ is power, $I$ is the current, $V$ is the voltage, and $f$ is the
frequency.  Since the operating voltage is constant at \SI{1.35}{\volt}, 
a linear relationship exists between $I$ and $f$.}
As a result,
we perform regression using linear least 
squares\ch{~\cite{legendre.book1805, gauss.book1809}} to fit the datasheet values 
to a quadratic model, and use this model to extrapolate the estimated \idd values
at \SI{800}{\mega\transfer\per\second}. 
We find that the datasheet values fit well to the \rev{linear model determined
through regression}.
For Vendor C, which has the worst fit out of our three vendors, 
the lowest $R^2$ value (which represents the goodness of fit)
across all IDD values is 0.9783.
Therefore, we conclude that our estimated \idd values at
\SI{800}{\mega\transfer\per\second} are accurate.

\chI{There are five types of \idd current values that we measure:
(1)~idle: \idd[2N], \idd[3N];
(2)~activate and precharge: \idd[0], \idd[1];
(3)~read and write: \idd[4R], \idd[4W], \idd[7];
(4)~refresh: \idd[5B]; and
(5)~power-down mode: \idd[2P1].}

\subsection{Idle (IDD2N/IDD3N)}
\label{sec:idd:2n-3n}
We start by measuring the idle (i.e., standby) current.  JEDEC
defines two idle current measurement loops:
(1)~\idd[2N], which measures the current consumed by the module when \emph{no}
banks have a row activated; and
(2)~\idd[3N], which measures the current consumed by the module when \emph{all}
banks have a row activated.

\begin{figure}[h]
  \centering
  \includegraphics[width=0.55\basewidth, trim=72 228 58 233, clip]{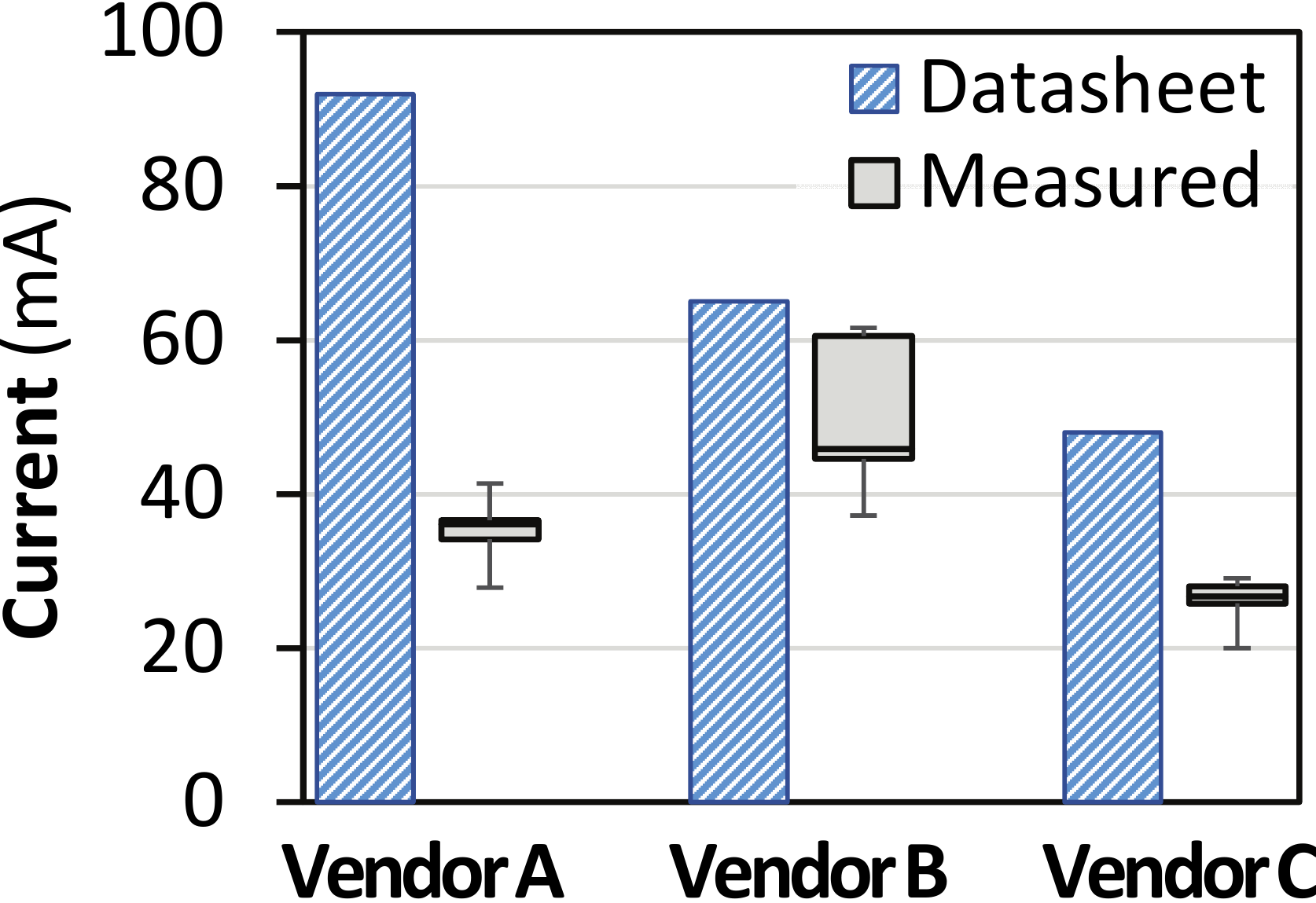}%
  \figspace%
  \includegraphics[width=0.42\basewidth, trim=55 168 57 172, clip]{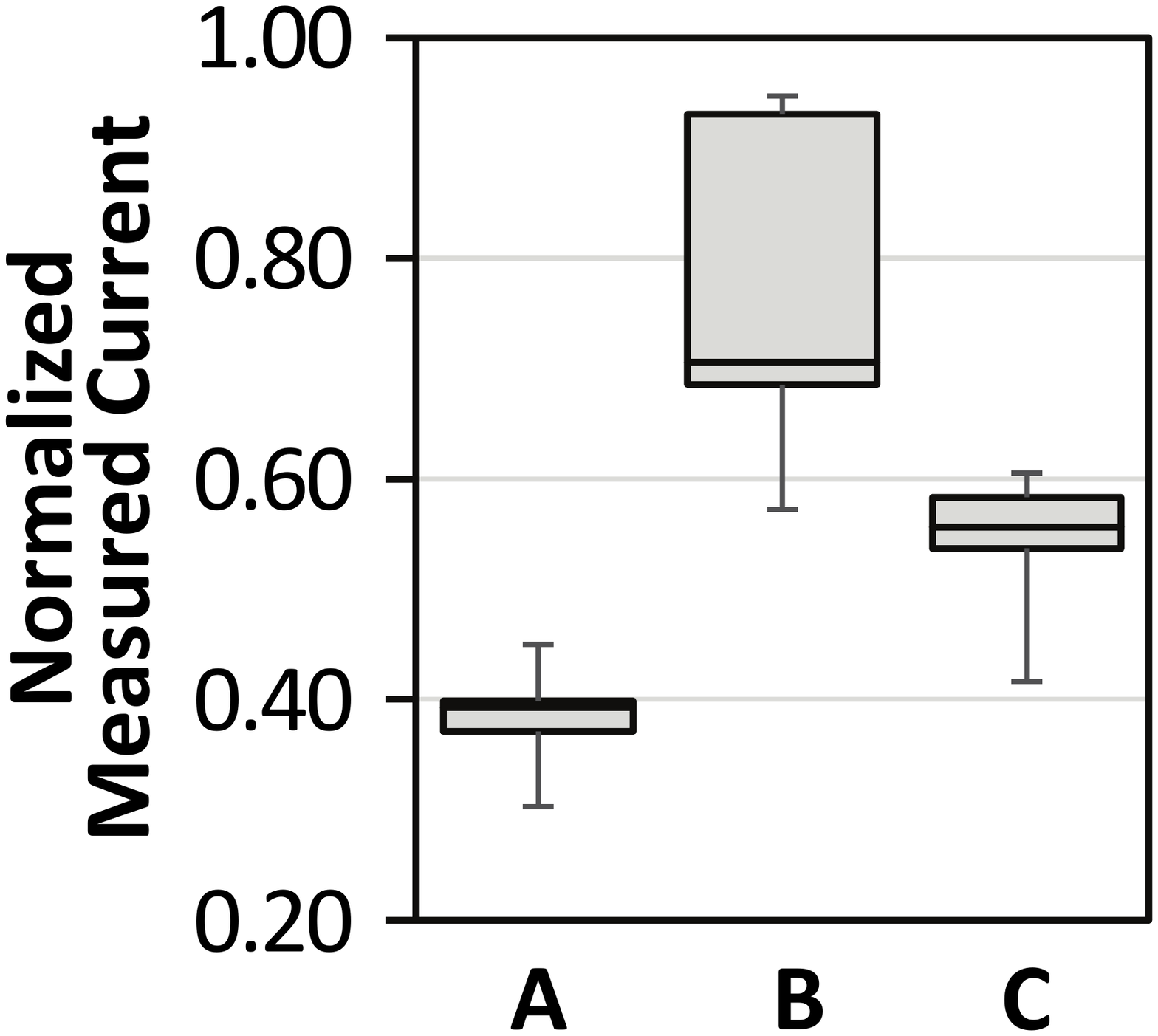}
  \caption{\idd[2N] current measurements (left),
    and current normalized to datasheet value (right).}%
  \label{fig:idd2n}
\end{figure}

Figure~\ref{fig:idd2n} shows the average current measured during the \idd[2N] loop.
We use \emph{box plots} to show the distribution across all modules from each 
vendor.  Each box illustrates the quartiles of the distribution, and the whiskers
illustrate the minimum and maximum values.
We make two key observations from this data.
First, \emph{there is non-trivial variation in the amount of current consumed
from module to module for the same DRAM \chI{vendor}}.  The amount of variation is 
different for each vendor, and the range normalized to the datasheet current
varies from 14.7\% for Vendor~A to 37.5\% for Vendor~B.  As the architecture
of the module remains the same for modules from the same \chI{vendor
(because we study a single part per vendor)}, we conclude that
these differences are \chI{a} result of manufacturing process variation.
Second, \emph{the measured currents are significantly lower than the datasheet
values}.  As we can see in the figure, DRAM vendors leave a \emph{guardband} 
(i.e., margin) \ch{in the reported \idd values}.  
The capacitors used for DRAM cells are very tall and narrow trenches~\cite{muller.iedm96},
which improves the chip density. Unfortunately, the very high \emph{aspect ratio}
(i.e., height over width) of the cells \ch{(e.g., $>70$ for modern DRAM
chips~\cite{hong.iedm10})} increases the difficulty of DRAM
lithography, and can result in \ch{significant} 
process variation. Vendors use the guardband to account
for the expected worst-case process variation \ch{in \idd values}.
Our average \idd[2N] measurement
\chII{is 38.3\%, 76.6\%, and \chIII{54.9\%} of the specified \idd[2N] current} for
Vendors~A, B, and C, respectively.

\begin{figure}[h]
  \centering
  \includegraphics[width=0.55\basewidth, trim=72 228 58 233, clip]{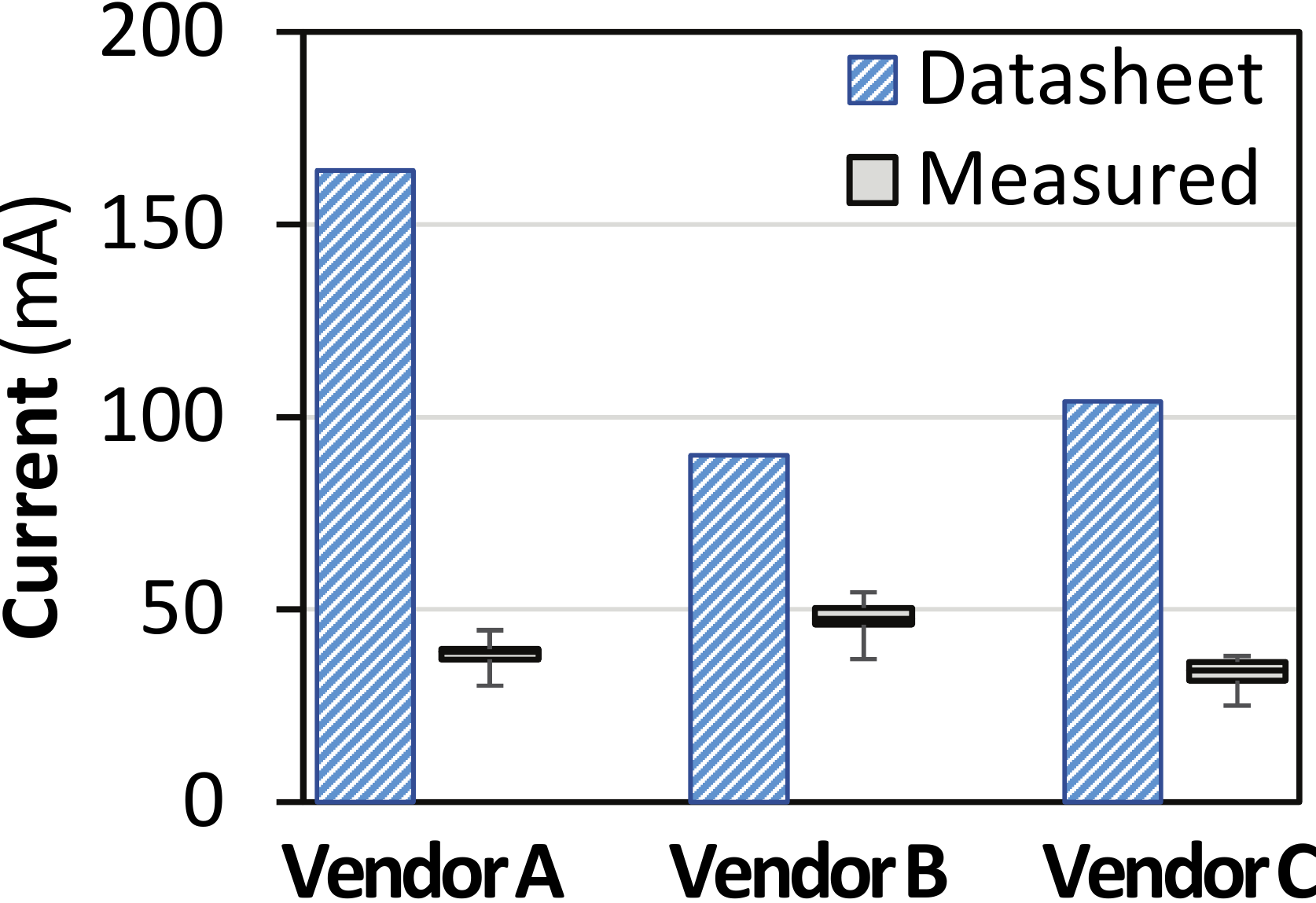}%
  \figspace%
  \includegraphics[width=0.42\basewidth, trim=55 168 57 172, clip]{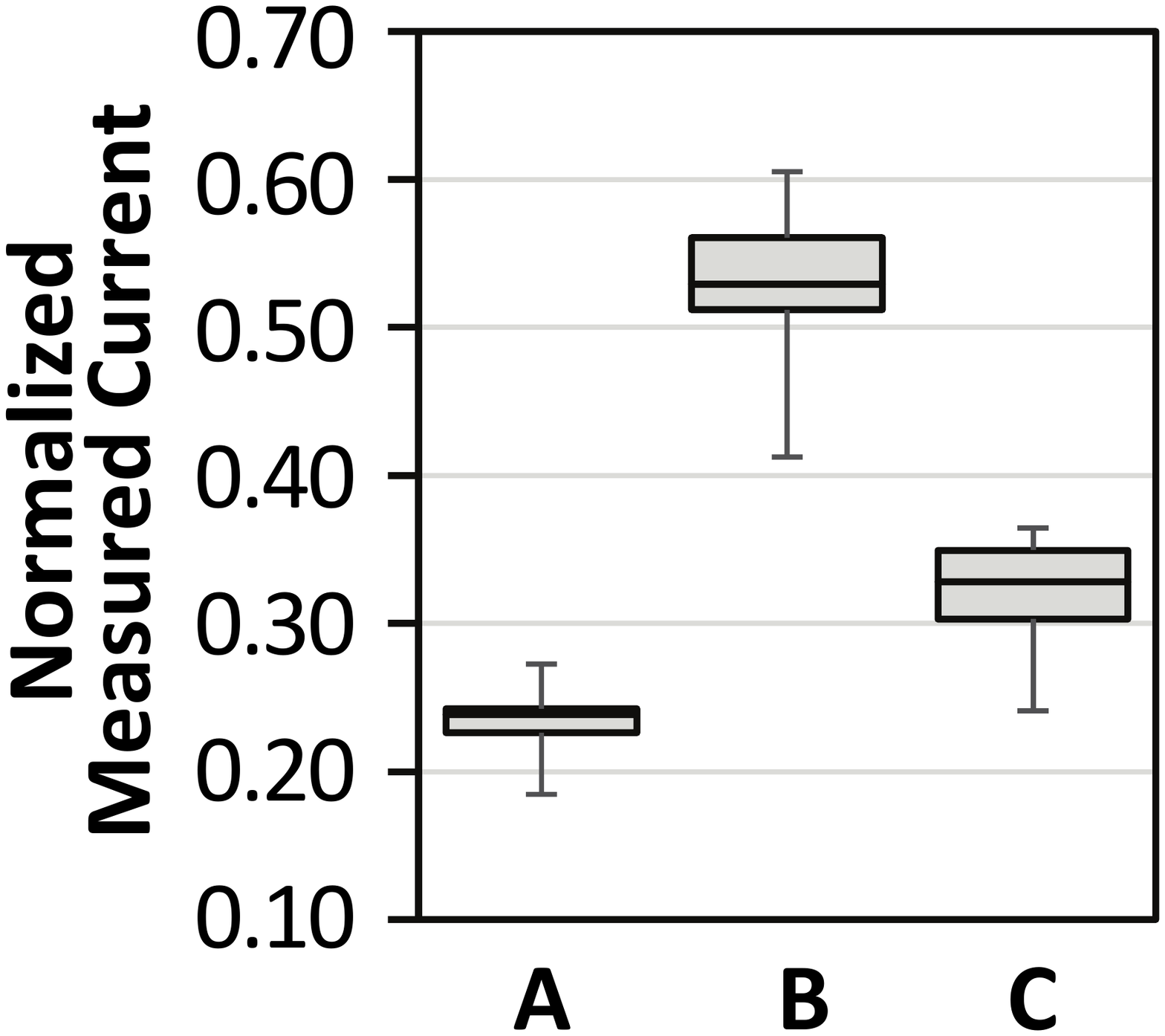}
  \caption{\idd[3N] current measurements (left),
    and current normalized to datasheet value (right).}%
  \label{fig:idd3n}
\end{figure}

We see the same trends for the current measured during the \idd[3N] loop,
as shown in Figure~\ref{fig:idd3n}.
\ch{The average measured} \idd[3N] current
\chII{is 23.4\%, 53.2\%, and \chIII{33.4\%} of the specified \idd[3N] current}
for Vendors~A, B, and C, respectively.
\ch{We observe that the full \emph{normalized} range of the measured current
\chI{(i.e., the difference \chII{in current} between the highest-current 
DRAM module and the lowest-current module)}
\chII{is 8.8\%, 19.3\%, and \chIII{12.4\%} of the specified current}, respectively for the three vendors.
The normalized range represents how much variation in current exists across
the modules that we tested for each vendor.}

\chI{We conclude that \chII{(1)~}the actual power consumed by real DRAM modules in the
idle state is much lower than the \idd[2N]/\idd[3N] values provided by the 
vendors, \chII{and
(2)~there is significant variation of these current values across parts
manufactured by a given vendor}.}

\subsection{Activate and Precharge (IDD0/IDD1)}
\label{sec:idd:0-1}
Next, we study the amount of current consumed during activate and precharge
operations.  Unfortunately, it is not possible to measure activation and precharge
current \chI{\emph{independently}} in real DRAM modules, as a second activation \emph{cannot}
take place before an already-activated row is precharged.  JEDEC defines two
measurement loops for activation and precharge:
(1)~\idd[0], which performs successive activate and precharge operations as
quickly as possible without violating DRAM timing parameters; and
(2)~\idd[1], which performs successive \ch{\{\emph{activate}, \emph{read}, 
\emph{precharge}\}} operations \chI{in a similar manner}.

\begin{figure}[h]
  \centering
  \includegraphics[width=0.55\basewidth, trim=72 228 58 233, clip]{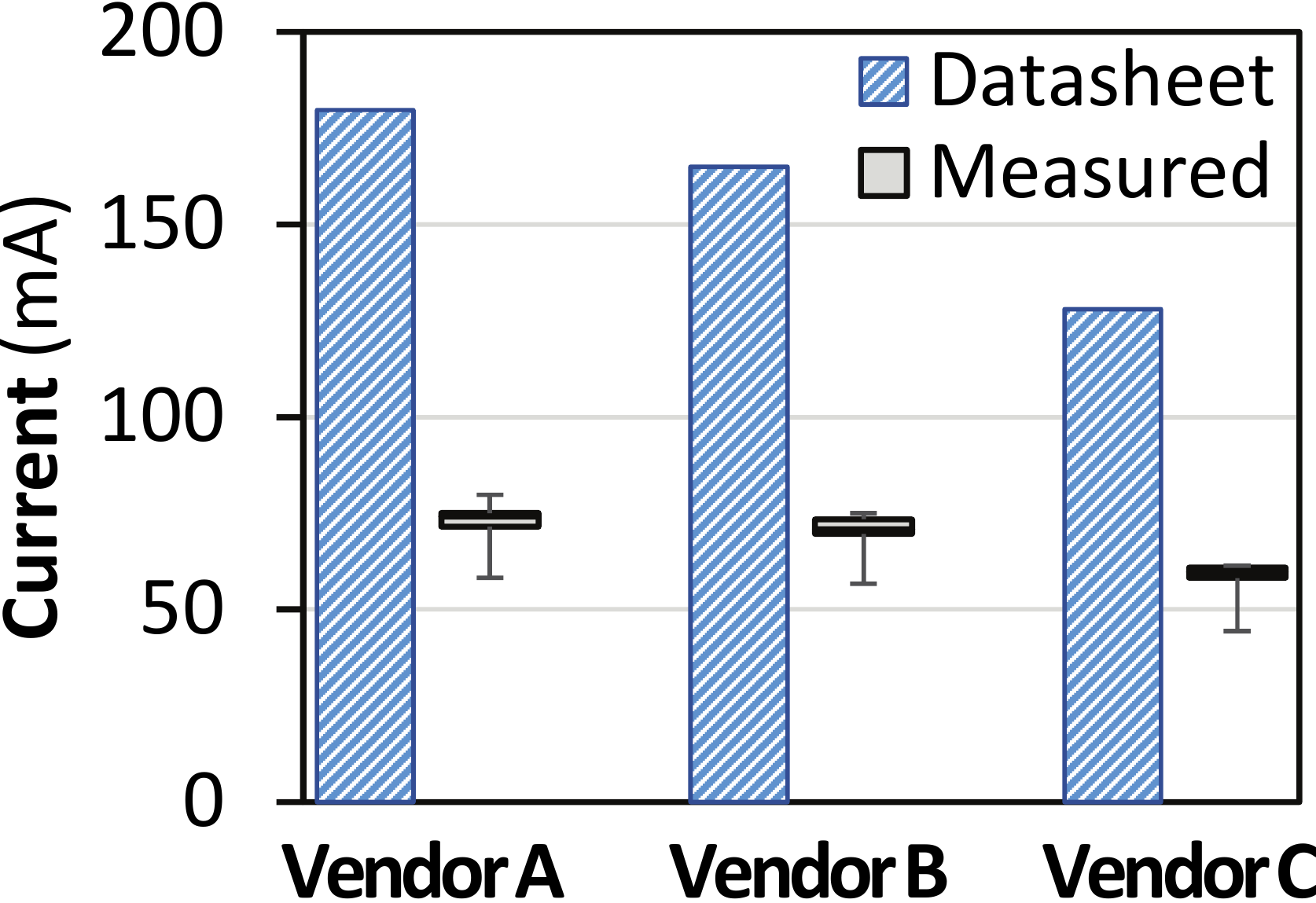}%
  \figspace%
  \includegraphics[width=0.42\basewidth, trim=55 168 57 172, clip]{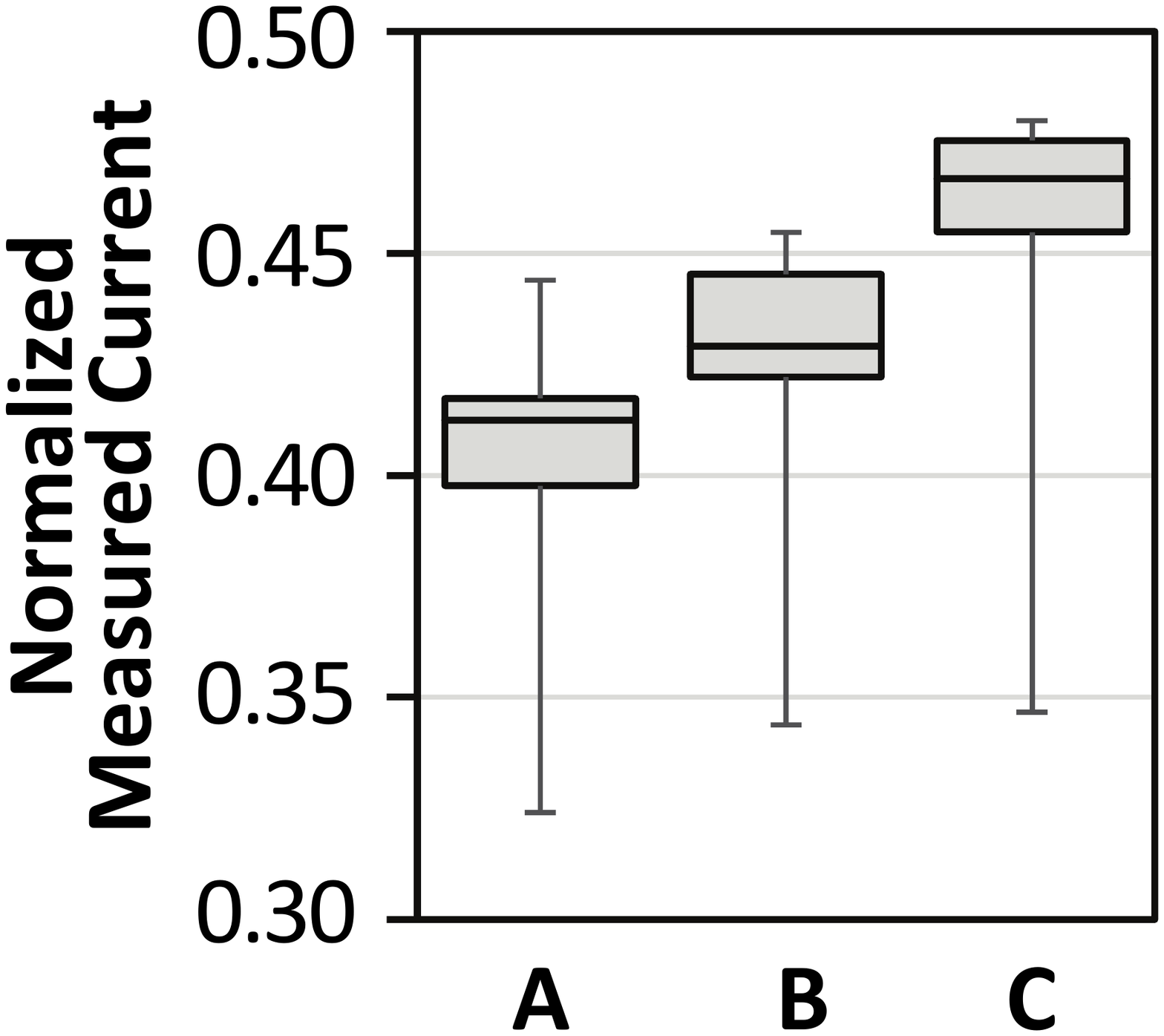}
  \caption{\idd[0] current measurements (left),
    and current normalized to datasheet value (right).}%
  \label{fig:idd0}
\end{figure}

\begin{figure}[h]
  \centering
  \includegraphics[width=0.55\basewidth, trim=72 228 58 233, clip]{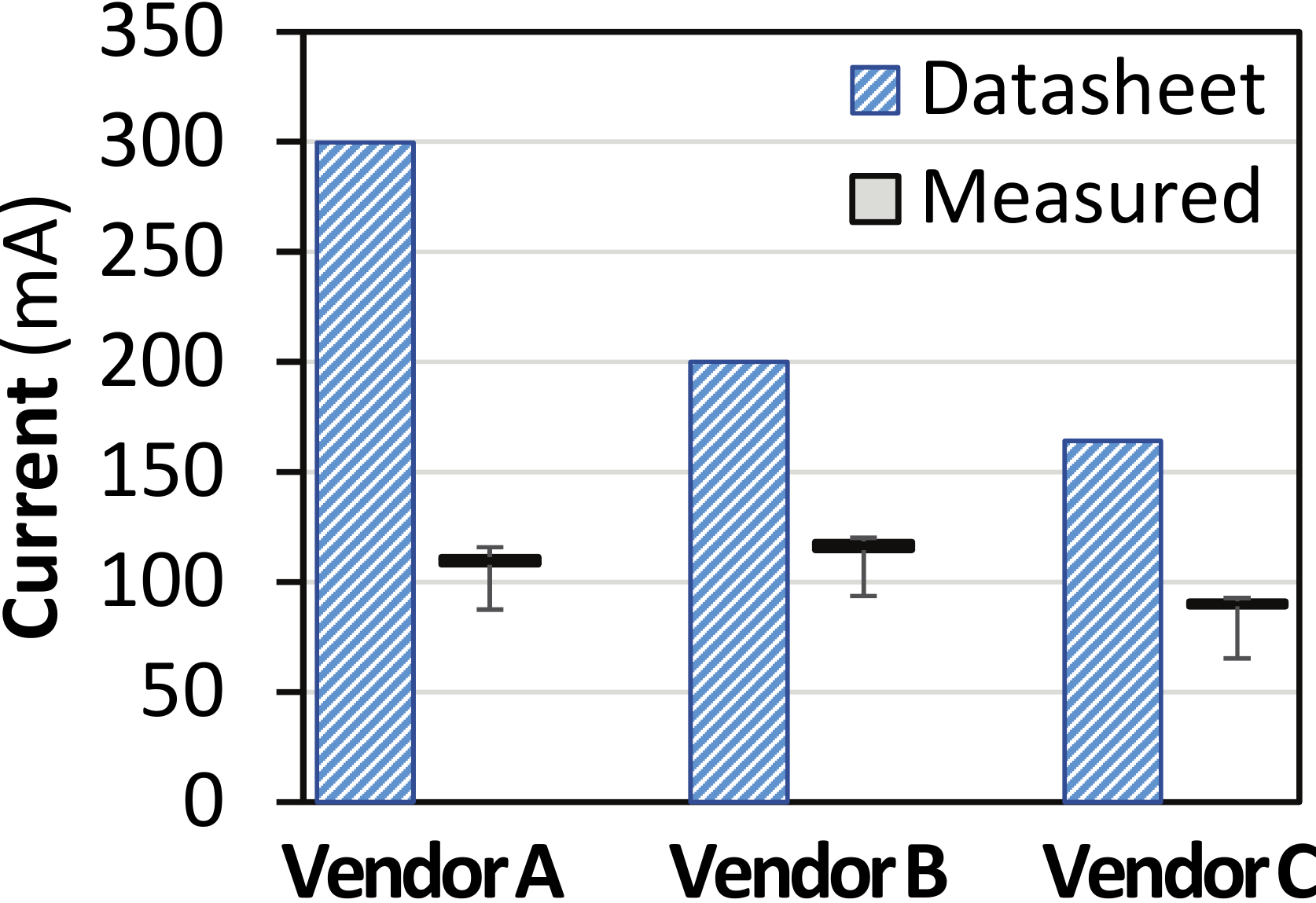}%
  \figspace%
  \includegraphics[width=0.42\basewidth, trim=55 168 57 172, clip]{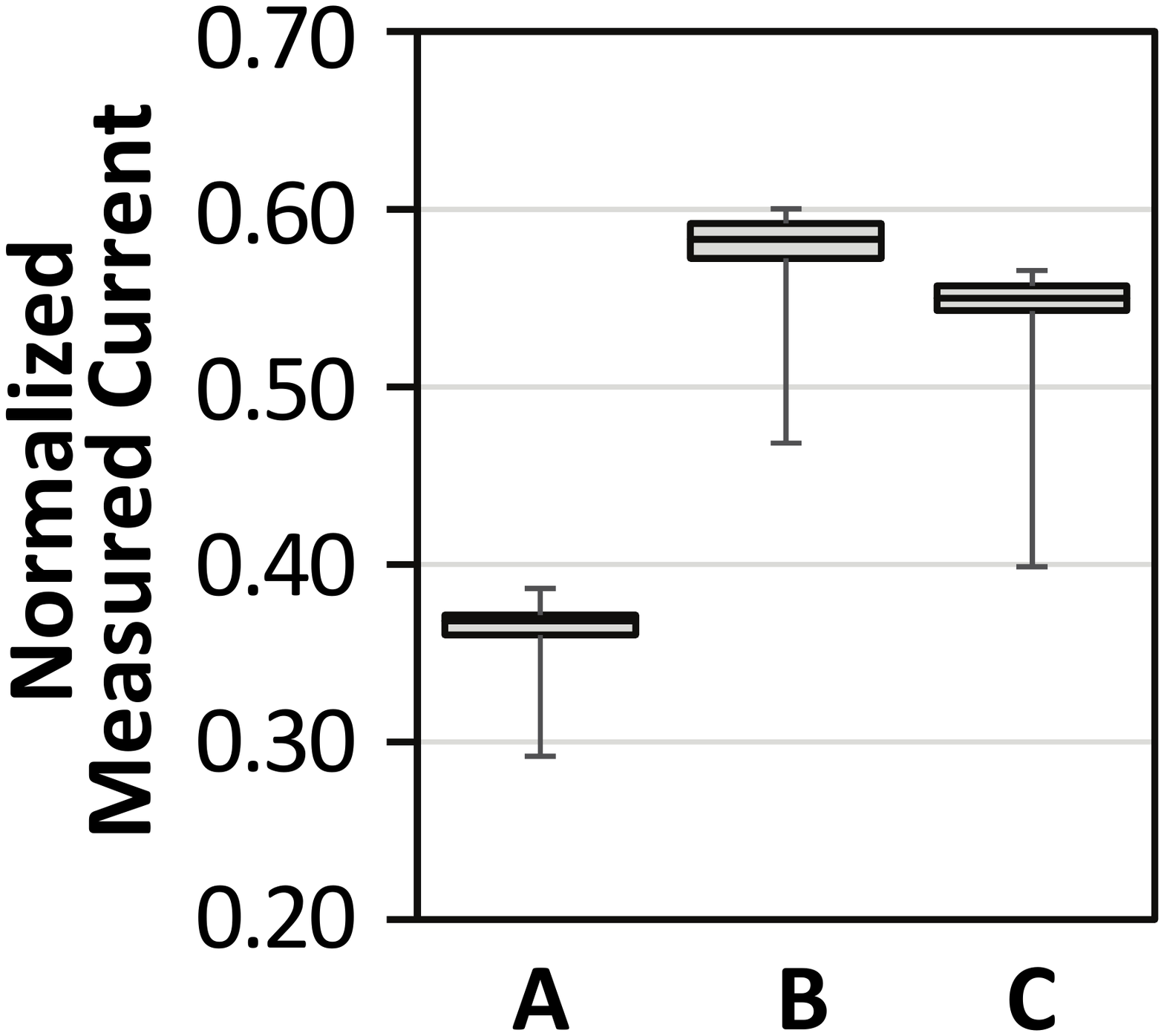}
  \caption{\idd[1] current measurements (left),
    and current normalized to datasheet value (right).}%
  \label{fig:idd1}
\end{figure}

Figure~\ref{fig:idd0} shows the average current measured during the \idd[0]
loop.
We make two key observations.
First, we again observe a large margin between the datasheet values and our
measurements, and find that the activation and precharge current consumption is
much lower than expected.  Our average \idd[0] measurement
\chII{is 40.2\%, 42.6\%, and \chIII{45.4\%} of the specified \idd[0] current}
for Vendors~A, B, and C, respectively.
Second, we find that the absolute amount of current consumed across all three
models is \ch{somewhat} similar despite the large difference in the datasheet
specification, with average current measurements of
\SI{72.2}{\milli\ampere}, \SI{70.4}{\milli\ampere}, and \chIII{\SI{58.1}{\milli\ampere}} for the three respective vendors.
We note very similar trends for \idd[1], as shown in Figure~\ref{fig:idd1},
with average current measurements of
\SI{107.4}{\milli\ampere}, \SI{114.9}{\milli\ampere}, and \chIII{\SI{87.9}{\milli\ampere}} for the three respective vendors

\chI{We conclude that the actual power consumed by real DRAM modules during the
activate and precharge operations is much lower than the \idd[0]/\idd[1] values 
provided by the vendors.}

\subsection{Read and Write (IDD4R/IDD4W/IDD7)}
\label{sec:idd:4r-4w-7}
\chII{We} study the amount of current consumed during read and write operations.
JEDEC defines three measurement loops for these operations:
(1)~\idd[4R], which performs back-to-back read operations
\ch{to open rows across \emph{all} eight banks};
(2)~\idd[4W], which performs back-to-back write operations
\ch{to open rows across \emph{all} eight banks}; and
(3)~\idd[7], which interleaves \ch{\{\emph{activate}, \emph{read}, 
\emph{auto-precharge}\}} operations across \ch{\emph{all}} eight banks.

\begin{figure}[h]
  \centering
  \includegraphics[width=0.55\basewidth, trim=72 228 58 233, clip]{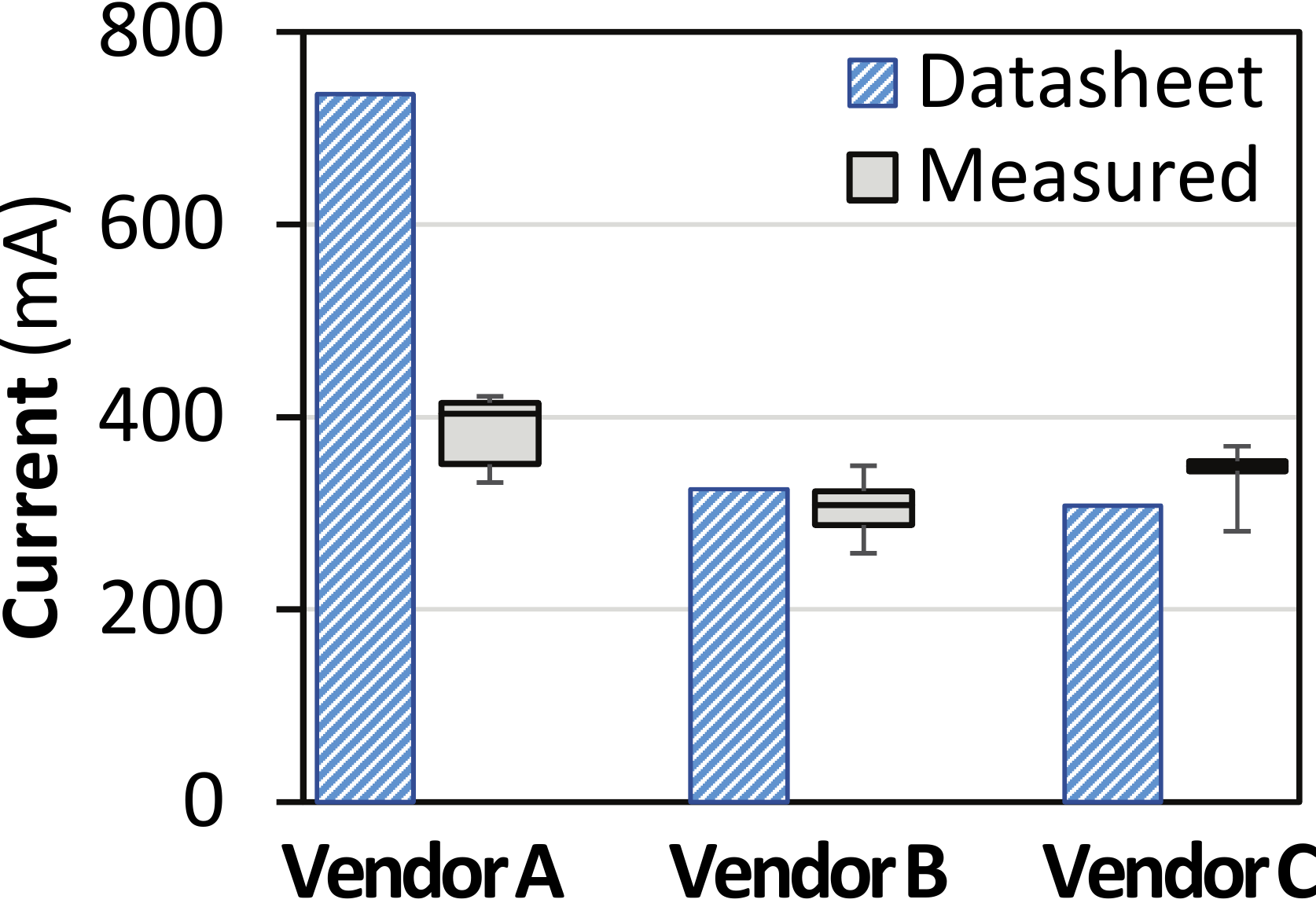}%
  \figspace%
  \includegraphics[width=0.42\basewidth, trim=55 168 57 172, clip]{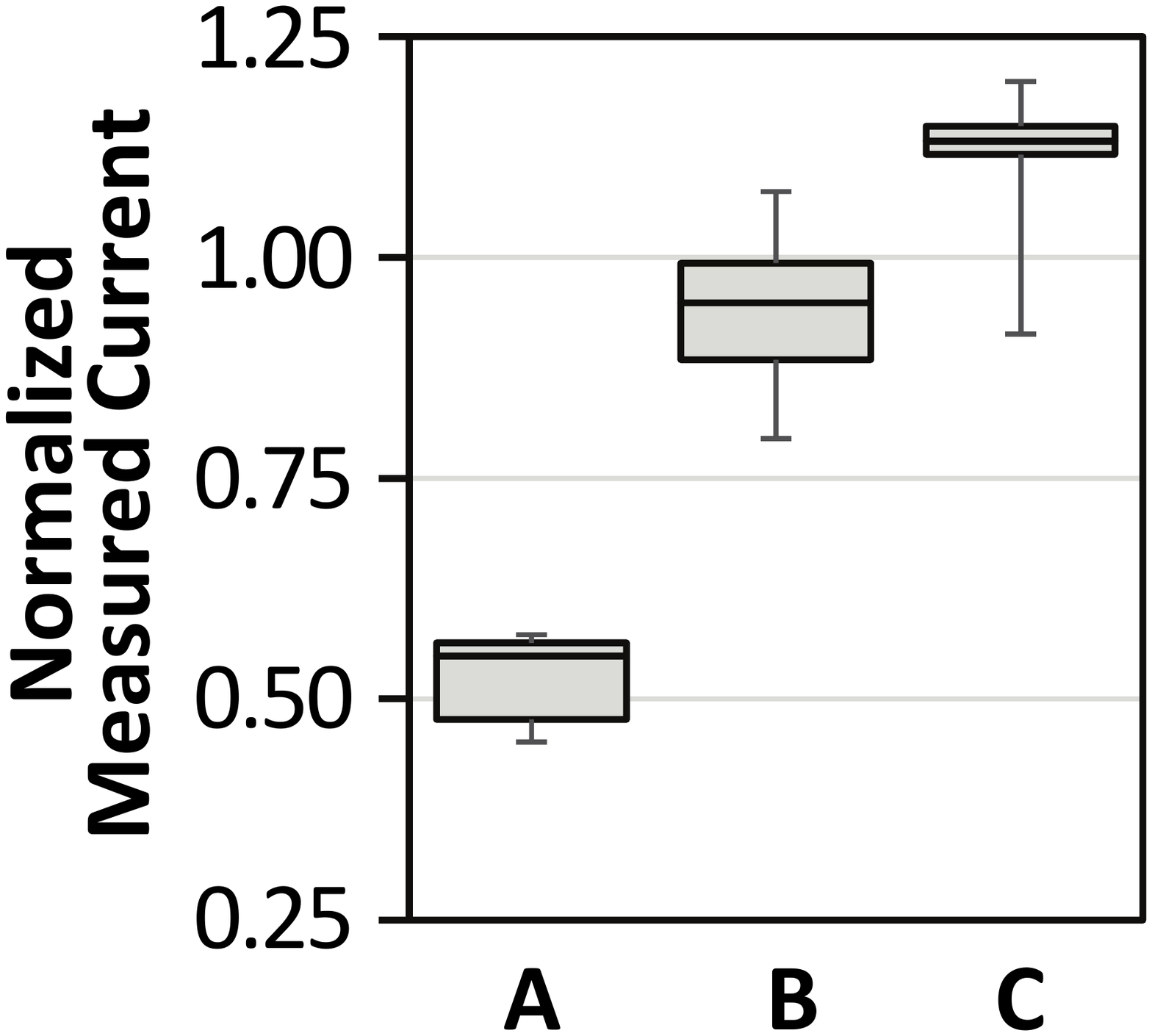}
  \caption{\idd[4R] current measurements (left),
    and current normalized to datasheet value (right).}%
  \label{fig:idd4r}
\end{figure}

Figure~\ref{fig:idd4r} shows the average current measured during the \idd[4R] loop.
\rev{As we observe from the figure, several of our current measurements actually
\emph{exceed} the value specified by vendors in the datasheets.  
In fact, for the 
modules from Vendor~C, the average current measured from the modules exceeds the
datasheet value by \chIII{11.4\%}, with a current of \chIII{\SI{343.5}{\milli\ampere}}.
These measurements \chI{should not be interpreted as a} lack of a margin for \idd[4R]
\chII{or an underestimation by the DRAM vendor}.
Instead, \chII{these measurements} represent a limitation of our FPGA measurement \ch{infrastructure.}
\ch{As part of the read operation, the DRAM module must drive the data values
across the memory channel.  To do so,}
a read operation selects a column from an open row \ch{of each DRAM chip
in the target rank}, and
uses the \emph{peripheral \ch{circuitry}} \ch{inside a DRAM chip}, which is
responsible for performing the external I/O \chI{(see Section~\ref{sec:bkgd:operations})}.  
While vendor specifications \ch{\emph{ignore}} the portion of the current used 
by the I/O driver \chI{in the \idd[4R] value}, 
our measurement infrastructure \chII{\emph{captures}} the \ch{I/O driver current}, 
which can account for a sizable portion of the total measured current.}
\chII{As a result, our measured current includes a portion that is not included 
by the DRAM vendors, causing some of our measured values to be larger than
\chIII{those} reported by vendor datasheets.}

\begin{figure}[h]
  \centering
  \includegraphics[width=0.95\basewidth, trim=62 297 62 300, clip]{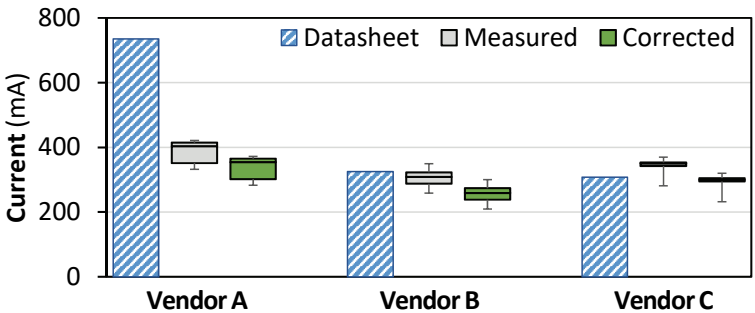}%
  \caption{\ch{\idd[4R] current measurements before and after correction for I/O driver current}.}%
  \label{fig:idd4r-corrected}
\end{figure}

\rev{\ch{We estimate the amount of the current consumed by the I/O driver (see
Section~\ref{sec:datadep:ones}), and \chII{subtract this amount} from our original \chII{IDD4R
measurement,} as shown in Figure~\ref{fig:idd4r-corrected}} \chII{(the \chIII{\emph{Corrected}} bars)}.
After this correction, the average \idd[4R] value
drops \chII{from 52.6\%, 94.7\%, and \chIII{111.4\%} of the specified \idd[4R] current to
45.9\%, 79.5\%, and \chIII{95.4\%}} for Vendors~A, B, and C, respectively.}
\ch{We observe that even with the corrections, the margins provided by Vendors~B and C for
\idd[4R] are much smaller than the margins for the other \idd values that we
measure.  This \chI{may be} because the read operation does \chI{\emph{not}} interact
directly with DRAM cells, which are susceptible to significant manufacturing
process variation, and predominantly makes use of the sense amplifiers,
peripheral logic, and I/O drivers.}

\begin{figure}[h]
  \centering
  \includegraphics[width=0.55\basewidth, trim=72 228 58 233, clip]{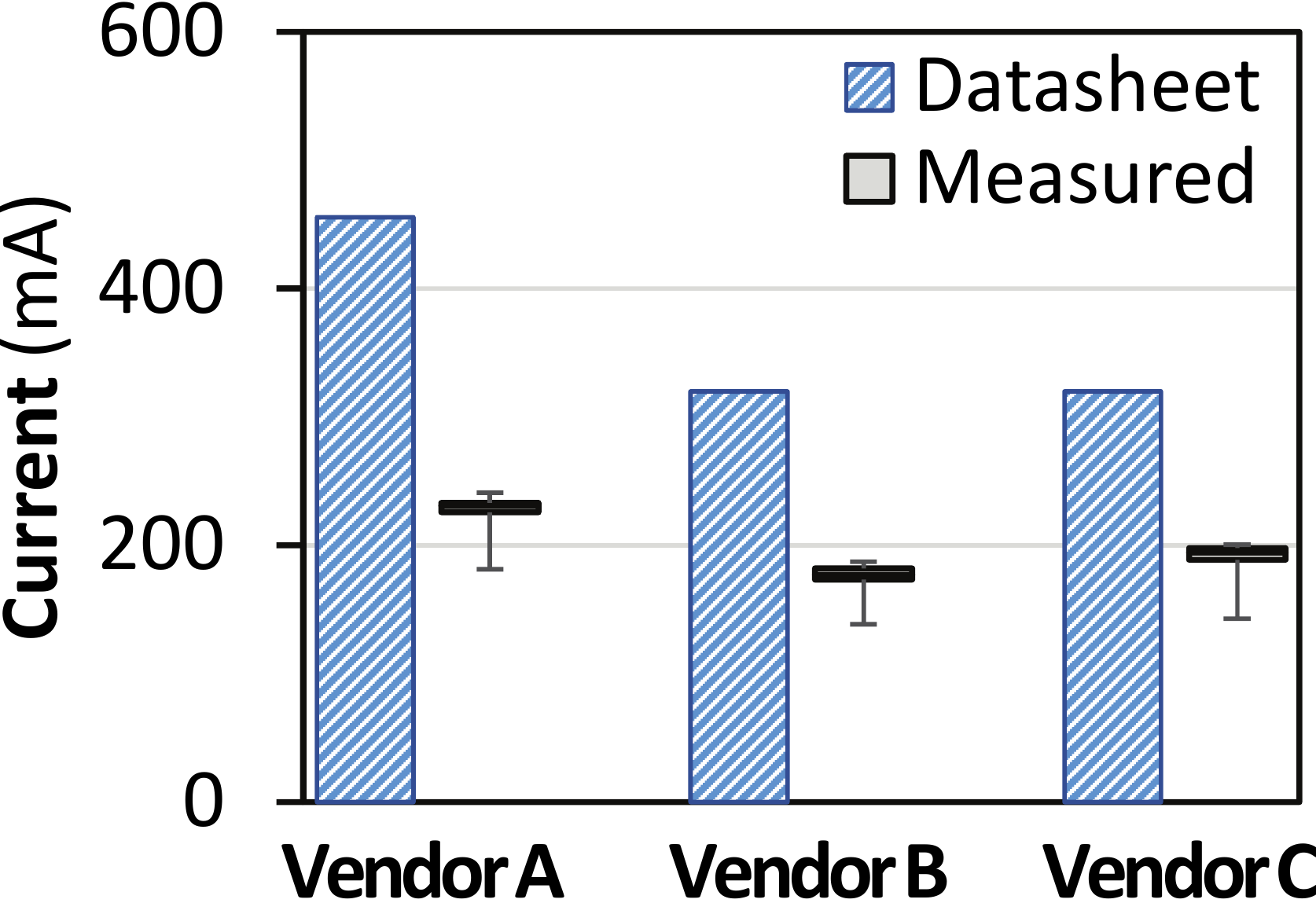}%
  \figspace%
  \includegraphics[width=0.42\basewidth, trim=55 168 57 172, clip]{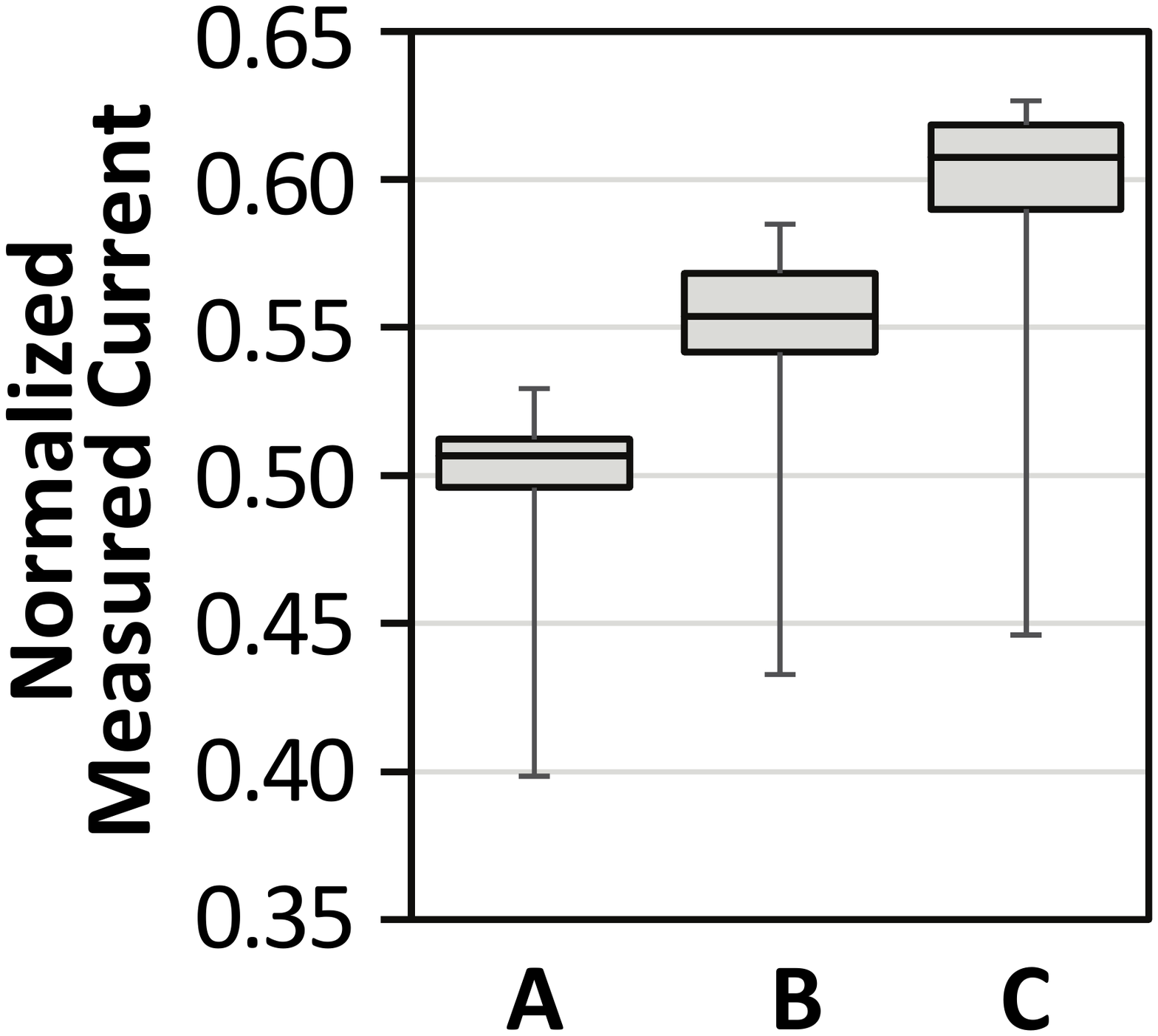}
  \caption{\idd[4W] current measurements (left),
    and current normalized to datasheet value (right).}%
  \label{fig:idd4w}
\end{figure}

\ch{Figure~\ref{fig:idd4w} shows the average current measured during the \idd[4W]
loop.  We observe that unlike the \idd[4R] results, our measurements for
\idd[4W] are much smaller than the datasheet \chIII{values}.  This is due to two reasons.
First, during} a write
operation, the peripheral circuitry within \chII{the DRAM chip does \emph{not}} need to drive current
across the memory channel, instead acting as a current sink.  \ch{Second, unlike} read 
operations, a write operation affects the charge stored within DRAM cells.
\chII{Because the DRAM cells are susceptible to significant manufacturing
process variation, the \idd[4W] numbers that are reported by the vendors 
include a large guardband to account for worst-case DRAM cells that
can consume much higher current than the typical cell.}
\chI{On average, the measured \idd[4W] current
\chII{is 49.1\%, 54.5\%, and \chIII{59.0\%} of the specified \idd[4W] current}
for Vendors~A, B, and C, respectively.}

\begin{figure}[h]
  \centering
  \includegraphics[width=0.55\basewidth, trim=72 228 58 233, clip]{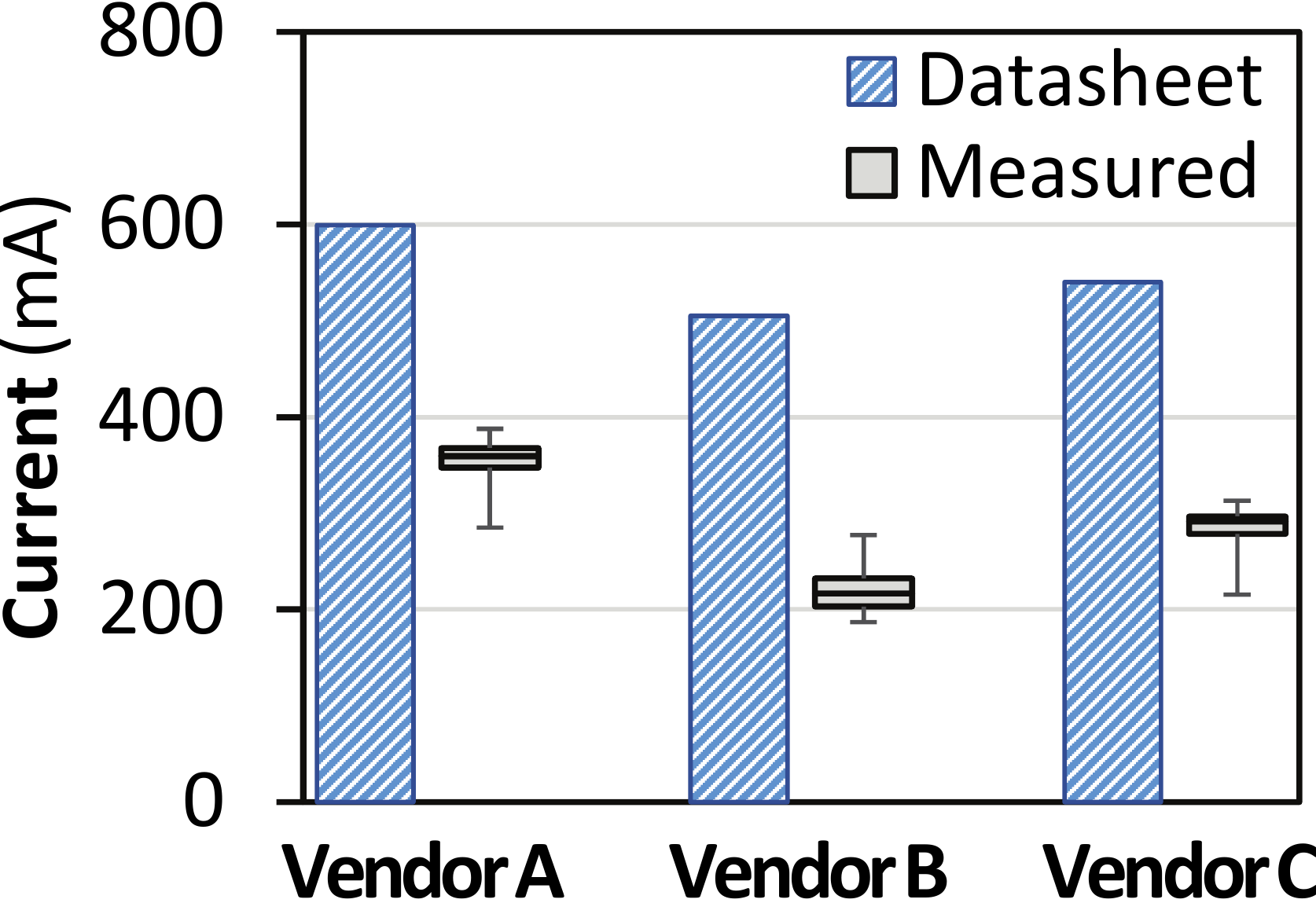}%
  \figspace%
  \includegraphics[width=0.42\basewidth, trim=55 168 57 172, clip]{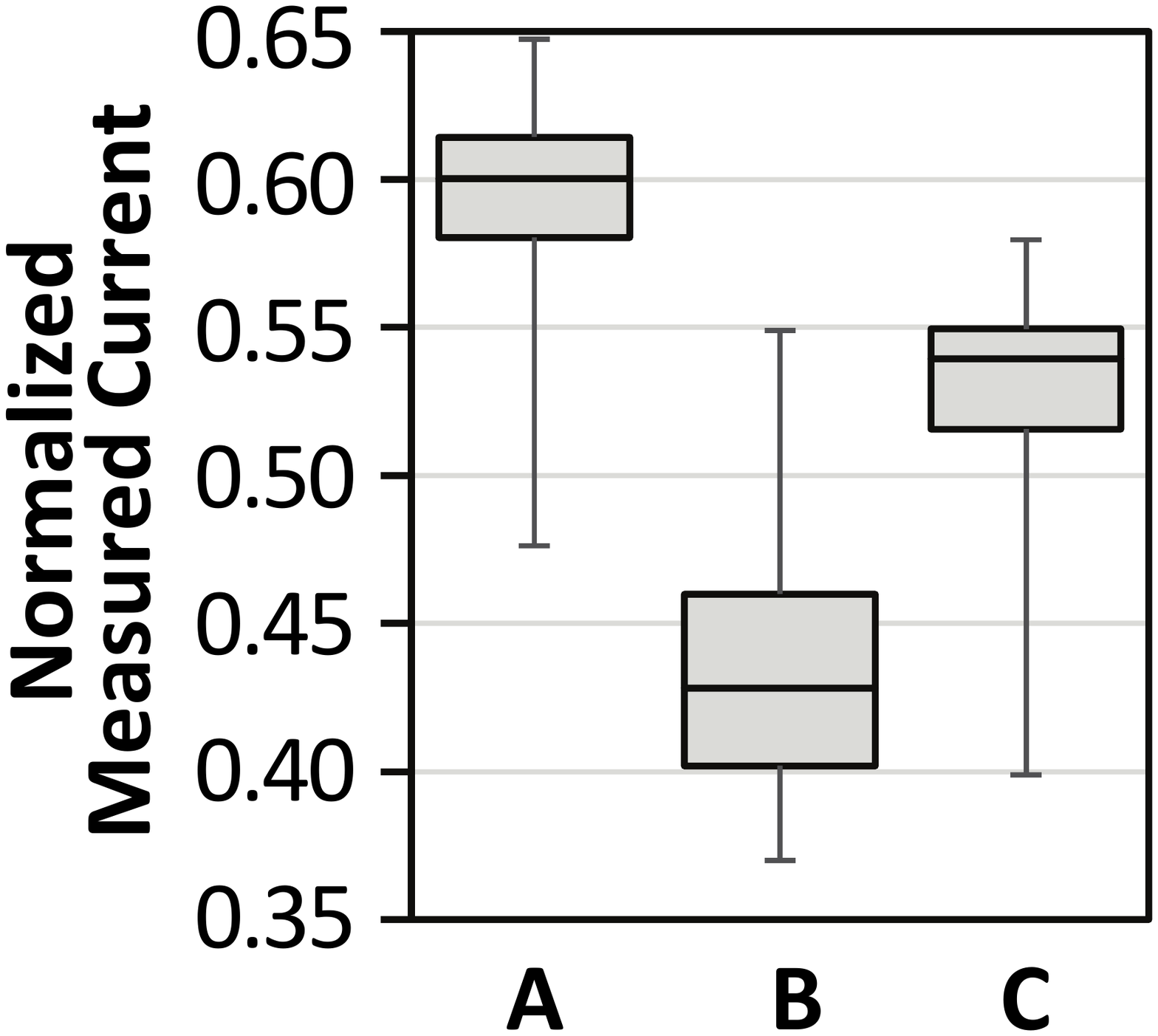}
  \caption{\idd[7] current measurements (left), and current normalized to datasheet value (right).}%
  \label{fig:idd7}
\end{figure}

Our \idd[7] measurements behave very similarly to \idd[0] and \idd[1],
but have a larger range, as shown in Figure~\ref{fig:idd7}.
As the read operations are interleaved with activate and precharge operations
to each bank, \ch{the \idd[7] measurement loop} \chI{accesses} the DRAM cells.
\ch{As a result, unlike \chII{what we observed for} \idd[4R], the measured \idd[7] values have large
margins compared to the datasheet}.  
\ch{The average measured \idd[7] current
\chII{is 58.4\%, 43.5\%, and \chIII{52.7\%} of the specified \idd[7] current}
for Vendors~A, B, and C,
respectively, and the full normalized range (i.e., the difference between the
highest-current module and the lowest-current module) is
\chII{10.1\%, 17.9\%, and \chIII{18.1\%},} respectively for the three vendors.}

\chI{We conclude that (1)~the measured read current is \emph{not} much lower 
than the datasheet value, 
even after we subtract the effect of the I/O driver current;
\chII{and
(2)~operations that access the cell array in addition to the peripheral circuitry
are likely to consume less current than the \chIII{specified datasheet values}}.}

\subsection{Refresh (IDD5B)}
\label{sec:idd:5b}
\begin{figure}[h]
  \centering
  \includegraphics[width=0.55\basewidth, trim=72 228 58 233, clip]{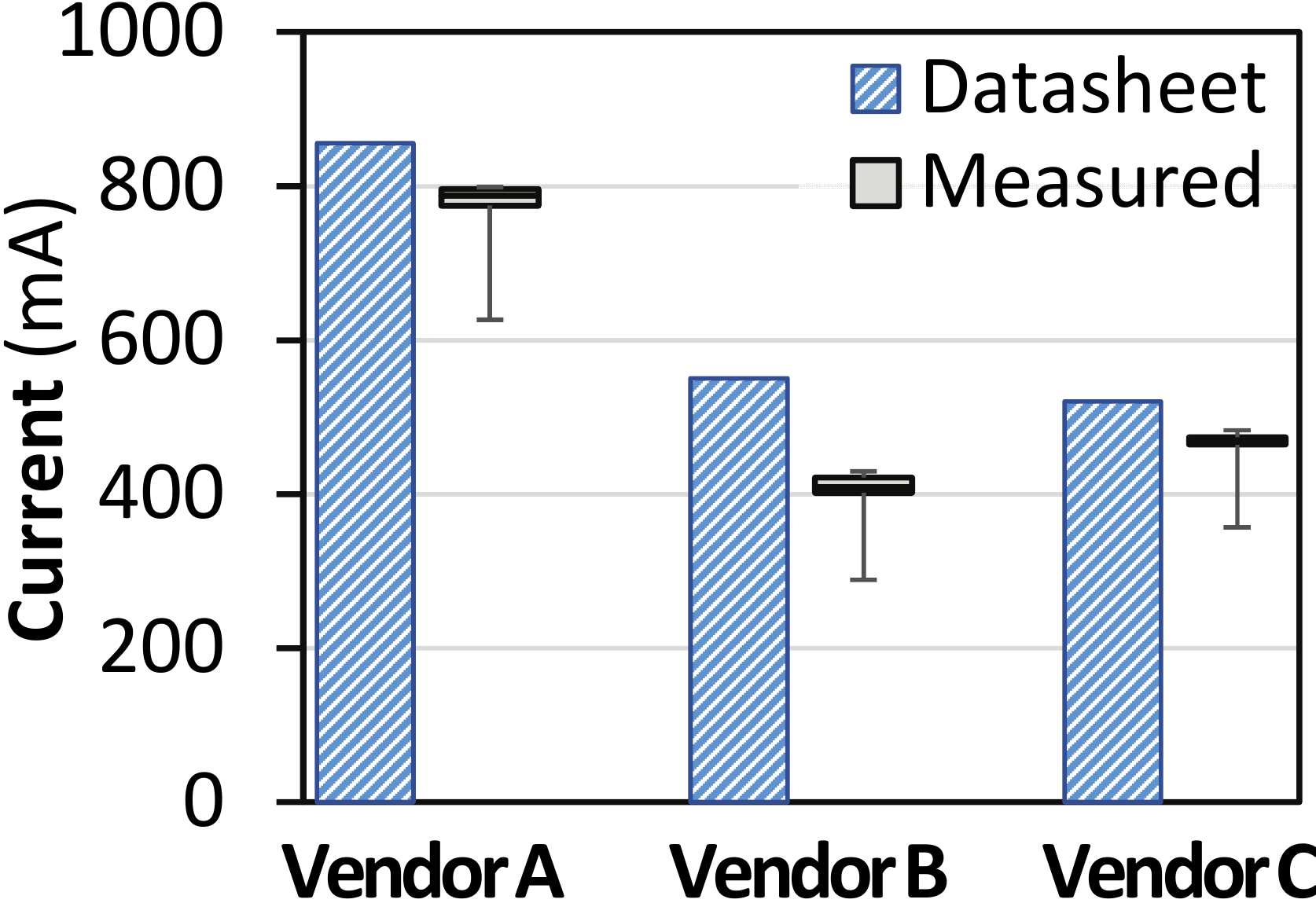}%
  \figspace%
  \includegraphics[width=0.42\basewidth, trim=55 168 57 172, clip]{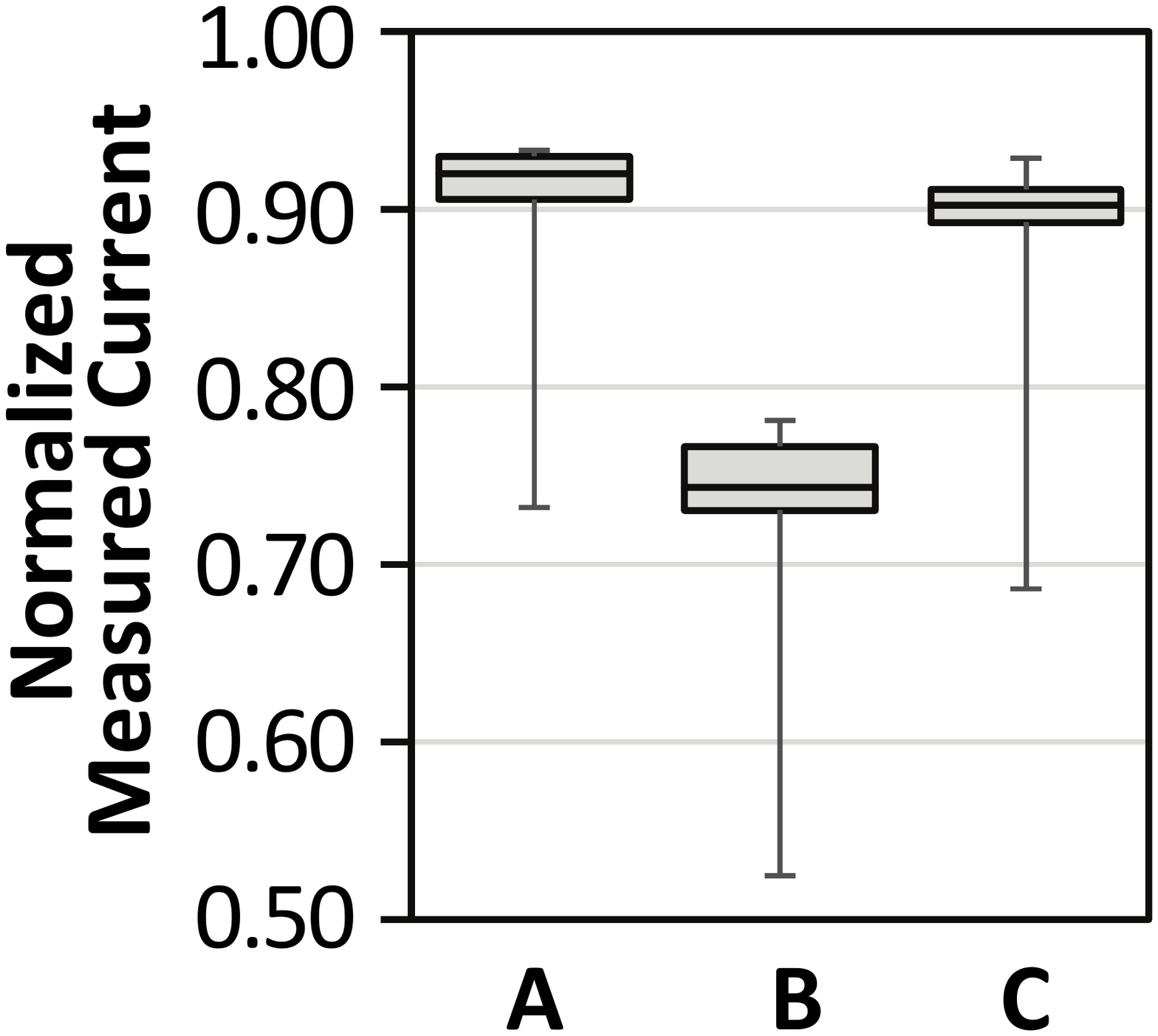}
  \caption{\idd[5B] current measurements (left), and current normalized to datasheet value (right).}%
  \label{fig:idd5b}
\end{figure}

Next, we study the amount of current consumed during refresh operations.
We study the \idd[5B] current measurement loop defined by JEDEC,
which performs a continuous burst of refresh commands.
Figure~\ref{fig:idd5b} shows the current measured during the \idd[5B] loop.
We note that the refresh current consumes the \chII{\emph{highest}} current of any of the 
operations that we have observed, and that the margin for refresh is \chI{\emph{not}}
as large as the idle current margin \chI{(see Section~\ref{sec:idd:2n-3n})}.  For Vendors~A, B, and C,
we observe average current consumption across all modules to be
88.6\%, 72.0\%, and \chIII{88.0\%} of the specified \idd[5B] current.
\ch{However, while the margin} is small,
the measured refresh current never exceeds the specified value.

\chI{We conclude that the measured refresh current is \emph{not} significantly
lower than the \chII{corresponding \idd[5B]} value in the datasheet.}

\subsection{Power-Down Mode (IDD2P1)}
\label{sec:idd:2p-2q-3p}
\chI{Last}, we study the impact of low-power modes in DRAM.  Modern DDR DRAM
architectures provide several modes to minimize power consumption during 
periods of low memory activity.  We focus on the \chI{\emph{fast power-down}} mode available in DDR3L
DRAM~\cite{ddr3l.jedec13, ddr3.jedec12}, which turns off the internal clock, decode logic, 
and I/O buffers, \ch{but keeps the delay-locked loop (DLL) circuit active}.
We study the \idd[2P1] measurement loop defined by JEDEC for \ch{the fast} power-down mode,
which measures current when \ch{\emph{no} bank is} active.\footnote{\ch{\chI{A
second mode, known as \emph{slow power-down}, turns off the DLL circuit \chII{\emph{in addition to}}
the internal clock, decode logic, and I/O buffers.} We are
unable to test the slow power-down mode, because our test infrastructure does 
\chII{\emph{not}} allow us to disable the DLL circuit.  As a result, we do not include results for the
\idd[2P0] measurement loop, which is designed to test the slow power-down mode.}}

\begin{figure}[h]
  \centering
  \includegraphics[width=0.55\basewidth, trim=72 228 58 233, clip]{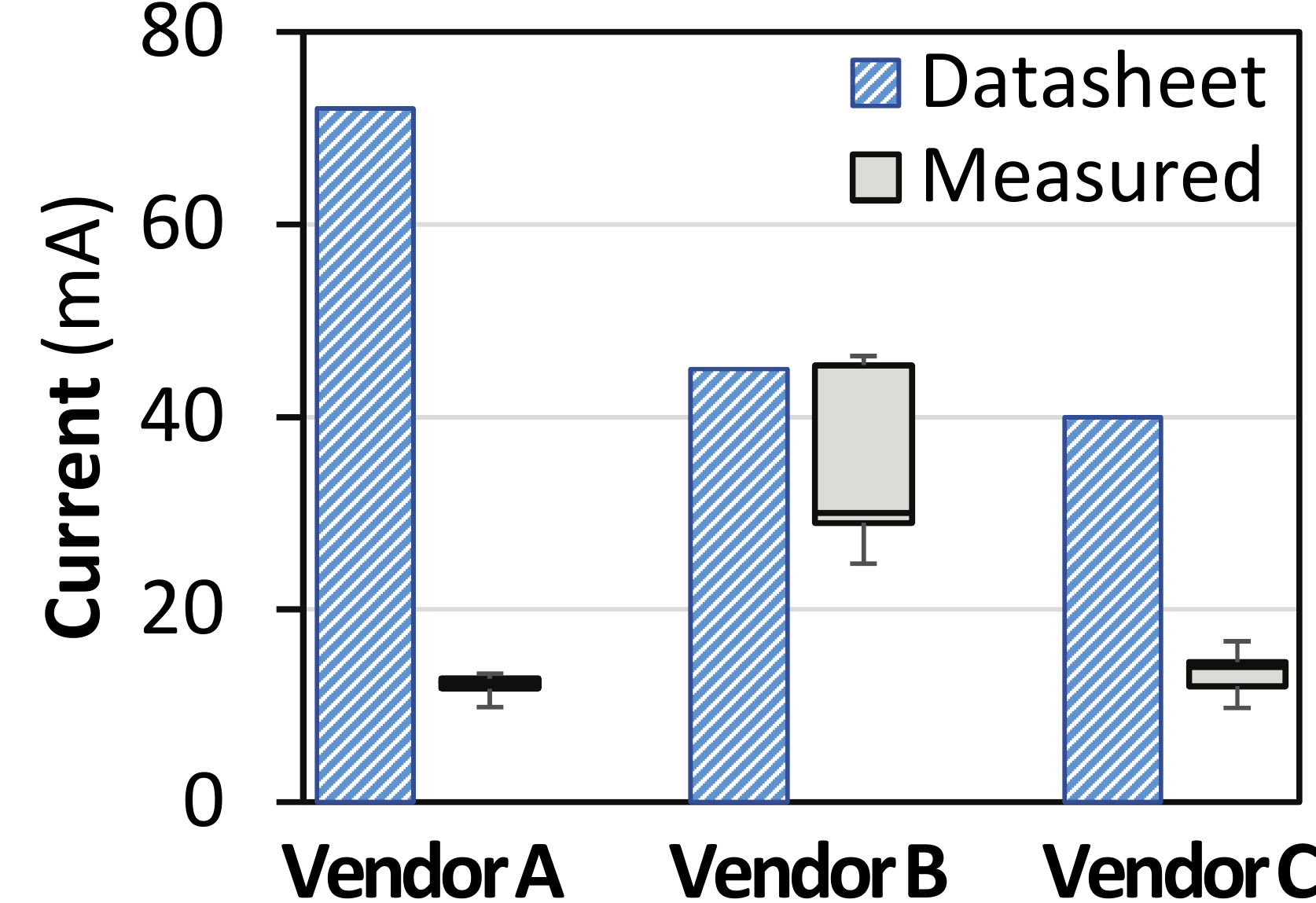}%
  \figspace%
  \includegraphics[width=0.42\basewidth, trim=55 168 57 172, clip]{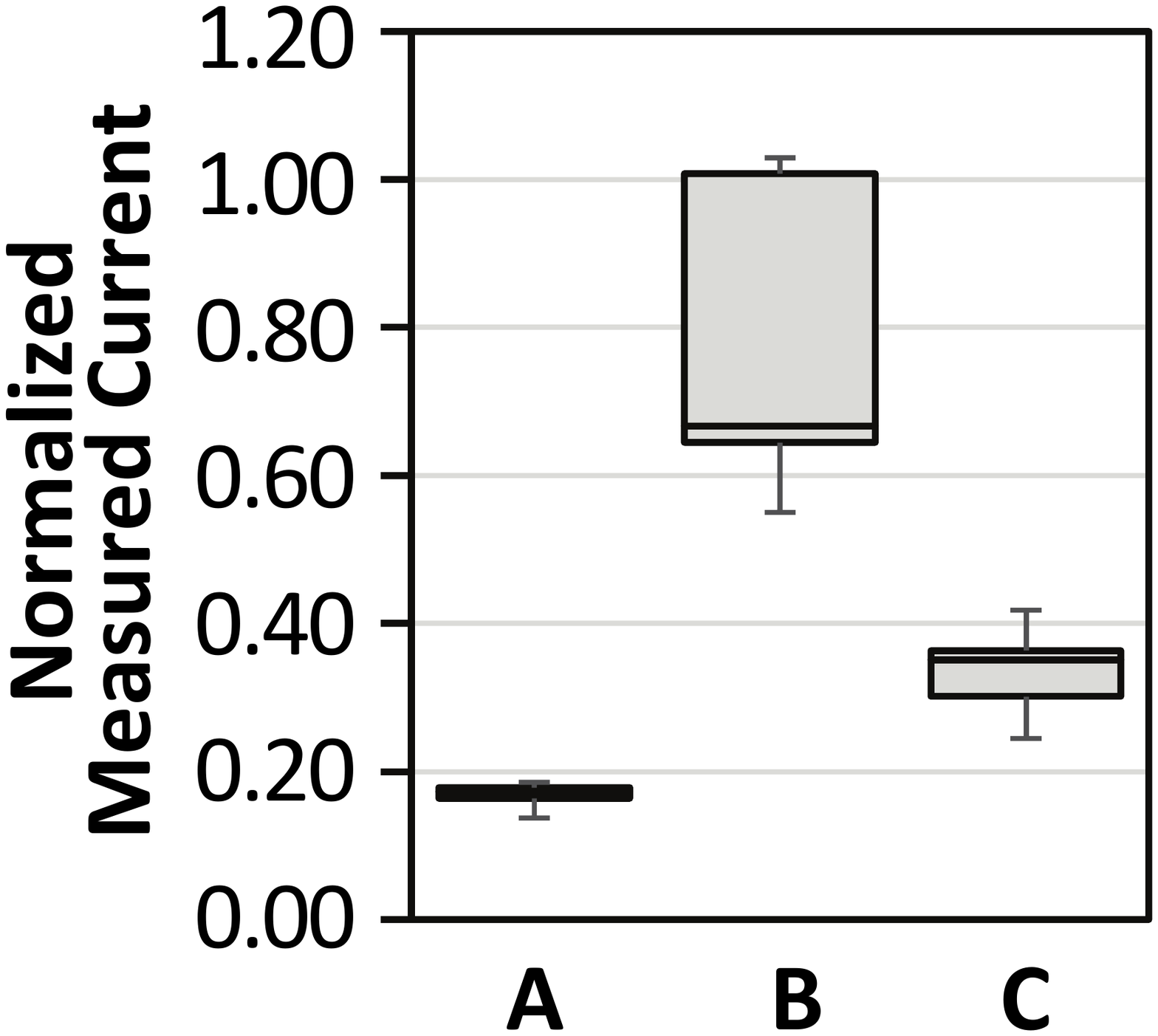}
  \caption{\idd[2P1] current measurements (left), and current normalized to datasheet value (right).}%
  \label{fig:idd2p1}
\end{figure}

Figure~\ref{fig:idd2p1} shows the current measured during the \idd[2P1] loop.
We observe that \chI{the} power-down mode is quite effective when no \ch{bank is} active,
\ch{reducing the current significantly compared to when no bank is active in
normal power mode (which we characterize above using the \idd[2N] measurement
loop, as shown in Figure~\ref{fig:idd2n}).}
For Vendors~A, B, and C, \ch{power-down mode reduces the current by} 
65.8\%, 30.6\%, and \chIII{48.7\%}, respectively \chI{(\chII{as} observed by comparing the
\chII{measured} values in Figure~\ref{fig:idd2p1} to those in Figure~\ref{fig:idd2n}.)}.
For Vendors~A and C, for whom power-down mode is highly effective,
the \ch{variation across modules} in power-down current is \chIII{small} as well, 
with a \ch{normalized range of \chII{4.8\% and \chIII{17.3\%} of the specified \idd[2P1]
current}, respectively}.  In contrast, the power-down mode
for Vendor~B consumes significantly greater power, and its current
ranges by as much as \chI{47.9\% of the specified \idd[2P1] current}, 
indicating a less efficient \ch{and more variation-prone} power-down implementation
\ch{than the implementations of Vendors~A and C}.

\chI{We conclude that the power-down mode is effective at reducing power,
but it has significant \chIII{power} variation across vendors.}

\subsection{General Observations}
\chII{Collectively examining} all of the \idd results that we present \chI{in 
Sections \ref{sec:idd:2n-3n}--\ref{sec:idd:2p-2q-3p}}, 
we make two key observations.

\chI{First}, we find that the majority of \idd values specified by the DRAM 
vendors are \chII{\emph{drastically}} different \ch{from} \chII{our} \ch{measured results.}
We believe that our measurements provide a more realistic vision of the
amount of power consumed by real-world modules \ch{than the 
\chII{vendor-specified} \idd values},
\chII{and demonstrate inter-vendor and intra-vendor module-to-module
variation in DRAM power consumption}.  
In fact, we find that power models based on the
\idd values \chIII{are oblivious to} many significant factors that can \ch{affect} the power
consumed by DRAM.  These power models \chIII{\emph{assume}} accesses to specific banks, rows,
and columns using specific data patterns, but as we show in Sections~\ref{sec:datadep}
and \ref{sec:var}, varying these factors can have a \chII{\emph{non-trivial}} impact on
the energy consumed by a \chII{DRAM} module.

\ch{\chI{Second}, \rev{\ch{we find that while there is a large difference in the 
\chI{datasheet-reported} \idd values
across our three vendors, the difference in \chI{\emph{measured}} current is much smaller
than the \idd values suggest for activate, read, write, and precharge operations}}.
For example, \chI{for the \idd[4R] value, the datasheets state that Vendor~A's
DRAM modules consume}
139\% more current than Vendor~C's modules (a total difference of 
\SI{427}{\milli\ampere}).
In reality, \chI{we find from our measurements that}
Vendor~A's modules consume only 26\% more current than
Vendor~C's modules \chI{on average} (a total difference of \SI{79}{\milli\ampere}).
We believe that \chI{this observation is} a result of all three vendors' modules
being manufactured using similar process technology nodes.  This is quite likely
given the fact that the modules were all manufactured around the same time
(see Table~\ref{tbl:dimms}).
Despite \chI{the use of similar process technology nodes by all three vendors}, 
Vendor~A's \idd values appear to include much larger margins
than the \idd values from Vendors~B and C.  This could reflect either
different levels of conservatism among the vendors, or}
could imply that \ch{modules from vendors that employ larger margins have} greater
variation (though we do \chII{\emph{not}} observe this in our experiments).


\section{Data Dependency}
\label{sec:datadep}

One major aspect that existing \chI{DRAM} power models ignore is the effect
that a \emph{data value} has on DRAM power consumption.
The \idd
measurement loops, as defined by JEDEC, use only the data patterns
0x00 and 0x33 (where the byte value is repeated for each byte within the cache
line).  While this allows for the tests to be standardized, it does \ch{\emph{not}}
offer any insight into whether or not current consumption depends on the
data value that is being read or written.

In this section, we design a set of measurements that determine whether or not
a relation exists between the data value and \ch{power} consumption, and analyze
the results of these measurements.  We break these studies down into two parts:
(1)~whether the \emph{number of ones} within a cache line \chI{impacts} the \ch{power}
consumed by DRAM (Section~\ref{sec:datadep:ones}), and
(2)~for a \emph{fixed number of ones}, whether the fraction of \emph{bits toggled}
within a cache line \chI{impacts} the \ch{power} consumed by DRAM (Section~\ref{sec:datadep:toggle}).
Based on our studies, we develop models for our modules that quantify how \ch{power}
consumption changes for each operation when we account for
\ch{(1)}~the number of ones \ch{in a cache line} and
\ch{(2)~the number of wires that experience toggling during a read or write
operation} (Section~\ref{sec:datadep:models}).

\subsection{Effect of Number of Ones in Data}
\label{sec:datadep:ones}

\begin{figure}[h]
  \centering
  \subfloat[\ch{Read command}]{%
    \includegraphics[width=0.49\basewidthwide, trim=88 245 77 233, clip]{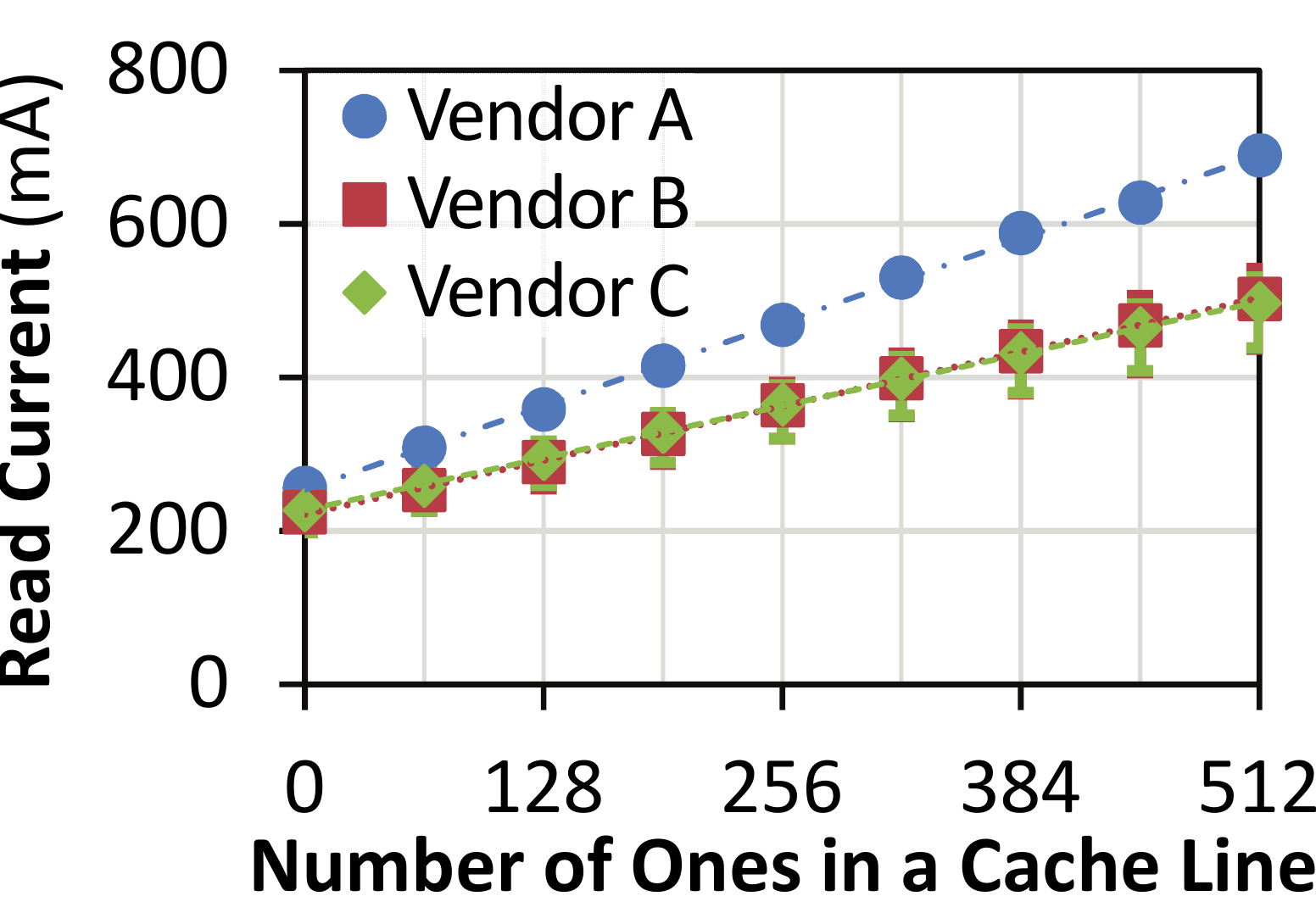}%
    \label{fig:ones_rowbuffer_rd}
  }%
  \figspace%
  \subfloat[\ch{Write command}]{%
    \includegraphics[width=0.49\basewidthwide, trim=88 245 77 233, clip]{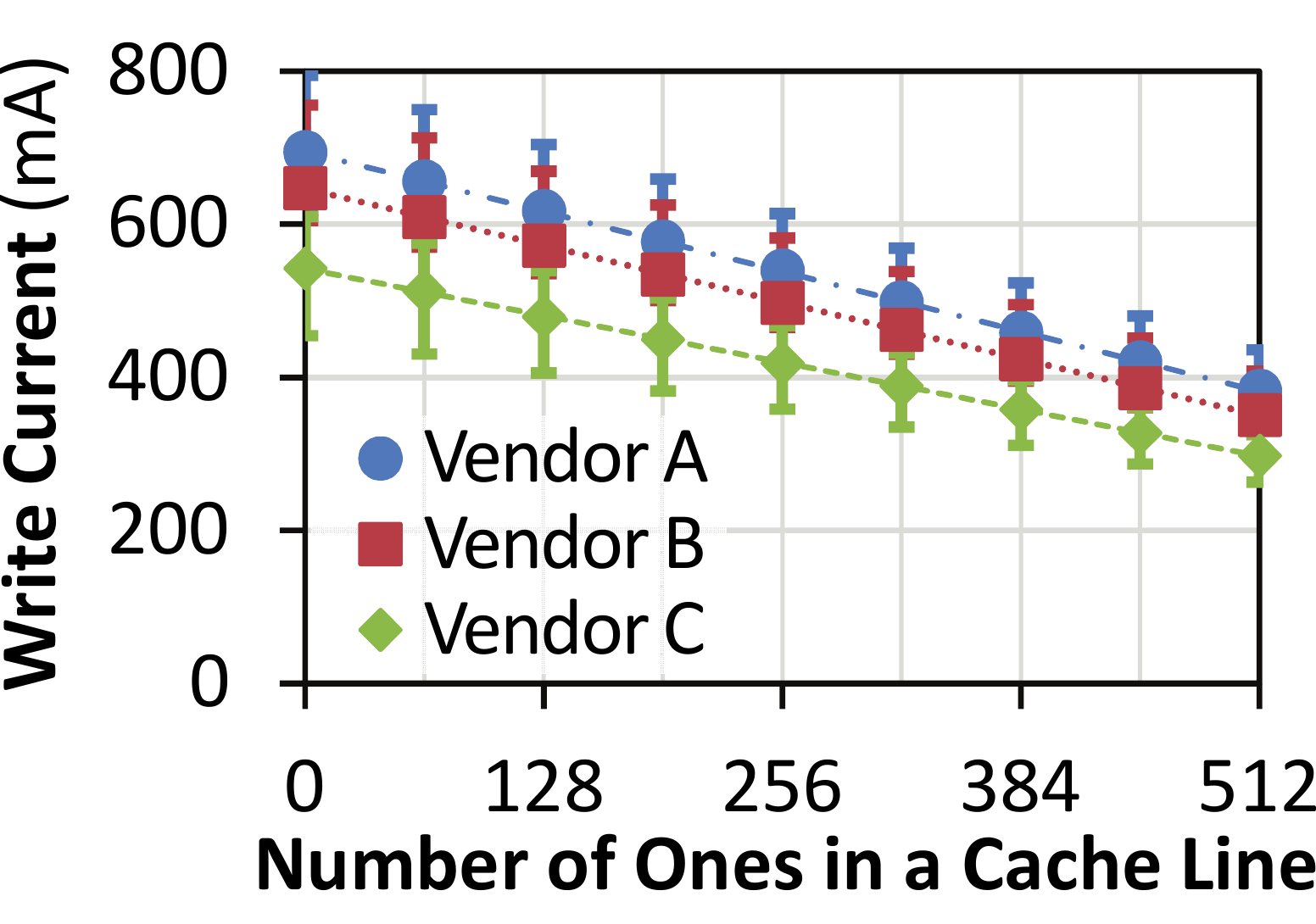}%
    \label{fig:ones_rowbuffer_wr}
  }%
  \caption{Effect of the number of ones on the read (left) and write (right) \ch{current drawn by DRAM}.
  Error bars indicate the 25th and 75th percentile measurements.}
  \label{fig:ones_rowbuffer}
\end{figure}

We start by exploring the relationship between the number of ones in a 
64-byte cache line (\chI{which consists of a number of columns read from DRAM,
as we discuss in Section~\ref{sec:bkgd:operations}}) and the power consumed.  In order 
to test this behavior, we select a set of rows that we \chI{would like} to test, and 
populate each column of the row with the same data pattern.  
We then repeatedly read data out of a \emph{single column in a single row}.
Figure~\ref{fig:ones_rowbuffer} shows how the current \ch{drawn by the 
DRAM module} (y-axis) changes as we
increase the number of ones in the cache line (x-axis), for both 
reads \ch{(Figure~\ref{fig:ones_rowbuffer_rd})} and 
writes \ch{(Figure~\ref{fig:ones_rowbuffer_wr})}.
We make two key observations from the figure.

\begin{figure}[h]
  \centering
  \subfloat[\ch{Read command}]{%
    \includegraphics[width=0.5\basewidthwide, trim=72 245 77 233, clip]{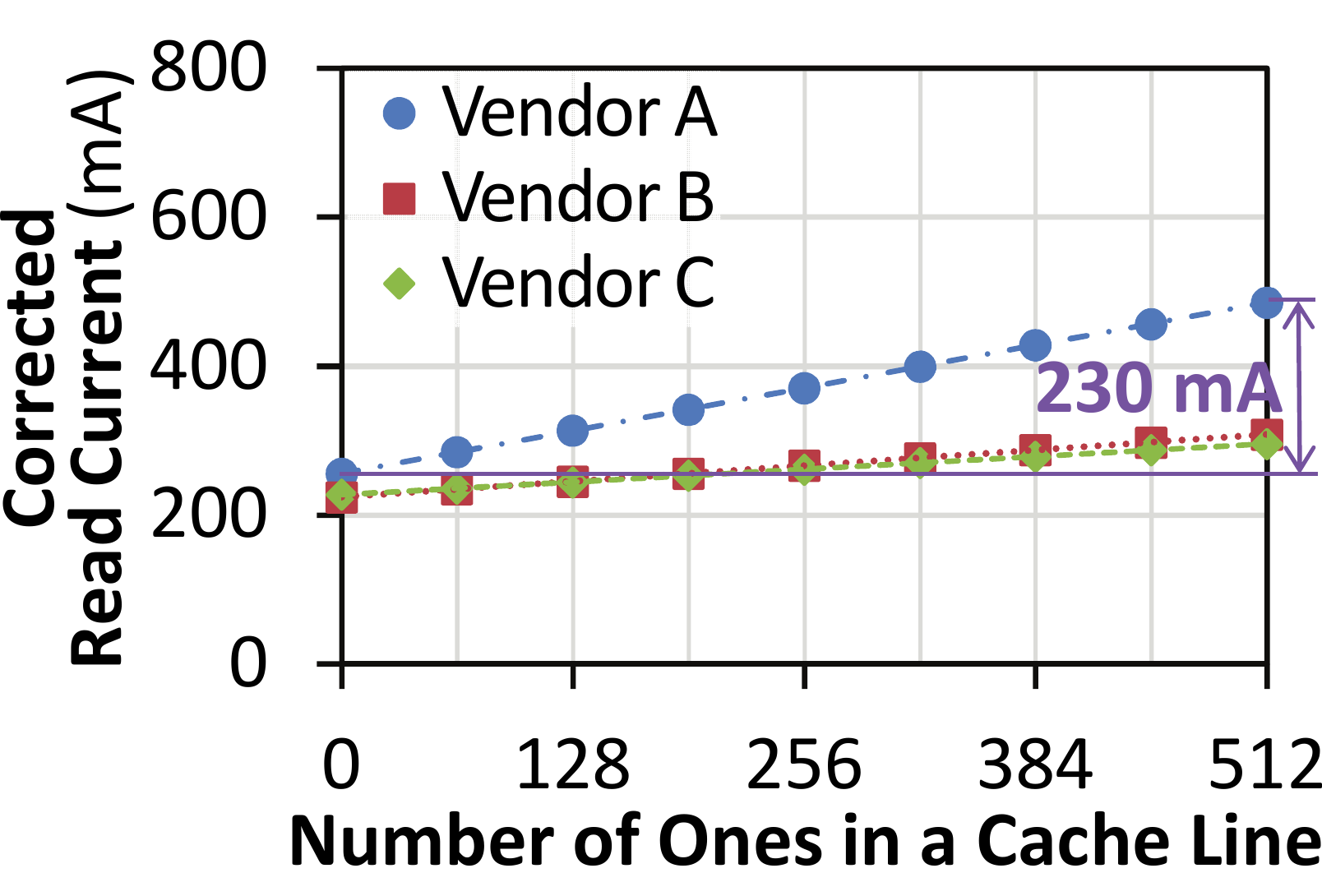}%
    \label{fig:ones_rowbuffer_rd_corrected}
  }%
  \figspace%
  \subfloat[\ch{Write command}]{%
    \includegraphics[width=0.5\basewidthwide, trim=72 245 77 233, clip]{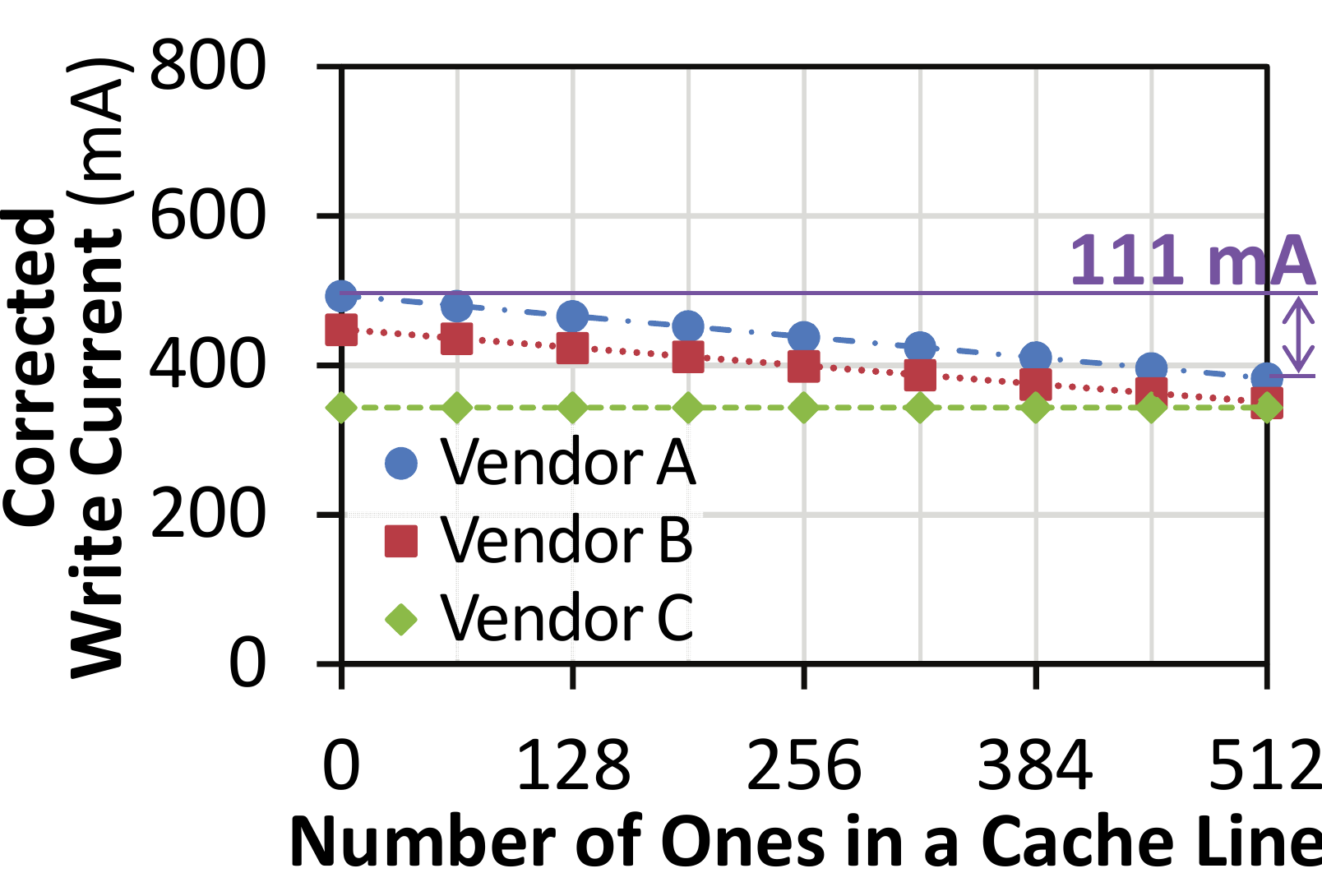}%
    \label{fig:ones_rowbuffer_wr_corrected}
  }%
  \caption{Effect of the number of ones on the read (left) and write (right) \ch{current drawn by DRAM},
  \chI{after subtracting a conservative estimate of the I/O driver current}.}
  \label{fig:ones_rowbuffer_corrected}
\end{figure}

First, \emph{as the number of ones in a cache line increases, the current required for a read operation
increases, while the current required for a write operation decreases}.  
\chI{The variation in power consumption with the number of ones is as much as
\SI{434}{\milli\ampere} for reads and \SI{311}{\milli\ampere} for writes (Vendor~A).}
There are two causes for this:
(1)~the I/O driver design, and
(2)~data-dependent power consumption within the DRAM \rev{module}.
\ch{As we discuss in Section~\ref{sec:bkgd:operations},
when data is transferred over the memory channel, each wire of the channel
is attached to two I/O drivers: one inside the DRAM module, and another inside
the memory controller.
\rev{\ch{Only one of the I/O drivers actively \emph{drives} current on the wire at a time, 
depending on 
(1)~the operation being performed, and
(2)~the bit value that is transferred on the wire.
When one of the I/O drivers is driving current on the wire, the other I/O driver
\emph{sinks} current~\cite{keeth-dram}.
For example, when the wire is transferring a bit value \emph{one} during a
\emph{read} operation from DRAM, the I/O driver in the DRAM module drives
the current on the wire, while the I/O driver in the memory controller sinks
current.
When the wire is transferring a bit value \emph{zero} during a \emph{read}
operation from DRAM, the I/O driver in the DRAM module sinks current, while
the I/O driver in the memory controller drives current.
The opposite is true for \emph{write} operations to DRAM:
the I/O driver in the memory controller drives current when the wire is 
transferring a bit value \emph{one} to DRAM, and the I/O driver in the 
DRAM module drives current when the wire is transferring a bit value 
\emph{zero} to DRAM.}}
Even if we eliminate the estimated current used by the I/O drivers (see
Section~\ref{sec:datadep:models}),} \chI{as shown in
Figure~\ref{fig:ones_rowbuffer_corrected},} we still
observe \chII{significantly} data-dependent power consumption, which can change the current by as
much as \SI{230}{\milli\ampere} \chI{for reads and \SI{111}{\milli\ampere} for
writes (Vendor~A)}.
\rev{While we cannot definitively identify the sources of data-dependent power
consumption within the DRAM, we suspect that other peripheral circuitry within
the DRAM, such as the bank select and column select logic (see
Section~\ref{sec:datadep:toggle}), may be responsible for the
data-dependent current behavior.}

Second, we observe that \emph{the relationship between the current consumption and
the number of ones is linear}.  
\ch{Note that this linear relationship is true \chII{\emph{even after}} we remove the 
effect of the I/O driver current on the current measurements shown in
Figure~\ref{fig:ones_rowbuffer}, \chII{as shown in Figure~\ref{fig:ones_rowbuffer_corrected}}.}
We use this linear relationship to build models
of the current consumption in Section~\ref{sec:datadep:models}.

We also perform tests to determine whether the data \chII{value} stored within a row affects
the activate and precharge current (not shown).  We find that there is \chI{\emph{no}} notable
variation for activate and precharge \chI{current consumption} based on the \chII{stored data value}.  
This is due to the way
\chI{in which} bitlines access a row during activation.  For each bitline, there is a 
corresponding reference bitline.
For a DRAM where a bit value 0 is
represented as \SI{0}{\volt} and a bit value 1 is represented as $V_{DD}$, the bitline and 
reference bitline
are precharged to voltage $V_{DD}/2$, allowing them to swing in either direction
based on the data held in the row.  
During activation, a cell swings (i.e., \emph{perturbs}) the
voltage of its bitline in the direction of the charge stored in the cell, and
the reference bitline is perturbed in the \chII{\emph{opposite}} direction.
\chII{Thus, for} every activation, there is always one line swinging up to $V_{DD}$,
and another line swinging down to 0V~\cite{keeth-dram}.  As a result, there
is little difference \chII{between} the power required for a \chII{cell that 
stores a zero and a cell that stores} a one
\chI{during activate and precharge operations}.

\subsection{Effect of Bit Toggling}
\label{sec:datadep:toggle}

\begin{figure}[h]
  \centering
  \includegraphics[width=\textwidth, trim = 38 172 38 212, clip]{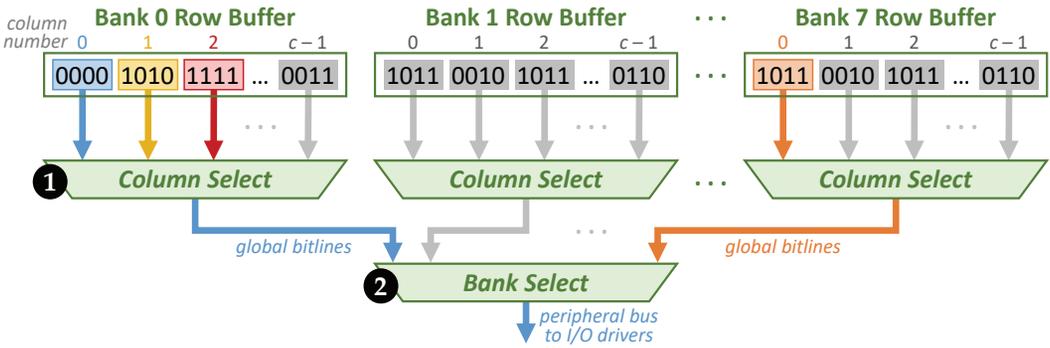}%
  \caption{\ch{Column and bank select logic organization.}}
  \label{fig:selectlogic}
\end{figure}

Next, we explore how interleaving \ch{memory requests} across multiple columns 
and banks affects the current \ch{drawn by DRAM}.
\ch{Figure~\ref{fig:selectlogic} shows a high-level overview of the logic used
in a DRAM chip to select the bank and column desired by a request.
Each bank contains \emph{column select} logic (e.g., \incircle{1} for
Bank~0 in Figure~\ref{fig:selectlogic}).
When a row is active in a bank, the entire contents of the row are latched in
a row buffer (see Section~\ref{sec:bkgd:org}).  The column select logic
chooses one column of data from the row buffer to output on the 
global bitline of the bank.
The DRAM chip then uses \chI{the} \emph{bank select} logic (\incircle{2}) to choose 
which global bitline contains the column that the chip \chI{should} send across the
memory channel.  The bank select logic sends this data across the
peripheral bus to the I/O drivers.}

\ch{\emph{Bit toggling} can occur when two back-to-back requests go to different 
banks and/or columns.  We can see where bit toggling occurs by using the 
example 4-bit select logic shown in Figure~\ref{fig:selectlogic}.
If we have two back-to-back read requests, where Request~W reads
Bank~0, Column~0, and Request~X reads Bank~0, Column~1, the
global bitline wires for Bank~0 first send binary value 0000 (for Request~W),
and then send binary value 1010 (for Request~X).
This causes two of the wires in the global bitline, and two of the wires
in the peripheral bus, to toggle from a bit value 0
to a bit value 1.
The number of wires that toggle is data dependent: if Request~X reads
Bank~0, Column~2 instead, all four wires of the global bitline
and all four wires of the peripheral bus toggle.
Requests~W and X are an example of \emph{column interleaving}, because
the requests go to two separate columns in the same bank.
Bit toggling can also take place during \emph{bank interleaving}, where
back-to-back requests go to different banks.  For example, if we have two
back-to-back read requests, where Request~Y reads Bank~0, Column~0,
and Request~Z reads Bank~7, Column~0, three of the peripheral bus
wires experience bit toggling, as the wires
first send binary value 0000 (for Request~Y) and then send binary value
1011 (for Request~Z).
Note that toggling does not always occur when requests are column- and/or
bank-interleaved.  For example, if one request reads Bank~7, Column~0,
and another request reads Bank~7, Column~2, the data on both the 
\chI{global bitlines} for Bank~7 and the peripheral bus does not change.}

We design \ch{a series of} tests that can capture the current consumed by 
\ch{the column and bank select logic},
and how \ch{bit} toggling \ch{at the column/bank select logic} affects current consumption.
\ch{We perform \chII{three} \chIII{types} of tests:}
\begin{enumerate}
  \item \ch{\emph{No Interleaving}:
    All requests access the \emph{same} bank and the \emph{same} column.}

  \item \ch{\emph{Column Interleaving}:
    Back-to-back requests access \emph{different} columns in the \emph{same} bank.}

  \item \ch{\emph{Bank+Column Interleaving}:
    Back-to-back requests access \emph{different} banks, and the column that is being
    accessed in a particular bank is \emph{different} \chI{from} the column that was accessed by the
    last access to that bank.}
\end{enumerate}
\ch{For each of the \chII{three} \chIII{types} above, we perform multiple tests, varying
(1)~the data pattern stored in or written to each column, and
(2)~whether the test consists of all read requests or all write requests.}

\ch{Note that when we change the data pattern being used, the change in current
is affected by two factors:
(1)~the increase in current due to bit toggling, and
(2)~the increase in current due to the number of ones in the data
(see Section~\ref{sec:datadep:ones}).}
\ch{As an example, \chI{consider} what happens during the \emph{column interleaving}
test for two different pairs of data values.  If we constantly alternate}
between reading data value 0x00 and data value 0xAA from two columns, 
50\% of the bitlines experience \chI{toggling}.
\ch{If we instead alternate}
between \ch{data value} 0x00 and \ch{data value} 0x0A, \ch{the toggle rate is} only 25\%. 
However, \ch{the column with data value 0xAA} also has two more bits set to `1' than
the \ch{the column with data value 0x0A, which requires more current, as we
discuss in Section~\ref{sec:datadep:ones}.}
In order to isolate the effect of \emph{only} bit \chI{toggling},
we calculate the total number of ones in the two reads, and subtract the
toggle-free current consumed when reading the same number of ones with
column interleaving.
\ch{For our example test where we alternate between a column with data value 0x00 
and another column with data value 0xAA (i.e., across both columns, there are
an average of two `1's per column),
we eliminate the impact of the number of ones by subtracting the
current consumed when we alternate between two columns that both
contain data value 0x88 (where each column has two `1's).}

\begin{figure}[h]
  \centering
  \subfloat[\chII{\emph{Column interleaving}}]{%
    \includegraphics[width=0.48\textwidth, trim=72 212 58 230, clip]{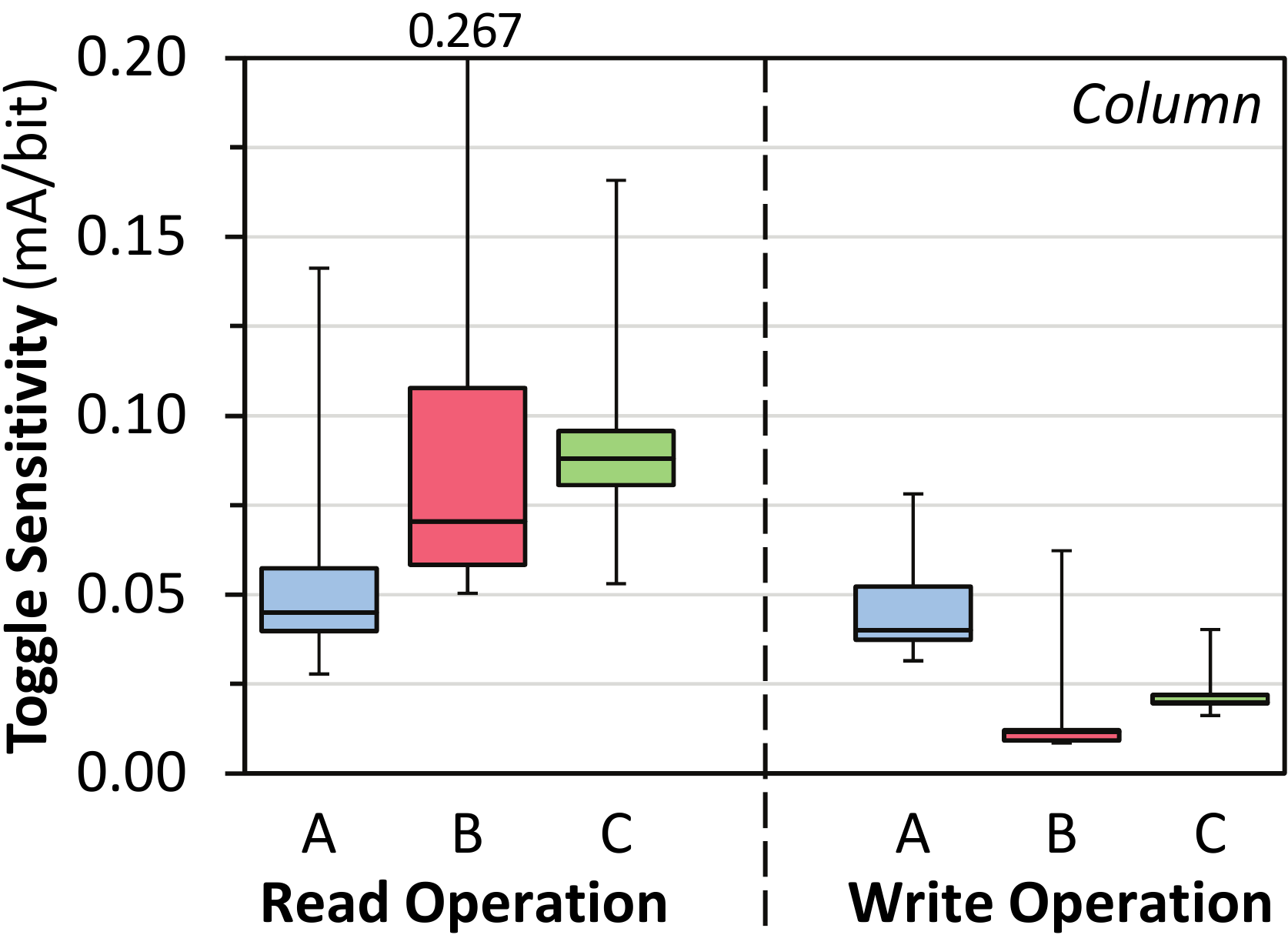}%
    \label{fig:toggle_column}
  }%
  \hfill%
  \subfloat[\chII{\emph{Bank+column interleaving}}]{%
    \includegraphics[width=0.48\textwidth, trim=72 212 58 230, clip]{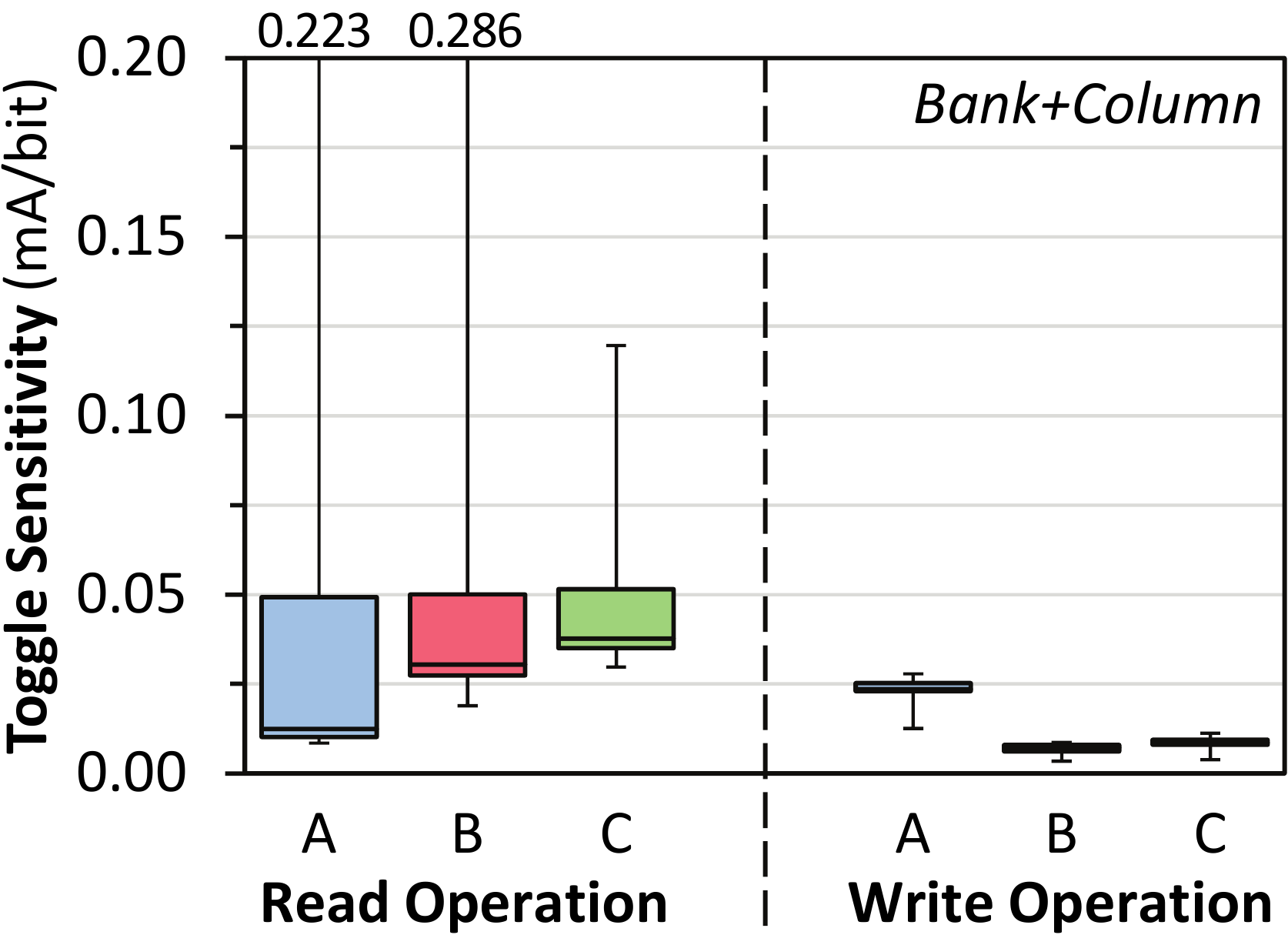}%
    \label{fig:toggle_bank_column}
  }%
  \caption{Effect of bit \chII{toggling
  on} read and write current consumption.}
  \label{fig:toggle}
\end{figure}

Figure~\ref{fig:toggle} shows \ch{how the measured current increases as
we increase the number of bits that are toggling, summarizing the increase
per bit across a wide range of data values for \chII{our
\emph{column interleaving} (Figure~\ref{fig:toggle_column}) and 
\emph{bank+column interleaving} (Figure~\ref{fig:toggle_bank_column}) tests}}.
As we found for the number of ones, the current
increases linearly as we increase the number of bits \ch{that are} toggling.  In Figure~\ref{fig:toggle},
we plot \emph{toggle sensitivity} on the y-axis, which shows the \emph{increase} in
current for each additional bit that is toggling, \chII{in terms of \si{\milli\ampere\per\bit}}.

We make two key observations from the figure.
First, \emph{the impact of bit toggling on DRAM current consumption}
\chI{(\chIII{\emph{up to}} \chII{a total of \SI{26}{\milli\ampere} for Vendor~A with column interleaving
when \chIII{\emph{all}} bits are toggling})} 
\emph{is much smaller than the impact of the number of ones} \chI{(\SI{230}{\milli\ampere}
for Vendor~A \chIII{when \chIII{\emph{all}} bits are set to ones}; see Section~\ref{sec:datadep:ones})}.
Second, \emph{toggling \ch{for the \emph{bank+column interleaving} test}
requires less current than toggling \ch{for the \emph{column interleaving} test}}.
We believe that both of these observations are due to the design of the select logic,
which we show in Figure~\ref{fig:selectlogic}.  When the DRAM selects another
column in the same bank, both
the wires between the column select logic and the bank select logic, and the
wires between the bank select logic and the I/O drivers, experience \ch{toggling}.  In contrast, \ch{when
DRAM} selects a different bank, only the wires of the peripheral bus experience \ch{toggling},
\chII{which reduces the bit toggling energy in the \emph{bank+column interleaving} test
compared to the bit toggling energy in the \emph{column interleaving} test}.

\ch{We conclude that there is a linear relationship between the current
consumed and the number of bits \chI{that toggle}, but that the amount \chI{of
current} consumed as a result of bit toggling is small,
\chII{especially when compared to the current consumption effect of the
number of ones in the data}.}

\subsection{Data Dependency Models}
\label{sec:datadep:models}

From our experiments in Sections~\ref{sec:datadep:ones} and \ref{sec:datadep:toggle},
we observe a linear relationship between the current consumed and the number of ones
in the cache line, as well as a linear relationship between the current consumed and
the number of bits \ch{that \chI{toggle due to} back-to-back read/write requests}.
As a result, we use linear least-squares regression
on our characterization data to develop quantitative models for this relationship,
in the \chII{following} form:
\begin{equation}
\chI{I_{total} = I_{zero} + \Delta I_{one} N_{ones} + \Delta I_{toggle} N_{toggles}}
\label{eq:model}
\end{equation}
where 
\chI{$I_{total}$} is the total current consumed (in \si{\milli\ampere}),
\chI{$I_{zero}$} is the current consumed when the cache line contains all zeroes, 
\chI{$\Delta I_{one}$} represents the extra current for each additional one in the cache line,
\chI{$N_{ones}$} is the number of ones in the cache line, 
\chI{$\Delta I_{toggle}$} represents the extra current for each additional bit that \ch{is toggling}, and
\chII{$N_{toggles}$} is the number of bits that were toggled,
We confirm the linear relationships of \chII{current with
(1)~}the number of ones in the cache line and 
\chII{(2)~}the number of \chI{bits toggled,} by using \ch{the} square of the Pearson correlation
coefficient~\cite{pearson.prsl1895}, commonly known as the $R^2$ value.
We find that across all of the modules that we measure, the $R^2$ value 
\chII{of these two correlations} is
never lower than 0.990.

\chII{Because different types of interleaving make use of different switching
circuitry, we require a separate set of model 
parameters to use in Equation~\ref{eq:model} for each
type of operation interleaving.}
Table~\ref{tbl:model} shows the average values of 
\chI{$I_{zero}$, $\Delta I_{one}$, and $\Delta I_{toggle}$}
in \si{\milli\ampere} for read and write
\chII{operations, with \chIII{\emph{column interleaving}},} for each module vendor.
We compare the output of \chII{our} model shown in Table~\ref{tbl:model}
to the average \ch{measured current of each data point \chII{shown} in
Figure~\ref{fig:ones_rowbuffer} (see Section~\ref{sec:datadep:ones}),}
and find that the percent error of our model never exceeds 1.40\%
\chII{(and the average percent error across all data points is 0.34\%)}.
Table~\ref{tbl:fullmodel} in the appendix shows the parameters used to
model other combinations of bank and/or column interleaving.
These data dependency models form a core component of our \chI{new}
measurement-based DRAM power model (see Section~\ref{sec:model}).

\begin{table}[h]
  \centering
  \small
  \caption{Parameters \ch{(\chII{$I_{zero}$, $\Delta I_{one}$, $\Delta I_{toggle}$}; in \si{\milli\ampere}) 
    used in Equation~\ref{eq:model}} to \chII{model} current consumption (\chI{$I_{total}$}) 
    due to data dependency when read/write operations are column interleaved.}%
  \label{tbl:model}%
  \vspace{-7pt}%
    \begin{tabular}{c||c|c|c||c|c|c}
        \hline
        \multirow{2}{*}{\bf Vendor} & \multicolumn{3}{c||}{\bf Read} & \multicolumn{3}{c}{\bf Write} \\
        \cline{2-7}
        & $I_{zero}$ (\si{\milli\ampere}) & $\Delta I_{one}$ (\si{\milli\ampere}) & $\Delta I_{toggle}$(\si{\milli\ampere}) & $I_{zero}$ (\si{\milli\ampere}) & $\Delta I_{one}$ (\si{\milli\ampere}) & $\Delta I_{toggle}$ (\si{\milli\ampere}) \\
        \hhline{=#===#===}
        A & 246.44 & 0.433 & 0.0515 & 531.18 & -0.246 & 0.0461 \\ \hline
        B & 217.42 & 0.157 & 0.0947 & 466.84 & -0.215 & 0.0166 \\ \hline
        C & 234.42 & 0.154 & 0.0856 & 368.29 & -0.116 & 0.0229 \\
        \hline
    \end{tabular}%
\end{table}

\chII{We conclude that our models accurately capture the relationship between 
data dependency and DRAM current consumption.}


\section{Characterizing Variation \protect\chI{of Current}}
\label{sec:var}

A significant limitation of existing \ch{DRAM power models~\cite{Jung, ChandrasekarRunTime, 
drampower, micron.2015}} is that they are based on IDD tests
that are performed \chI{by DRAM vendors} on only a fixed set of banks and rows 
at room temperature\ch{~\cite{ddr3.jedec12}}.
While these fixed conditions ensure \ch{repeatability}, \chI{the resulting
existing models} do \ch{\emph{not}} capture \chI{power consumption}
variation across banks, rows, or temperature.  In this section, we perform a
series of experiments to characterize
(1)~\emph{structural variation}, where current may vary based on the bank, row, or column selected
due to the circuit-level design of the DRAM \ch{chip} (Section~\ref{sec:var:structural}), and
(2)~whether the \emph{operating temperature} of DRAM affects
its current consumption \ch{(Section~\ref{sec:var:temp})}.
The results we have presented throughout the paper so far already capture
a third type of variation: \emph{process variation}.  As we \ch{have shown} using 
box plots \ch{(Figures \ref{fig:idd2n}--\ref{fig:idd2p1})},
different modules \chI{of the same part, from the same vendor,} exhibit a non-trivial 
amount of current variation for tests that target the same bank(s), row(s), and column(s).

\subsection{Structural Variation \protect\chI{of Current}}
\label{sec:var:structural}
\label{sec:var:banks}
\label{sec:var:rows}
\label{sec:var:cols}

Each module consists of \ch{a number} of hierarchical structures (i.e., banks, rows,
and columns) that are connected together to provide density and parallelism.
Due to the need to maximize density and optimize the \ch{physical} chip layout, 
there may be low-level differences among some components.  For
example, in the Open Bitline architecture~\cite{inoue.ijsc1988},
a DRAM array is broken up into \emph{subarrays}\chI{~\cite{SALP, 
seshadri.micro13, LISA, Chang}}, and pairs of subarrays
share common row buffer structures.  Subarrays in the middle of the bank share
structures with \chI{\emph{both}} neighbors, but subarrays placed at the edge share structures
with only \chI{\emph{one}} neighbor.  Likewise, due to variation in the distance between different
rows in a subarray and the logic required to access the row (e.g., wordline select logic,
row buffers), there can be significant variation in the latency of different
rows\ch{~\cite{lee.sigmetrics17, chang.sigmetrics2016}}.  We now study if such
\ch{structural variation} factors impact the current consumed by DRAM.
We consider variation to be
structural in nature only when we observe the same trend repeated in each of the 
modules we study \chI{from the same vendor}.
\ch{We consider all other variation to be due to manufacturing process variation,
and do not report it in this section.}

\subsubsection{Structural Variation Across Banks}

\begin{figure}[h]
  \centering
  \includegraphics[width=0.8\basewidth, trim=232 255 225 252, clip]{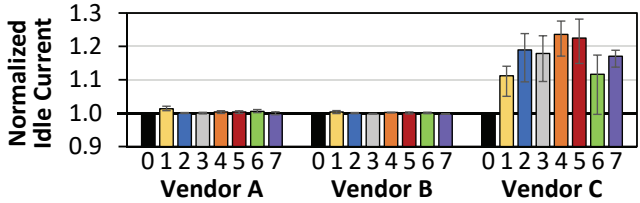}%
  \caption{Idle current variation across banks when one bank is active, normalized to current 
  for Bank~0.  \chIII{Error bars indicate the 25th and 75th percentile measurements.}}%
  \label{fig:bankvar_idle}
\end{figure}

We first characterize \ch{current variation} \emph{across banks within the 
same module}.  Figure~\ref{fig:bankvar_idle} shows the \chI{\textbf{idle current}} consumed when 
we \chIII{keep} Row~0, which contains all zeros, \chIII{open (i.e., activated)} in each bank.
\chIII{In the figure,}
the average \ch{measured} current of each bank is normalized to the average \ch{measured} current consumed 
by Bank~0, \chIII{where Row~0 is activated and contains all zeroes,} for each vendor. 
We observe that modules from Vendors~A and B show 
\chI{little to} no \ch{inter-bank} variation in their idle current
consumption, but modules from Vendor~C have significant variation.
Depending on which bank is activated, the current can vary by as much as 23.6\%.
\ch{We find that these results hold for other rows that we test.}
\chIII{As we discuss in detail below, we hypothesize that Vendor~C's
DRAM chip \chIII{organization} contains structural differences across each bank,
resulting in the current variation that we observe.}

\begin{figure}[h]
  \centering
  \includegraphics[width=0.8\basewidth, trim=232 255 225 252, clip]{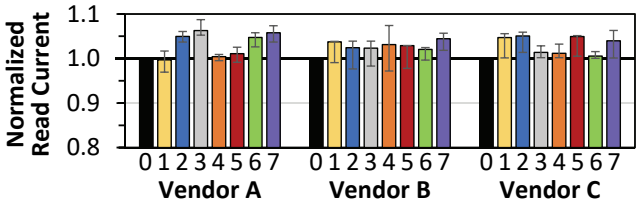}%
  \caption{Read current variation across banks, normalized to current 
  for Bank~0.  \chIII{Error bars indicate the 25th and 75th percentile measurements.}}%
  \label{fig:bankvar_idd4r}
\end{figure}

\begin{figure}[h]
  \centering
  \includegraphics[width=0.8\basewidth, trim=232 255 225 252, clip]{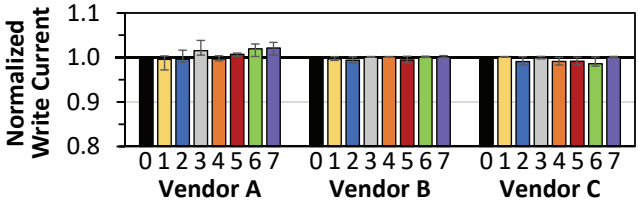}%
  \caption{\ch{Write current variation across banks, normalized to current 
  for Bank~0.  \chIII{Error bars indicate the 25th and 75th percentile measurements.}}}%
  \label{fig:bankvar_idd4w}
\end{figure}

\chIII{Next, we} characterize the variation in \chI{\textbf{read current}} for each bank within the 
same module.  Figure~\ref{fig:bankvar_idd4r} shows the average current consumed when
we repeatedly read Column~0 from Row~0 for each bank, normalized to the 
average \chI{measured} current consumed for Bank~0 for each vendor.
\chIV{For these experiments, Row~0 contains all zeroes.}
We observe that, in this case, \chI{\emph{all}} of the modules exhibit variation.
The variation for Vendor~C does not match the variation trend observed in
Figure~\ref{fig:bankvar_idle}.
We perform the same experiments for writes (\ch{Figure~\ref{fig:bankvar_idd4w}}), but find no notable
variation for any of our modules.
\chIII{This indicates that the structural variation is a result of components
that are used during read operations but not during write operations.}
\chIV{We observe similar variation trends when we repeat the experiments with
different data values in Row~0.}

Based on our results, we \ch{hypothesize} that for modules from Vendors~A and B, the
variation is a result of structural variation in the I/O driver circuits used to
read data, as the I/O \ch{drivers in the DRAM module drive current on the DRAM
channel only} during a read operation.  
As Vendor~C's modules show variation in the idle state and
during read operations, but the variation trends do not match, we conclude that
there are multiple sources for the variation that we observe, which include
I/O driver variation.
\ch{Overall, we find that structural variation exists across banks, but that the
pattern of variation is highly dependent on the vendor, due to differences in the
DRAM architecture from vendor to vendor.  Unfortunately, without access to
detailed information about the underlying DRAM architecture of each part
(which is information proprietary to DRAM vendors\chI{~\cite{lee.sigmetrics17,
PARBOR}}), we are \chI{currently} unable to pinpoint the exact
sources of this \chI{structural} variation.}

\subsubsection{Structural Variation Across Rows}

\begin{figure}[h]
  \centering
  \includegraphics[width=0.83\basewidth, trim=220 235 226 232, clip]{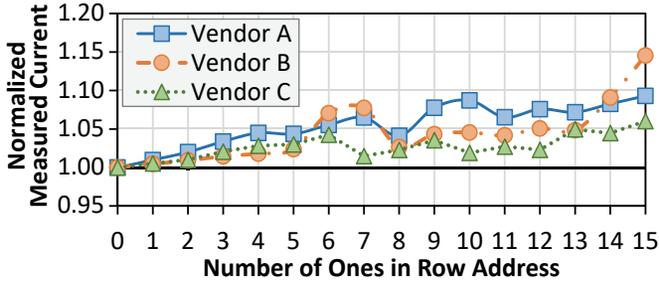}%
  \caption{Relationship between activation current and number of
ones in row address, normalized to activation current for Row~0.}%
  \label{fig:rowvar}
\end{figure}

Next, we characterize current variation \emph{across rows within
the same bank}.  For each module, we measure and compare the current consumed 
when we repeatedly activate and precharge 512~different rows.\footnote{We do not study the power 
consumed by read and write operations across different rows, as these \ch{operations
are \chIII{\emph{not}} performed on the cells themselves.  Instead, reads and writes} operate on data 
\ch{that is \chIII{already}} in the row buffer, 
\ch{and the same row buffer is used by} all of the rows in a bank.
\ch{Thus, the read and write operations use the same \chIII{hardware structures} regardless of
the row being accessed.}}
We find that there is systematic structural variation in 
each of our modules.  We observe that the \emph{current consumed by each row
increases with the number of ones in the row address}.  
Figure~\ref{fig:rowvar} shows this trend, where we average together the current
consumed by rows that contain the same number of ones in their row address,
and plot the average current sorted by the number of ones in the address
(on the x-axis).
As the figure shows, modules from Vendors~A and B show a correlation
between the number of ones in the row address and the current.  For
modules from Vendor~B, a row with 15~ones in its address consumes 14.6\% more
current than a row with all zeroes in its address.
\ch{Modules}
from Vendor~C show a similar trend, but exhibit a much smaller 
slope, and thus less variation, than modules from Vendors~A and B.

We hypothesize that there are two potential sources \ch{of} this variation.
\chIII{First, due to the way that rows are organized within the DRAM cell array,
rows with more ones in their row addresses are \chIV{\emph{more}} likely to be
physically \chIV{\emph{further}} away from the sense amplifier and column select 
logic~\cite{lee.sigmetrics17}.  As a result, longer segments of wires must be
driven for operations to rows with more ones in their row addresses.
Second, the row decoding logic uses some or all of the row address bits
to enable the wordline of the row that is activated.  The row decoding logic
may consume more energy when a greater number of row address bits are
set to one.}
We cannot, however, confirm these hypotheses without knowing the
circuit-level implementation of the \chIII{internal logic of \chIV{each DRAM chip}}, 
which is \ch{information proprietary to DRAM vendors}.

\subsubsection{Structural Variation Across Columns}

\chIII{Last}, we characterize the \ch{current variation} \emph{across columns
within the same row}.  \chIII{To this end, for each column in an activated row,}
\chIII{we measure the current consumed when we repeatedly read from or write to
that column, and compare this with the current consumed when we repeatedly
read from or write to Column~0.  (We do not show these comparisons for brevity.)
We find that for both read and write operations, there is no notable variation in
current consumption from one column to another.
We hypothesize that this lack of variation is because read and write operations
to different columns in a row make use of the \emph{same} global bitlines, 
bank select logic, peripheral bus, and I/O drivers (see Figure~\ref{fig:peripheral}
in Section~\ref{sec:bkgd:operations}).}
\chIII{Thus, we conclude} that there is
\chIII{\emph{no}} significant \chIII{source of} structural \ch{current variation} between columns.

\subsection{Variation \protect\chI{of Current} Due to Temperature}
\label{sec:var:temp}

Prior work has shown that DRAM latency and refresh rates can be affected by the
temperature at which DRAM operates\chI{~\cite{chang.sigmetrics2016, 
chang.pomacs2017, Patel, lee.hpca15, Liu, kim.hpca18}}.
To investigate if a relationship exists between operating temperature and
the current consumed by DRAM, we repeat all of our experiments at \SI{70(1)}{\celsius}.

From our experiments, we do \ch{\emph{not}} observe any measurable current variation
due to temperature (results not shown for brevity).
We believe that this is a limitation of our \ch{DRAM testing} infrastructure.  In DRAM, the
main source of temperature-related effects is the change in charge leakage.
At higher temperatures, the charge stored within a DRAM cell leaks more 
rapidly\chII{~\cite{mukundan.isca13, Liu, Patel, Chang, kim.hpca18, liu.isca2012}}.
As a result, \ch{refresh operations must either (1)~}restore a greater amount of
charge into the cell \ch{or (2)~be performed more frequently}, 
to make up for the additional charge that leaked over the
same amount of time.  As our test infrastructure continually
\chI{iterates over a loop of DRAM commands}, the DRAM cells are continually accessed, and
do \ch{\emph{not}} have enough time to leak charge~\cite{Liu}.
Thus, our measurements \ch{\emph{cannot}} capture the impact of charge leakage without
extensive modifications \chI{to the SoftMC design}.
We leave 
\chII{such modifications to SoftMC, and the resulting characterization of how
charge leakage due to temperature affects DRAM power consumption, to
future work.}


\section{Generational Trends}
\label{sec:gen}

The \chI{results we} have presented so far \ch{examine} the power consumed by modules
manufactured in recent years, using the latest process technologies developed 
for DRAM.
As is the case with \chII{microprocessors}, end users and system designers have grown
accustomed to \ch{reduced} power consumption when new process technologies are
used.  For DRAM, users and designers \ch{currently rely} on datasheet
current specifications to estimate the amount of power savings
\ch{from one generation to another}.
In this section, we compare the power reduction trends indicated by the datasheet
\ch{values} with the \emph{actual} power \ch{savings}, as measured using our 
infrastructure.

We study changes in power consumption across DRAM generations for Vendor~C.
In addition to the modules listed in Table~\ref{tbl:dimms}, we have access to a
number of older modules manufactured by Vendor~C.  \chI{Table~\ref{tbl:gen_dimms}
summarizes select properties of these modules.}  
\chI{We test modules of two older parts,}
with \chI{one of the parts} manufactured in 2011, and \chI{the
second part} manufactured in 2012.  In comparison, the Vendor~C modules studied
thus far in this paper were manufactured in 2015.

\begin{table}[h]
  \centering
  \small
  \caption{Properties of older DDR3L modules from Vendor~C.}%
  \label{tbl:gen_dimms}%
  \vspace{-7pt}%
    \setlength{\tabcolsep}{.45em}
    \begin{tabular}{cccccc}
        \toprule
       {\bf Number} & {\bf Total Number} & \textbf{Timing} (ns) &
        {\bf Assembly} & {\bf Supply} & {\bf Max. Channel} \\
        {\bf of Modules} & \bf{of Chips} & (t\textsubscript{RCD}/t\textsubscript{RP}/t\textsubscript{RAS}) & {\bf Year} & {\bf Voltage} & {\bf Frequency} (MT/s)  \\
        \midrule
        3 \chI{SO-DIMMs} & 24 & 13.75/13.75/35 & 2011 & \SI{1.35}{\volt} & 1333 \\
        4 \chI{SO-DIMMs} & 32 & 13.75/13.75/35 & 2012 & \SI{1.35}{\volt} & 1600 \\
        \bottomrule
    \end{tabular}%
\end{table}

\begin{figure}[h]
  \centering
  \subfloat[{\idd[2N]} \ch{(idle/standby)}]{%
    \includegraphics[width=0.48\basewidthwide, trim=66 238 90 238, clip]{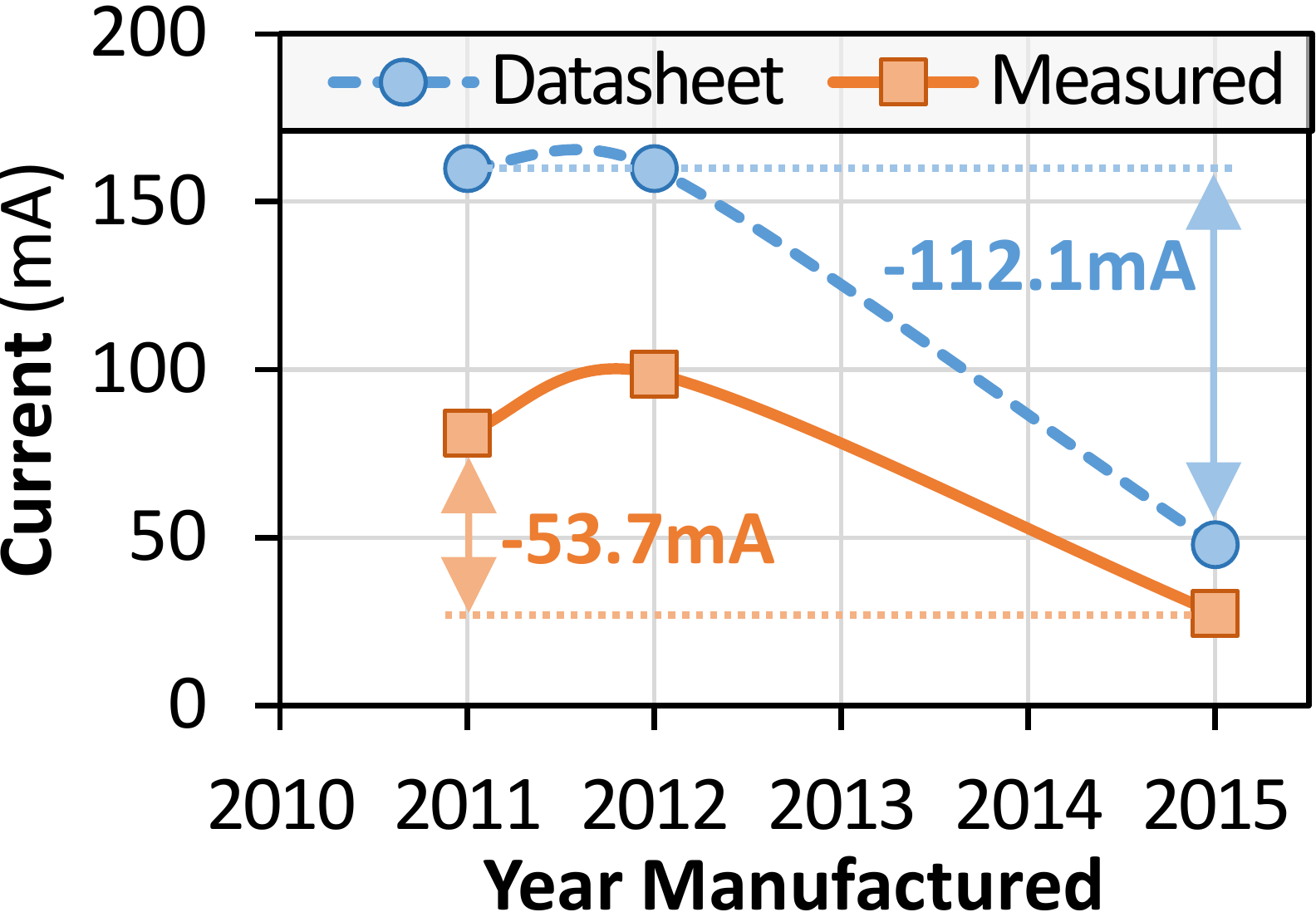}%
    \label{fig:gen-idd2n}
  }%
  \figspace%
  \subfloat[{\idd[0]} \ch{(activate and precharge)}]{%
    \includegraphics[width=0.48\basewidthwide, trim=66 238 90 238, clip]{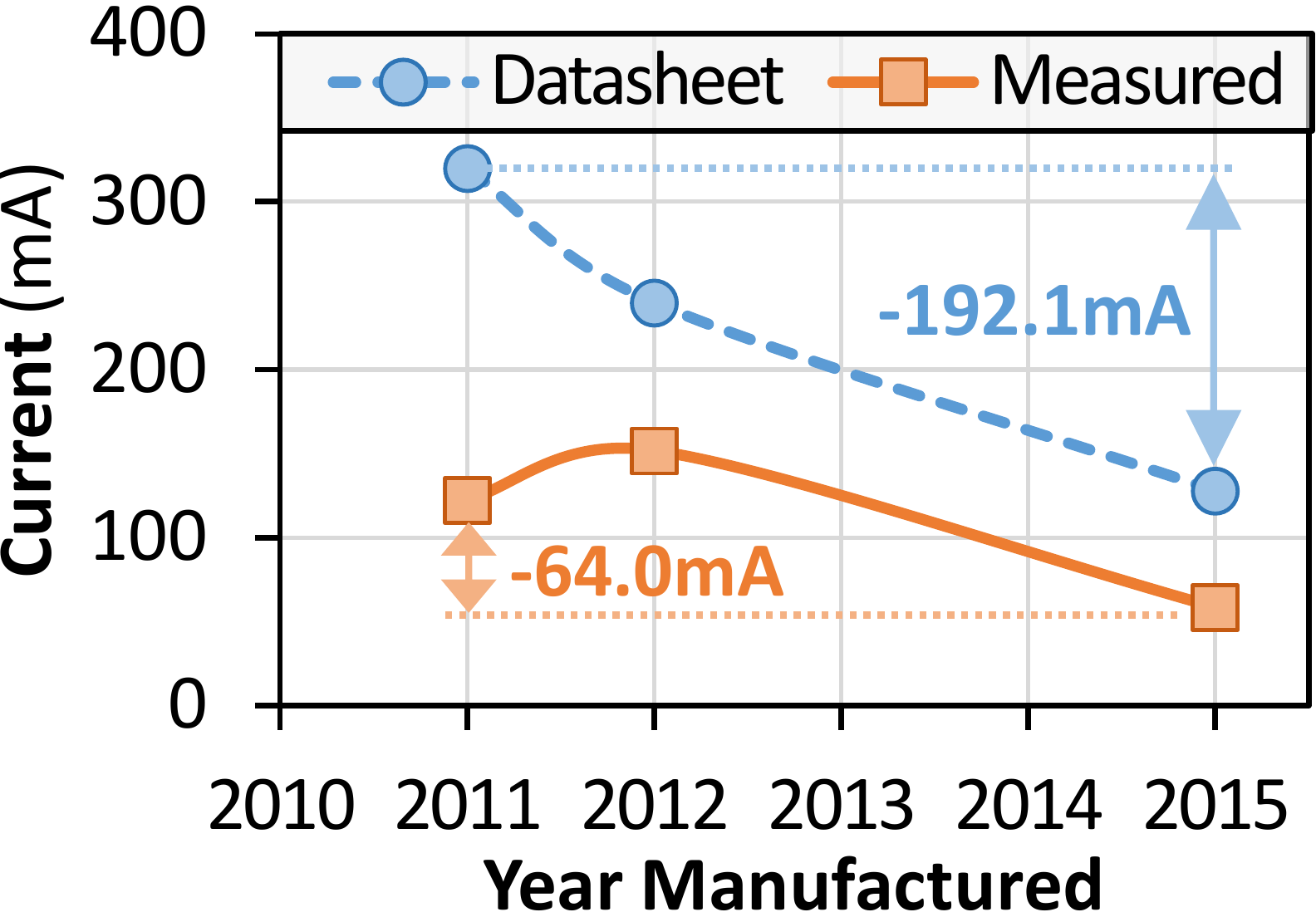}%
    \label{fig:gen-idd0}
  }%
  \\ 
  \subfloat[{\idd[4R]} \ch{(read)}]{%
    \includegraphics[width=0.48\basewidthwide, trim=66 238 90 238, clip]{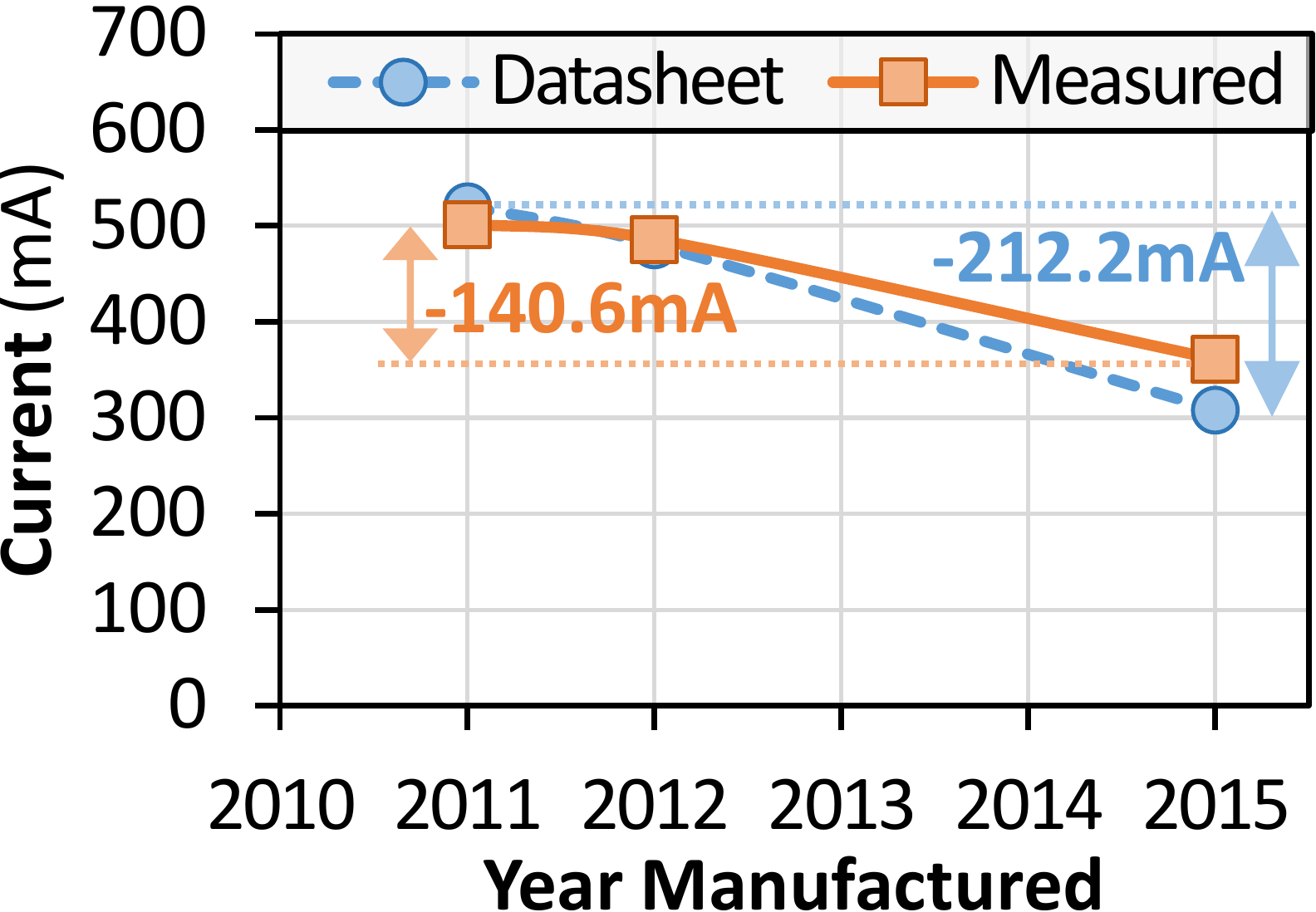}%
    \label{fig:gen-idd4r}
  }%
  \figspace%
  \subfloat[{\idd[4W]} \ch{(write)}]{%
    \includegraphics[width=0.48\basewidthwide, trim=66 238 90 238, clip]{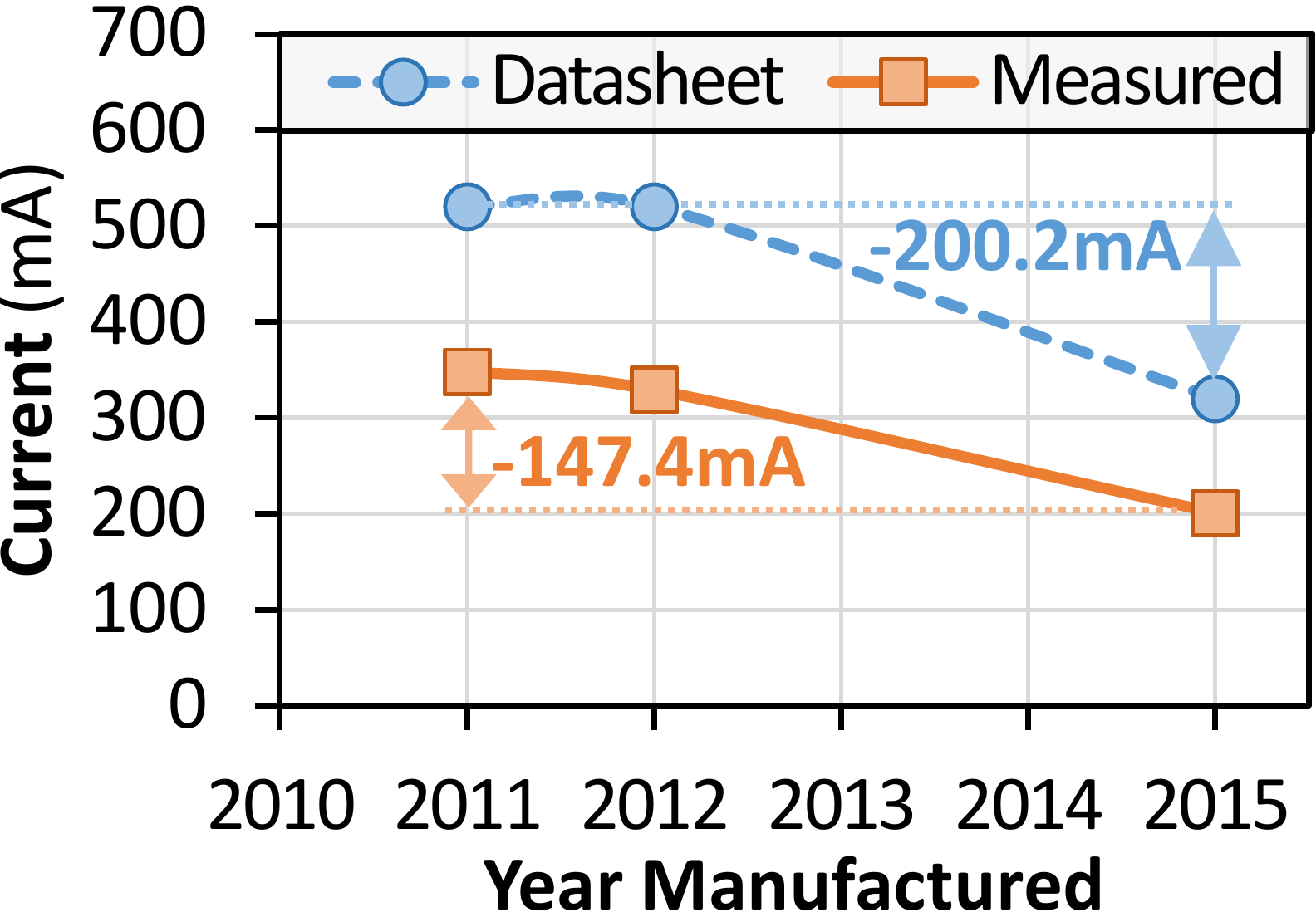}%
    \label{fig:gen-idd4w}
  }%
  \caption{\ch{Generational trends in} \idd measurements.}%
  \label{fig:gen}
\end{figure}

To compare the change in power consumption \ch{across generations}, we measure the \idd values for 
each module.  Figure~\ref{fig:gen} shows four of these \idd values, representing
\chI{idle/standby (\idd[2N]),} activate and precharge (\idd[0]), read (\idd[4R]), and
write (\idd[4W]) \ch{currents}.  If we study the expected savings from the datasheet \ch{values}
(dotted blue lines), we see a general downward trend as modules move to newer
process technologies (we plot the year of manufacture along the x-axis),
\chI{indicating that the power consumption should have decreased}.
Based on our measurements, we make two key observations.

First, we observe that for each \idd value, \emph{the actual power 
saved by switching to a newer-generation module, \ch{as measured by using
our infrastructure,} is significantly lower than 
the savings predicted by the datasheet}.  For example, based on the datasheet, the
\chII{\emph{expected}} decrease in \idd[0] current (Figure~\ref{fig:gen-idd0}) for moving
from a module manufactured in 2011 to a module manufactured in 2015 is \SI{192.1}{\milli\ampere}.
In comparison, we measure an \chII{\emph{actual}} decrease of \SI{64.0}{\milli\ampere}, representing only 33.3\% of the expected
savings.  For read and write operations, the difference is less drastic, but 
still \ch{statistically} significant.  Using \idd[4W] (Figure~\ref{fig:gen-idd4w}) as an example,
we see that the expected decrease from the datasheet \chII{is} \SI{200.2}{\milli\ampere}, but the
decrease measured from the actual \chII{DRAM} modules is \SI{147.4}{\milli\ampere}, \ch{or 73.6\%} of the
expected savings.

Second, we observe that in the case of \idd[4R] (Figure~\ref{fig:gen-idd4r}),
while the read power consumed by older-generation modules was within the
\idd[4R] value in the datasheet, the lower-than-expected savings have caused 
the measured current to \emph{exceed} the expected current based on the datasheet. 
This is in part due to the \chII{fact that the} I/O driver current 
\chII{is included in our read current measurement due to \chIII{the
design of our infrastructure},}
\chII{but the I/O driver current} is \ch{\emph{not}} \chII{included as part of} the 
vendor-specified current.  
\chI{As we discuss in Section~\ref{sec:idd:4r-4w-7}, the inclusion of the
I/O driver current is a limitation of our measurement infrastructure, 
but we can \chIII{eliminate} the I/O
driver current \chIII{by applying correction mechanisms}.  
\chII{Since the I/O driver current is a constant value for all of our
measurements, the amount by which the measured read current \emph{decreases}
across generations is \emph{not} affected by the I/O driver current.}}
\chII{We} find that the actual measured power savings for \idd[4R]
\ch{is \chII{only} 66.3\% of the expected savings \chIV{reported in the datasheets}}
\chIII{(a measured decrease of \SI{140.6}{\milli\ampere} vs.\ 
an expected decrease of \SI{212.2}{\milli\ampere})}.

From our observations, we conclude that the power \chI{reduction} from DRAM scaling 
\ch{is} \chII{\emph{not}} as significant as expected from the datasheet \chI{values provided 
by DRAM vendors}. With almost half of
the total system power now consumed by DRAM\ch{~\cite{Lefurgy, ware.power7, 
david.icac11, holzle.book09, malladi.isca12, yoon.isca12, elmore.tr16, paul.isca15}},
system designers may not \ch{be able to} obtain the total system power savings they
had expected by transitioning to newer DRAM models, which could \chI{adversely} affect the
amount of power and/or battery that is provisioned for a system.


\section{\protect\ch{Summary of Key Findings}}
\label{sec:summary}

\ch{We have presented extensive experimental characterization results and
analyses of DRAM power consumption that capture a wide range of properties
affecting the power \chI{consumption of real modern DRAM devices}.  
We summarize our findings in four key conclusions:}

\begin{enumerate}
  \item
  \ch{The current consumed by real DRAM modules \emph{varies significantly} from the
  current specified in datasheets by the vendors, across all IDD values that 
  we measure (Section~\ref{sec:idd}).}
  \chII{We comprehensively show that there is \chIII{significant} inter-vendor and intra-vendor 
  module-to-module variation in DRAM power consumption.}

  \item
  \ch{DRAM power consumption strongly \emph{depends on the data value}
  that is read \chI{from or written to the DRAM chip}, but does not
  strongly depend on \chI{the amount of} bit toggling (Section~\ref{sec:datadep}).}

  \item
  \ch{There is significant \emph{structural variation} \chI{of power consumption
  within a DRAM chip}, where the current varies
  based on which bank or row is \chII{accessed} in \chI{the DRAM chip}
  (Section~\ref{sec:var}).}

  \item
  \ch{Across successive process technology generations, the actual
  power \chI{reduction} of DRAM is \emph{much lower} than the savings indicated by the
  vendor-specified IDD values in the datasheets (Section~\ref{sec:gen}).}
\end{enumerate}

\section{\protect\ch{VAMPIRE:} Modeling DRAM Power}
\label{sec:model}

In order to \ch{overcome} the shortcomings of existing power models,
we use \ch{the new} observations \ch{from our rigorous experimental
characterization \chI{of real DRAM chips} (Sections \ref{sec:idd}--\ref{sec:var})} to build the 
\emph{Variation-Aware model of Memory Power Informed by Real Experiments}
(VAMPIRE).  \chI{To our knowledge,} VAMPIRE is the first 
real-measurement-based power model for DRAM.
By using our actual measurements from our characterization, VAMPIRE predicts
a realistic value for DRAM power consumption, with \ch{little error}.  
We validate VAMPIRE against
microbenchmarks executed on our power measurement infrastructure to \chI{test and} ensure
its accuracy (see Section~\ref{sec:model:validation}).

\ch{VAMPIRE takes in DRAM command traces, similar to the command traces
used by DRAMPower~\cite{drampower, ChandrasekarRunTime}.  Each line of the
command trace contains 
(1)~the command name;
(2)~the target rank, bank, row, and column for the command, if applicable; and
(3)~for \chII{read and write} commands, the 64-byte data that is \chII{read from
or} written to DRAM.
By annotating the data alongside the write command, VAMPIRE can determine
data-dependent power consumption.
VAMPIRE also supports traces that do \chI{\emph{not}} include the written data \chII{values:
users can instead manually input a certain distribution for the fraction 
of ones and the amount of bit \chIII{toggling, which VAMPIRE uses to
approximate the effect of data dependency on power consumption.}}

VAMPIRE consists of three core components: 
(1)~read and write power modeling, which incorporates data-dependent behavior;
(2)~idle/activation/precharge power modeling; and
(3)~structural variation modeling.
The first model component, read and write power modeling,
uses the data-dependency-aware current models that we
develop in Section~\ref{sec:datadep:models}.
These current models incorporate the
\chI{change in power consumption due to
(1)~the number of bits set to one in the data,
(2)~bit toggling, and
(3)~switching between different banks and columns}.
The second model component, idle/activation/precharge power modeling, uses
the measurements from Section~\ref{sec:idd} to capture the actual power
consumed for \ch{all DRAM commands aside from read and write.} 
\chI{This includes the power consumed by} 
\ch{activate operations, precharge operations,} refresh operations and
power-down modes, and when a DIMM is idle.
The third model component, structural variation modeling, uses \ch{our
characterization} results from Section~\ref{sec:var} to adjust the
\chI{estimated} current based on which bank and row are accessed.

VAMPIRE outputs a separate power value for each vendor, and 
can account for process variation by outputting a range of
power values based on \ch{the impact of variation, as estimated from
the variation that we capture \chI{in our experimental characterization}}.
We plan to integrate VAMPIRE into several memory \chI{system}
simulators (e.g., gem5~\cite{gem5}, DRAMSim2~\cite{rosenfeld-cal2011},
Ramulator~\cite{kim.cal15, ramulator.github}, NVSim~\cite{dong.tcad12}),
and will open-source the model~\cite{vampire.github}.

\subsection{Model Validation}
\label{sec:model:validation}

\rev{\ch{We use a new series of experimental measurements, which were \chI{\emph{not}} used 
to construct VAMPIRE, to \chI{\emph{validate}} the accuracy of our model and 
compare \chI{it} with the accuracy of two 
popular state-of-the-art \chI{DRAM power} models: the Micron power calculator~\cite{micron.2015} 
and DRAMPower~\cite{drampower, ChandrasekarRunTime}.
Both models are based off of worst-case \idd values \chI{reported in 
vendor datasheets}, and \chI{\emph{neither} of them models}
\chII{most process variation, data-dependent power consumption, or structural variation}.
We use the extrapolated \idd values that we calculate in Section~\ref{sec:idd}
as parameter inputs into both the Micron power calculator and DRAMPower.}}

\rev{\ch{In the validation experiments, we execute the following sequence of commands:
\{\emph{activate}, $n \times$\emph{read}, \emph{precharge}\},
where we sweep various values of $n$ between 0 and 764.\footnote{\ch{We ensure that
all DRAM timing constraints are met for each experiment.}}
Each read operation reads a cache line where all bytes of the cache line
contain the data value 0xAA.  All reads are performed to Bank~0, Row~128,
and back-to-back reads are interleaved across different columns (see
Section~\ref{sec:datadep:toggle}.
For the validation experiments, we measure the power consumption of 
22~DRAM modules (8~modules for Vendor~A,
and 7~modules each for Vendors~B and C), 
where the modules of each vendor are selected randomly from the
50~modules listed in Table~\ref{tbl:dimms}.
We generate traces that capture the behavior of each experiment, and then
feed them into each of the DRAM power models that we evaluate.}}

\rev{\ch{Figure~\ref{fig:validation} shows the \chI{\emph{mean absolute 
percentage error}} (MAPE) \chII{across each \chIII{of our validation experiments},}
for each \chI{DRAM power} model 
compared to the measured current of each vendor's DRAM
modules.  We make three observations from the figure.
First, the Micron power model has a very high error across all three vendors,
with a MAPE of 160.6\%, averaged across all three vendors.
The Micron model significantly overestimates the power consumption of
DRAM, as prior work has shown~\cite{ChandrasekarRunTime, Jung}, \chI{since it
does \emph{not} accurately model a number of important phenomena}, such as 
the fact that when only one bank is active, the DRAM module consumes 
much less power than when all banks are active~\cite{Jung}.
Second, DRAMPower has a MAPE of 32.4\%, averaged across all three
vendors.  While this is lower than the error \chI{of} the Micron power model,
DRAMPower still has high error rates for Vendors~A and B.  This is 
because DRAMPower does \chI{\emph{not}} capture the impact of our four observations
(see Section~\ref{sec:summary}).  \chI{The largest source of DRAMPower's
high error rates is its inability to accurately model}
(1)~the impact of data dependency on the power consumed during each read 
operation; and
(2)~the much lower power consumed by activate and precharge operations
and during idle time, as compared to the \chII{\idd[0] and \idd[2N]} values.
Third, VAMPIRE has a MAPE of only 6.8\%, averaged across all three vendors.
Unlike the Micron power model and DRAMPower, VAMPIRE has a low MAPE for
all three vendors (with the highest per-vendor MAPE being 7.1\%).}}

\begin{figure}[h]
  \centering
  \includegraphics[width=\basewidthwide, trim=66 302 60 309, clip]{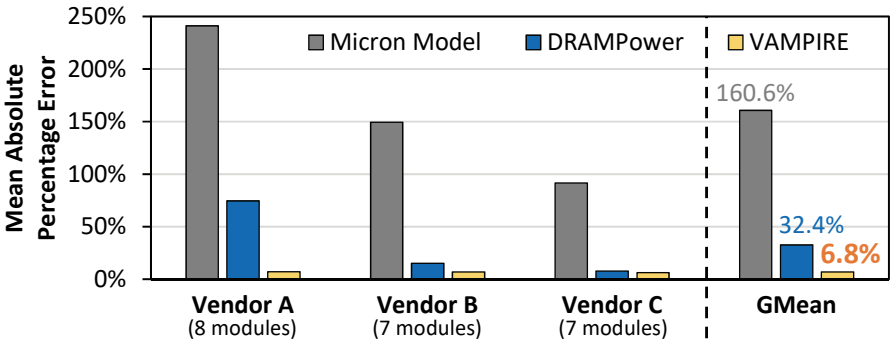}%
  \caption{\rev{\ch{Mean absolute percentage error of state-of-the-art DRAM power models
  and of VAMPIRE, compared to real measured DRAM power.}}}%
  \label{fig:validation}
\end{figure}

\rev{\ch{We conclude that VAMPIRE is significantly more accurate than state-of-the-art
DRAM power models, because it incorporates \chI{our new observations on 
(1)~the \chII{large differences} between the real measured DRAM power and the
vendor-provided \idd values,
(2)~data-dependent DRAM power consumption, and
(3)~the impact of structural variation on DRAM power consumption.}}}

\subsection{\ch{Evaluating DRAM Power Consumption with Large Applications}}
\label{sec:studies:meth}

\ch{In addition to the measurement-based validation, we compare the power consumption
reported by VAMPIRE to the power consumption reported by 
DRAMPower~\cite{drampower, ChandrasekarRunTime} (the best state-of-the-art
DRAM power model) when we simulate the memory access behavior of
real applications.  Unfortunately, we cannot compare the reported power consumption
numbers to real DRAM power measurements, due to the inability of SoftMC to
interactively execute command traces from full applications~\cite{hassan.hpca17}.
Instead, we measure the \emph{relative error} of DRAMPower, which does not
capture several aspects of DRAM power consumption, \chI{with respect} to VAMPIRE, which
captures all of the \chII{key} observations that we \chII{make
based on} our \chI{experimental} characterization.}

For each \chI{DRAM power} model, we determine the power consumed by a single channel of DDR3L
memory \chI{while executing applications on a single CPU core.}
Table~\ref{tbl:config} shows the system configuration that we simulate.
\chI{We evaluate} 23~applications from the SPEC CPU2006 suite~\cite{henning.can06}.
\ch{We} use Pin~\cite{pintool} to record the last level cache misses generated by 
each application.  
\ch{We fast-forward each application past its} initialization phase, and collect 
\ch{a memory trace} for a representative 100~million instruction \ch{portion.}  
We generate the input for each model by executing the memory trace on
Ramulator\ch{~\cite{kim.cal15, ramulator.github}}, an open-source DRAM 
\chI{system} simulator.  We modify
Ramulator to output the correct format of DRAM commands for VAMPIRE,
\chI{which includes data values for write commands}.

\begin{table}[h]
  \centering
  \caption{Evaluated system configuration.}
  \label{tbl:config}
  \vspace{-7pt}
  \footnotesize
    \setlength{\tabcolsep}{.6em}
    \begin{tabular}{ll}
        \toprule
        \multirow{1}{*}{\bf Processor} & x86-64 ISA, \chI{one core,} \SI{3.2}{\giga\hertz}, 128-entry instruction window \\
        \midrule
        \textbf{Cache} & L1: \SI{64}{\kilo\byte}, 4-way associative; L2: \SI{2}{\mega\byte}, 16-way associative \\
        \midrule
        \textbf{Memory Controller} & \multirow{1}{*}{64/64-entry read/write request \ch{queues},
            FR-FCFS~\cite{rixner.isca00,zuravleff.patent97}} \\
        \midrule
        \multirow{1}{*}{\bf DRAM} & DDR3L-800~\cite{ddr3l.jedec13}, 1 channel, 1 rank/8 banks per channel\\
        \bottomrule
    \end{tabular}
\end{table}

Figure~\ref{fig:modelerror} shows the relative error for the Micron power model 
\ch{compared to VAMPIRE}.
We show the relative error as a box plot, where the box represents the quartiles
of the output for each application.  We observe that there is significant error
in \ch{DRAMPower} compared to VAMPIRE.
\ch{The average relative errors of DRAMPower for Vendors A, B, and C
are 58.3\%, 45.0\%, and 33.5\%, respectively.}
From the figure, we observe that the error actually varies significantly from 
application to application.
\ch{This is because relative error of each application is highly dependent on
the application's memory access behavior.  In general, DRAMPower reports
a much higher power consumption value than VAMPIRE for applications that are
memory intensive, and reports a lower power consumption value than VAMPIRE
for applications that are not memory intensive.
We conclude that the properties of DRAM power consumption that are missing from
DRAMPower greatly affect its reported power consumption.}
\chI{VAMPIRE, by accurately modeling key properties that affect DRAM power
consumption, provides much more accurate estimates of
DRAM power consumption by large applications.}

\begin{figure}[h]
  \centering
  \includegraphics[width=0.95\basewidth, trim=55 318 55 318, clip]{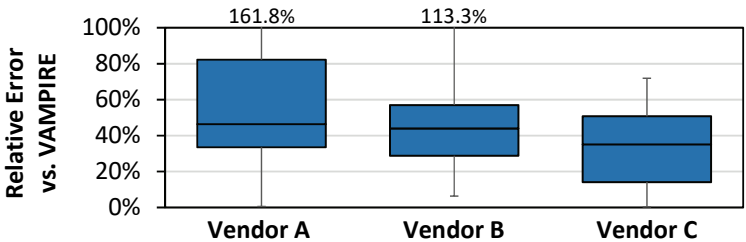}%
  \caption{Box plots showing the distribution of the relative error for 
  \chIII{the DRAMPower model compared to VAMPIRE}, across the applications that we simulate.
  \chII{Each box illustrates the quartiles of the distribution of the relative error
  for each application, and the whiskers illustrate the minimum and maximum error values,
  \chIII{across all applications evaluated}.}}%
  \label{fig:modelerror}
\end{figure}

\subsection{\protect\rev{\chI{Example Applications of VAMPIRE}}}

\rev{A large fraction of the error that we observe in existing power models is 
likely the result of missing data dependency and variation characteristics in
these models.  
By capturing these characteristics in detail, VAMPIRE allows researchers,
architects and system designers to accurately \ch{evaluate} 
\chI{and take into account} the \chI{various} sources of
power consumption in modern DRAM modules.
This has several implications for DRAM architectures and system
designs, of which we present three examples.
First, taking advantage of the systematic structural variation that VAMPIRE captures 
across different banks and rows in a DRAM module, a system designer can
rewrite the virtual memory manager in the operating system to optimize physical
page allocation \ch{for low energy consumption}.  Instead of treating all physical
page locations within DRAM as equal, the virtual memory manager could allocate
data that is accessed more frequently to those physical pages that reside in
banks and rows \ch{that} consume less power.
Second, VAMPIRE's insights can be used to determine when to schedule
power-down modes for DRAM.
VAMPIRE provides \ch{accurate} information on the actual power saved
during power-down mode \ch{by a DRAM module}, and on the \ch{actual}
power required to wake the module back up
to full-power mode, allowing designers and architects to accurately predict
whether there is enough time spent in power-down mode to amortize the
\ch{performance and power} \chI{overheads} of powering down and waking up the module.
Third, \ch{VAMPIRE's model of data-dependent power consumption} can
be used to design alternate data encodings that \ch{reduce}
power consumption \chI{within \chII{a DRAM chip}}.  We discuss one such data encoding
in Section~\ref{sec:studies:encoding}, and show how it takes advantage
of the \ch{data-dependent} behavior captured by VAMPIRE to reduce DRAM energy.
\chI{We believe there are many other use cases of VAMPIRE, and leave it to
future work to uncover such other use cases.}}

\section{Case Study: Data Encoding}
\label{sec:studies}
\label{sec:studies:encoding}

VAMPIRE enables a wide range of
studies that were \chI{\emph{not}} possible using existing models,
\ch{because it captures} characteristics such as data dependency and structural
variation \ch{that the existing models do \chI{\emph{not}} take into account}.  
\ch{An example of such a study is exploring how the DRAM power consumption
changes for different \emph{data encodings}, \chI{i.e., the
mechanisms with which the \chII{memory controller and/or DRAM module}}
\chIII{transform} the cache line data values that \chI{are 
stored} in the DRAM module.}
In this section, we examine the potential of \ch{cache line encodings} that
exploit the data dependency of \chI{DRAM} power \ch{consumption.}

Prior studies on specialized data encodings for DRAM\chI{~\cite{Wilson, Ekman, stan.tvlsi95, stan.glsvlsi95, hollis.tcasii09, pekhimenko.hpca16}} have largely focused on minimizing the amount of bit toggling that takes place on the off-chip memory channel (see Section~\ref{sec:related:encoding}).  When a 64-byte cache line is transmitted along the 64-bit memory channel, the data is split up into eight \emph{bursts}, and is sent one burst at a time.  This can increase DRAM power consumption due to bit toggling across different bursts \emph{from the same cache line}.  A number of studies\chI{~\cite{pekhimenko.hpca16, udipi.hipc09, bojnordi.micro13, beckmann.micro03, stan.tvlsi95, stan.glsvlsi95, hollis.tcasii09}} have shown the increased power consumption due to this inter-burst bit toggling.
\chI{In contrast to all these prior studies, we} study the data dependency and bit toggling that takes place \ch{\emph{within}} \chI{the DRAM chip}, where the bit toggling occurs across \emph{different cache lines} \ch{(see Section~\ref{sec:datadep:toggle})}.  As we discuss in Section~\ref{sec:datadep}, the amount of power consumed during read and write operations depends on the number of ones in the cache line, and on the number of bits that \ch{are toggled} \chI{within DRAM}.

\subsection{Encodings Studied}
\label{sec:studies:encoding:list}

We examine how four different \ch{cache-line-level} data encodings can affect \ch{DRAM} power consumption
\ch{when they are applied to the data before the data is \chI{written to DRAM}}:
\begin{itemize}

  \item \emph{Baseline}: \ch{The data is not encoded before being transferred to DRAM.}

  \item \emph{Base-Delta Immediate} (BDI)~\cite{pekhimenko.pact12}: We apply BDI compression to the data.  Prior work\chI{~\cite{pekhimenko.hpca16}} has shown that many compression algorithms, including BDI, \ch{can} consume more power \ch{than \emph{Baseline}} as a result of the bit \ch{toggling} that takes place on the memory channel.  

  \item \emph{Optimized}: \ch{This per-byte encoding scheme encodes the most-
frequently-used byte values using the least number of ones in the encoded byte value.} 
For each \ch{application}, we sort all possible byte values (i.e., 0--255) based on their frequency of occurrence.\footnote{Note that \chI{our goal is to perform} limit studies to gauge the potential of \chI{data encodings that exploit} DRAM power variation.  As such, 
we \chI{do not assess} the practicality of 
\ch{sorting all byte values that are read and written by an application}.
We leave such considerations for future work, which we hope will develop a \chI{more} practical encoding mechanism.}  
The byte values are then \ch{assigned \chI{to encoded values} such} that the most frequent byte value is assigned to the encoded value with the least number of
ones (i.e., an encoded value of zero), and the least frequent byte value is assigned to the encoded value with the most ones.
This \chI{encoding has two advantages.  First, it} 
minimizes the number of ones used for the \ch{application}, which \ch{is likely to} reduce the read power, 
\ch{as the power consumed by a read operation increases when the data that is 
read contains a greater number of ones}.
\chI{Second, it reduces the probability of toggling many bits at the same time
between transmissions of different cache lines.}
\ch{\chI{On the other hand, one drawback of this encoding scheme}
is that this increases the power consumed by write operations,
as the data-dependent power consumption of writes has an inverse relationship
with the number of ones as that of reads.}

  \item \emph{Optimized with Write Inversion} (OWI): We develop a variation of our \emph{Optimized} encoding, to minimize DRAM power consumption for \chI{\emph{both}} reads and writes.  Our measurements in Section~\ref{sec:datadep} show that read power increases with the number of ones in the cache line, and write power \emph{decreases} with the number of ones, due to I/O driver power and data-dependent power consumption within DRAM.
In order to maximize \ch{the} power savings, we assume that cache lines \ch{that are to be} written to DRAM are first transformed using our Optimized encoding, and then \chI{\emph{inverted} (i.e., bitwise complemented)} by the memory controller.  Once the \ch{data that is to be written} passes through the I/O drivers and peripheral circuitry within the module, \ch{the DRAM chip inverts} \chI{(\chII{i.e.,} bitwise complements)} the data before it is written to the DRAM cells.

\end{itemize}

\rev{The overhead to implement the \ch{\emph{Optimized} and \emph{OWI}} encodings is \ch{small}.
We assume that \ch{both encoding mechanisms use} \SI{256}{\bit}$~\times~$\ch{\SI{8}{\bit}}
lookup tables in each DRAM chip.  We use \ch{CACTI 7.0~\cite{cacti7} to 
estimate the area and latency of the lookup table using a \SI{22}{\nano\meter}
manufacturing process technology, and find that each table requires only \SI{0.0024}{\milli\meter\squared}
of area, and can perform a lookup in \SI{0.134}{\nano\second}.}
The lookup tables are hard-coded, and \ch{we conservatively assume that the lookup 
adds one DRAM cycle of latency}.
\chII{This \chIII{latency is similar to that incurred by}} channel encodings such as data bus inversion\chI{~\cite{stan.tvlsi95, stan.glsvlsi95, hollis.tcasii09}},
\chII{which is \chIII{an optional feature in}} LPDDR4 memory~\cite{lpddr4.jedec17}.}

We use the same simulation methodology described in Section~\ref{sec:model}, 
and develop a tool to encode the data recorded by Pin\chI{~\cite{pintool}}
and Ramulator\chI{~\cite{kim.cal15, ramulator.github}} for each encoding.
To account for the performance overhead of the \ch{\emph{Optimized} and \emph{OWI}} encodings,
we add one \ch{DRAM} cycle to each read and write operation in Ramulator.
\chI{We account for the additional energy consumed \chIII{for encoding data using
\emph{Optimized} and \emph{OWI}}.}

\subsection{Evaluation}
\label{sec:studies:encoding:eval}

Figure~\ref{fig:encoding} shows the average energy consumed \chI{inside
DRAM for} each encoding, normalized to the energy consumption of \emph{Baseline}, 
for each \ch{vendor}.  We make three observations from the figure.
\chI{First}, the \emph{OWI} encoding, which optimizes the number of ones for both
reads and writes, achieves significant savings.
\ch{\emph{OWI} reduces the energy consumption over \emph{Baseline}}
by 12.2\% on average, and up to 28.6\%.  
Second, our \emph{Optimized} encoding provides no tangible reduction in
energy, as the increase in write current energy \ch{due to data dependency}
cancels out any savings from optimizing the read current.
\chI{Third}, data compressed with BDI consumes no more energy than uncompressed data.
While compression may increase the amount of energy consumed on the memory 
channel, we see no notable effect \ch{\emph{within}} DRAM, due to the relatively minor 
impact bit toggling has on the current \ch{observed in our real chip
measurements} (see Section~\ref{sec:datadep:toggle}).

\begin{figure}[h]
  \centering
  \includegraphics[width=0.8\basewidth, trim=81 215 70 204, clip]{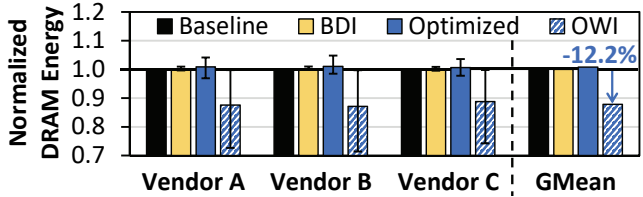}%
  \caption{Average energy consumption for various \ch{cache-line-level data} encodings, normalized to energy consumed with \emph{Baseline} encoding.
  \ch{Error bars show maximum and minimum normalized energy values across all applications.}}%
  \label{fig:encoding}
\end{figure}

We conclude that \ch{(1)~}DRAM energy can \ch{be reduced} by using a power-aware
\chI{and DRAM-data-dependence-aware} data encoding \chII{mechanism};
\ch{and (2)~}VAMPIRE, our new power model, enables the exploration of the energy 
consumption of the encodings we have examined, as well as the \chI{new} encodings that we hope 
future research will develop.

\section{Related Work}
\label{sec:related}

Several DRAM power and energy models exist. Most of these models are 
derived from the \idd \chI{(current)} \ch{values} provided 
by DRAM manufacturers \ch{in their datasheets for worst-case power consumption}.  
\chI{To our} knowledge, our work is the first to 
(1)~\rev{\ch{characterize the power consumption of a large number of real DRAM modules
from three major DRAM vendors, across a wide variety of tests;}}
(2)~demonstrate how \ch{\idd values} often deviate significantly from actual DRAM power consumption
\ch{and contain large guardbands};
\chI{(3)~comprehensively demonstrate the inter-vendor and intra-vendor 
module-to-module variation of DRAM power consumption;}
\ch{\chI{(4)~show that} DRAM power consumption depends on
data values and structural variation;}
\ch{\chI{(5)}~show that DRAM power consumption has \chI{\emph{not}} decreased as much as
vendor specifications indicate over successive generations; and
\chI{(6)}~develop a DRAM power model that accounts for the impact of 
\idd value guardbands, data dependency, and structural variation on
the power that is consumed.}

\subsection{Architectural Power Models}

The Micron DRAM power model~\cite{micron.2015} uses \ch{\idd values,} command count, execution time, and timing parameters from datasheets to calculate power consumption. However, \chI{prior works~\cite{Chandrasekar, Jung} have shown that} \ch{the Micron power model} does \ch{\emph{not} 
(1)~account for any additional time that may elapse between two DRAM commands~\cite{Chandrasekar},
(2)~model an open-\chI{row} policy~\cite{Chandrasekar} \chI{or more sophisticated
row buffer management policies, or}
(3)~properly account for the power consumed when the number of active banks changes~\cite{Jung}.}
\chI{As we discuss in Section~\ref{sec:model:validation}, we find that the 
Micron power model also does \emph{not} \chIII{take into account
typical-case DRAM power consumption (\chIV{which, as we show in Section~\ref{sec:idd},}
is much lower than the \idd values specified by vendors in datasheets),}
data-dependent power consumption, or the impact of structural variation on
power consumption in DRAM.}

DRAMPower\chI{~\cite{Chandrasekar, drampower}} is an open-source tool that can be used at the command level and transaction level to estimate power consumption. It allows DRAM command traces to be logged and also has an optional command scheduler, which emulates a memory controller. While DRAMPower \ch{accounts for \chI{three of the} major characteristics that are missing from the Micron power model \chI{(the additional time between commands, alternative row buffer management policies, and the number of active banks)}},
it is still predominantly based upon the worst-case \idd estimates extracted from datasheets.
With the exception of (1)~the change in active background power based on
the number of banks open~\cite{Mathew, Jung} and (2)~\idd values 
\chI{and module-to-module variation} for a small number \ch{(10)} of tested DRAM
modules~\cite{drampower}, the power models and input parameters employed by
DRAMPower are \emph{not} based off of measured data, and \chI{they} do \ch{\emph{not} take into account} 
data dependency, \chI{a comprehensive notion of module-to-module variation,
or structural variation} in DRAM power.
\rev{To our knowledge, the limited measurements from real devices used in DRAMPower 
are performed \chI{on a} small
number of DRAM modules, do not capture any variation trends, and do not
span across multiple vendors.}
As we show in \ch{Section~\ref{sec:model:validation}, 
because it is still based off of worst-case \idd values, DRAMPower has a 
mean absolute percentage error of 32.4\% compared to real measured
power in the DRAM chips that we test.}
In contrast, VAMPIRE, our power model, is based fully off of measured data, and captures
\ch{the impact of data dependency and structural variation on power consumption,
providing high accuracy \chI{(\chII{only} 6.8\% error, as shown in Section~\ref{sec:model:validation})}}.

CACTI-D~\cite{Cacti} and Vogelsang~\cite{vogelsang.micro10} use circuit-level models to characterize the power consumed by DRAM.  The peripheral circuitry used in CACTI-D\chI{~\cite{Cacti}} is largely based on SRAM caches, and does not accurately reflect the design of DRAM peripheral logic.  Vogelsang\chI{~\cite{vogelsang.micro10}} uses \ch{transistor}-level models to predict future trends in \chII{DRAM} power consumption, but the models are calibrated to \ch{datasheet} \idd values, and do not capture 
\chI{data dependency, module-to-module variation, or structural variation}.
\chI{Similarly, the PADRAM model~\cite{Lebeck} models a subset of RDRAM\chII{~\cite{rambus.rdram.website}} components
and vendor specifications to develop a DRAM power model.  Aside from not capturing
real DRAM power behavior, the PADRAM model is designed for RDRAM modules, and is not fully compatible
with the \chII{modern} DDR SDRAMs that we characterize.}

Wattch~\cite{Wattch}, McPAT~\cite{McPAT}, and the model by Fan et al.~\cite{Fan} are models designed 
to capture the power dissipation and energy consumption of modern processors.  While these models
can include a DRAM power component, this component again relies on the \idd datasheet values.

\subsection{Low-Power DRAM}

Prior works propose models, chip designs, and architectures to reduce DRAM power consumption. 
A number of works~\cite{chang.pomacs2017, david.icac11, Choi} \ch{study} how to reduce the voltage and/or frequency of DRAM to lower
power consumption.
Several prior works exploit \ch{low-power modes} to increase the time spent in a low-power state~\cite{Fan, Lebeck, kandemir.date2007,luz.dac2002,kandemir.iccad2001, Anagnostopoulou, Deng1,Deng2, Aggarwal, Bi, Diniz, Amin, Lyuh, stuecheli.isca2010,stuecheli.micro2011,kandemir.cases2004}.
Row-buffer-aware DRAM designs\ch{~\cite{sudan.asplos2010,kaseridis.micro2011, SALP}} optimize data placement to increase row buffer hits and, thus, reduce the energy spent on row activation and precharge. 
A number of DRAM architectures reduce DRAM power by activating only a fraction of the row~\cite{Chatterjee, Cooper-Balis, Udipi, Zhang},
\ch{a fraction of the bitlines~\cite{Tiered-Latency}, or a fraction of the DRAM chips on a module~\cite{Ware, Yoon, Zheng},
by reducing the access latency\chI{~\cite{lee.hpca15, lee.sigmetrics17,
LISA, seshadri.micro13, Hassan}}},
\chII{or by reducing the operating frequencies of some layers in a 3D-stacked DRAM chip~\cite{lee.taco16}}.

Many works study eliminating margins designed for worst-case DRAM modules to improve energy efficiency.  \ch{Various works~\cite{Liu,Agrawal,Bhati,Khan,Lin,liu.isca2012,Ohsawa,Patel,Qureshi,PARBOR, Wilkerson, Venkatesan}} reduce DRAM refresh power by characterizing cell retention times and reducing unnecessary refresh operations.  \ch{Multiple} works~\cite{chang.sigmetrics2016,LISA,Hassan,SALP,lee.hpca15,lee.sigmetrics17,Tiered-Latency,mukundan.isca13,ChandrasekarRunTime,Shin} make DRAM more energy efficient by reducing the latencies of DRAM operations.

\chII{Low-power DDR (LPDDR)~\cite{lpddr3.jedec15, lpddr4.jedec17} is a 
family of DRAM architectures designed by JEDEC for use in low-power systems
(e.g., mobile devices).  LPDDR architectures employ major design changes over
conventional DDR memory architectures.  Two such changes include the
use of a low-voltage-swing I/O interface (\chIII{which, in LPDDR4 DRAM,} consumes 40\% less I/O power 
than DDR4 DRAM~\cite{Choi}), and the inclusion of additional low-power modes 
that make use of \chIII{lower supply voltage levels that are} not available in DDR memory.}

\ch{None of these works characterize DRAM power \chI{consumption}, and their ideas are orthogonal to ours.}

\subsection{Experimental DRAM Characterization}

\ch{Various experimental studies}\chII{~\cite{ChandrasekarRunTime,hassan.hpca17,Jung,Meza,Schroeder,Sridharan,Sridharan2,Matthias,lee.sigmetrics17,
Liu, lee.hpca15, chang.sigmetrics2016, chang.pomacs2017, Patel, PARBOR, Wilkerson,
khan.micro17, kim.isca14, kim.hpca18, Khan, Qureshi, mutlu.date17, hwang.asplos12}} characterize DRAM reliability, data retention, and latency by measuring \ch{characteristics of real} DRAM chips, but \ch{they} do not measure power consumption. Kim et al.~\cite{kim.wwc2000} study DRAM energy consumption under different processor cache configurations, but do not study
how different DRAM operations contribute to energy consumption in modern DRAM
devices.
Jung et al.\chIII{~\cite{Jung}} experimentally characterize a limited subset of \idd values to study the effect of the number of open banks on active background power, but do not characterize \ch{any other aspect} of DRAM power consumption. 
\chIII{A subsequent work by Jung et al.~\cite{Matthias} demonstrates a power measurement platform 
for DDR3 DRAM, but does not (1)~comprehensively report power consumption across
all DRAM operations; or
(2)~study the effects of process variation, data dependency, or structural variation on power consumption.}
In Section~\ref{sec:idd}, we measure a comprehensive \ch{set} of \idd values, and we show how the power consumption \chI{differs and} varies significantly from the \ch{values provided in the datasheets}.

\subsection{Compression and Encoding Schemes}
\label{sec:related:encoding}

Prior works \chI{propose} using compression and encoding schemes for caches\chII{~\cite{Yang, Alameldeen, Dusser, Chen, pekhimenko.pact12, vijaykumar.isca15, pekhimenko.hpca15}} and memory\chI{~\cite{Wilson, Ekman, pekhimenko.micro13}}. Most of these are pattern-based schemes and perform compression at a word granularity. 
Base-Delta-Immediate compression\chI{~\cite{pekhimenko.pact12, pekhimenko.micro13}} performs compression at a cache line granularity, by identifying cache lines where \chI{the data value of} each word in the line \chI{is} within a small range of values.
\chII{Data bus inversion (DBI)~\cite{stan.tvlsi95, stan.glsvlsi95, hollis.tcasii09}
is an encoding that reduces the power consumed by the memory channel,
by inverting the data transmitted during each data burst when the data contains
more zeroes than ones.  DBI is an optional feature that can be enabled in 
LPDDR4 memories~\cite{lpddr4.jedec17}.}

\chI{These prior works either do not study the impact of compression encodings on DRAM
power consumption, or do not study the impact of implementing \chII{different} encodings within the
DRAM \chII{chip}.  In Section~\ref{sec:studies:encoding}, we study how a new power-aware cache line
encoding \chIII{mechanism} can reduce energy consumption in a DRAM \chII{chip} by exploiting our observations on data-dependent DRAM power consumption.}


\section{Conclusion}
\label{sec:conclusion}

DRAM power consumption \ch{is a critical} issue in contemporary computer 
systems, as \ch{DRAM} now accounts for as much as half of the total system
\ch{power consumption}.
\chI{While there is a pressing need to invent new low-power DRAM architectures,
existing DRAM power models do \emph{not} accurately model the power
consumed by DRAM modules, limiting researchers' understanding of the sources of
power consumption.  The existing DRAM power models are inaccurate because
they rely only on vendor-specified current measurements, and do \emph{not}
capture several important characteristics of power consumption that are present
in real DRAM devices.}

To address the shortcomings of existing DRAM power models, we \ch{first perform \rev{an}}
extensive \emph{experimental characterization} of the power consumed by real
state-of-the-art DDR3L DRAM devices.
We measure the current consumed by \dimmcnt~DRAM modules from three major 
vendors, and make four key new observations that previous models did not
capture:
(1)~the actual current consumed \ch{deviates} significantly from the vendor 
\ch{specifications in the datasheets};
(2)~the data value \ch{that is} read from or written to DRAM significantly impacts 
power consumption;
(3)~\ch{power consumption} varies significantly based on which bank or row of a DRAM module
is being \ch{accessed}; and
(4)~across \ch{successive} process technology generations, the actual power \chI[reduction} is
often much lower than the savings indicated by vendor specifications.

Based on our real device measurements and analysis, we build VAMPIRE,
a new, accurate DRAM power model.
VAMPIRE enables studies that could \ch{\emph{not}} be performed using prior 
\chI{DRAM power} models.
For example, we show that a \ch{new} power-aware data encoding \ch{scheme}
can reduce DRAM power consumption by an average of 12.2\% (up to 28.6\%).
We will release VAMPIRE and all of our raw measurement data 
online\ch{~\cite{vampire.github}}.  \chI{We} hope 
that the findings in this work and our new power model will inspire
new research directions, new ideas, and rigorous \chI{and more accurate} evaluations in
power-and energy-aware \chI{memory system} design.

\section*{Acknowledgments}
We thank our shepherd Thomas Wenisch, the anonymous reviewers, and SAFARI
\chI{Research Group} members for feedback. 
\ch{Thanks to Naveen Kakarla for his assistance with the experimental
validation of VAMPIRE.}
We acknowledge the \chII{generous support of our industrial partners
(Google, Intel, Microsoft, NVIDIA, Samsung, and VMware)} and the
United States Department of Energy. This research was supported in part by the 
\ch{Semiconductor Research Corporation and the 
National Science Foundation} (grants 1212962 and 1320531).


\bibliographystyle{IEEEtranS}
\bibliography{references}

\begin{thebibliography}{100}
\providecommand{\url}[1]{#1}
\csname url@samestyle\endcsname
\providecommand{\newblock}{\relax}
\providecommand{\bibinfo}[2]{#2}
\providecommand{\BIBentrySTDinterwordspacing}{\spaceskip=0pt\relax}
\providecommand{\BIBentryALTinterwordstretchfactor}{4}
\providecommand{\BIBentryALTinterwordspacing}{\spaceskip=\fontdimen2\font plus
\BIBentryALTinterwordstretchfactor\fontdimen3\font minus
  \fontdimen4\font\relax}
\providecommand{\BIBforeignlanguage}[2]{{%
\expandafter\ifx\csname l@#1\endcsname\relax
\typeout{** WARNING: IEEEtranS.bst: No hyphenation pattern has been}%
\typeout{** loaded for the language `#1'. Using the pattern for}%
\typeout{** the default language instead.}%
\else
\language=\csname l@#1\endcsname
\fi
#2}}
\providecommand{\BIBdecl}{\relax}
\BIBdecl

\bibitem{Aggarwal}
N.~Aggarwal, J.~F. Cantin, M.~H. Lipasti, and J.~E. Smith, ``{Power-Efficient
  DRAM Speculation},'' in \emph{HPCA}, 2008.

\bibitem{Agrawal}
A.~Agrawal, A.~Ansari, and J.~Torrellas, ``{Mosaic: Exploiting the Spatial
  Locality of Process Variation to Reduce Refresh Energy in On-Chip eDRAM
  Modules},'' in \emph{{HPCA}}, 2014.

\bibitem{Alameldeen}
A.~R. Alameldeen and D.~A. Wood, ``{Adaptive Cache Compression for
  High-Performance Processors},'' in \emph{{ISCA}}, 2004.

\bibitem{Amin}
A.~M. Amin and Z.~A. Chishti, ``{Rank-Aware Cache Replacement and Write
  Buffering to Improve DRAM Energy Efficiency},'' in \emph{{ISLPED}}, 2010.

\bibitem{Anagnostopoulou}
V.~Anagnostopoulou, S.~Biswas, H.~Saadeldeen, A.~Savage, R.~Bianchini, T.~Yang,
  D.~Franklin, and F.~T. Chong, ``{Barely Alive Memory Servers: Keeping Data
  Active in a Low-Power State},'' \emph{{ACM JETC}}, 2012.

\bibitem{bakhoda.ispass09}
A.~Bakhoda, G.~Yuan, W.~W.~L. Fung, H.~Wong, and T.~M. Aamodt, ``{Analyzing
  CUDA Workloads Using a Detailed GPU Simulator},'' in \emph{ISPASS}, 2009.

\bibitem{beckmann.micro03}
B.~M. Beckmann and D.~A. Wood, ``{TLC: Transmission Line Caches},'' in
  \emph{MICRO}, 2003.

\bibitem{bhati.islped13}
I.~Bhati, Z.~Chishti, and B.~Jacob, ``{Coordinated Refresh: Energy Efficient
  Techniques for DRAM Refresh Scheduling},'' in \emph{{ISLPED}}, 2013.

\bibitem{Bhati}
I.~Bhati, Z.~Chishti, S.~Lu, and B.~Jacob, ``{Flexible Auto-Refresh: Enabling
  Scalable and Energy-Efficient DRAM Refresh Reductions},'' in \emph{{ISCA}},
  2015.

\bibitem{Bi}
M.~Bi, R.~Duan, and C.~Gniady, ``{Delay-Hiding Energy Management Mechanisms}
  for {DRAM},'' in \emph{HPCA}, 2010.

\bibitem{gem5}
N.~Binkert, B.~Beckmann, G.~Black, S.~K. Reinhardt, A.~Saidi, A.~Basu,
  J.~Hestness, D.~R. Hower, T.~Krishna, S.~Sardashti, R.~Sen, K.~Sewell,
  M.~Shoaib, N.~Vaish, M.~D. Hill, and D.~A. Wood, ``{gem5: A Multiple-ISA Full
  System Simulator with Detailed Memory Model},'' \emph{CAN}, vol.~39, June
  2011.

\bibitem{bojnordi.micro13}
M.~N. Bojnordi and E.~\.{I}pek, ``{DESC: Energy-Efficient Data Exchange Using
  Synchronized Counters},'' in \emph{MICRO}, 2013.

\bibitem{Wattch}
D.~Brooks, V.~Tiwari, and M.~Martonosi, ``{Wattch: A Framework for
  Architectural-Level Power Analysis and Optimizations},'' in \emph{{ISCA}},
  2000.

\bibitem{cai.hpca17}
Y.~Cai, S.~Ghose, Y.~Luo, K.~Mai, O.~Mutlu, and E.~F. Haratsch,
  ``{Vulnerabilities in MLC NAND Flash Memory Programming: Experimental
  Analysis, Exploits, and Mitigation Techniques},'' in \emph{HPCA}, 2017.

\bibitem{cai.dsn15}
Y.~Cai, Y.~Luo, S.~Ghose, E.~F. Haratsch, K.~Mai, and O.~Mutlu, ``{Read Disturb
  Errors in MLC NAND Flash Memory: Characterization and Mitigation},'' in
  \emph{DSN}, 2015.

\bibitem{cai.hpca15}
Y.~Cai, Y.~Luo, E.~F. Haratsch, K.~Mai, and O.~Mutlu, ``{Data Retention in MLC
  NAND Flash Memory: Characterization, Optimization, and Recovery},'' in
  \emph{HPCA}, 2015.

\bibitem{cai.iccd12}
Y.~Cai, G.~Yalcin, O.~Mutlu, E.~F. Haratsch, A.~Cristal, O.~Unsal, and K.~Mai,
  ``{Flash Correct and Refresh: Retention Aware Management for Increased
  Lifetime},'' in \emph{ICCD}, 2012.

\bibitem{cai.itj13}
Y.~Cai, G.~Yalcin, O.~Mutlu, E.~F. Haratsch, A.~Cristal, O.~Unsal, and K.~Mai,
  ``{Error Analysis and Retention-Aware Error Management for NAND Flash
  Memory},'' \emph{Intel Technol. J.}, May 2013.

\bibitem{cai.procieee17}
Y.~Cai, S.~Ghose, E.~F. Haratsch, Y.~Luo, and O.~Mutlu, ``{Error
  Characterization, Mitigation, and Recovery in Flash-Memory-Based Solid-State
  Drives},'' \emph{Proceedings of the IEEE}, 2017.

\bibitem{cai.book18}
Y.~Cai, S.~Ghose, E.~F. Haratsch, Y.~Luo, and O.~Mutlu, ``{Reliability Issues
  in Flash-Memory-Based Solid-State Drives: Experimental Analysis, Mitigation,
  Recovery},'' in \emph{Inside Solid State Drives (SSDs)}, 2nd~ed.\hskip 1em
  plus 0.5em minus 0.4em\relax Springer Nature, 2018.

\bibitem{cai.date12}
Y.~Cai, E.~F. Haratsch, O.~Mutlu, and K.~Mai, ``{Error Patterns in MLC NAND
  Flash Memory: Measurement, Characterization, and Analysis},'' in \emph{DATE},
  2012.

\bibitem{cai.date13}
Y.~Cai, E.~F. Haratsch, O.~Mutlu, and K.~Mai, ``{Threshold Voltage Distribution
  in MLC NAND Flash Memory: Characterization, Analysis, and Modeling},'' in
  \emph{DATE}, 2013.

\bibitem{cai.iccd13}
Y.~Cai, O.~Mutlu, E.~F. Haratsch, and K.~Mai, ``{Program Interference in MLC
  NAND Flash Memory: Characterization, Modeling, and Mitigation},'' in
  \emph{ICCD}, 2013.

\bibitem{cai.sigmetrics14}
Y.~Cai, G.~Yalcin, O.~Mutlu, E.~F. Haratsch, O.~Unsal, A.~Cristal, and K.~Mai,
  ``{Neighbor-Cell Assisted Error Correction for MLC NAND Flash Memories},'' in
  \emph{SIGMETRICS}, 2014.

\bibitem{ChandrasekarRunTime}
K.~Chandrasekar, S.~Goossens, C.~Weis, M.~Koedam, B.~Akesson, N.~Wehn, and
  K.~Goossens, ``{Exploiting Expendable Process-Margins in DRAMs for Run-Time
  Performance Optimization},'' in \emph{{DATE}}, 2014.

\bibitem{Chandrasekar}
K.~Chandrasekar, B.~Akesson, and K.~Goossens, ``{Improved Power Modelling of
  DDR SDRAMs},'' in \emph{{DSD}}, 2011.

\bibitem{drampower}
K.~Chandrasekar, C.~Weis, Y.~Li, S.~Goossens, M.~Jung, O.~Naji, B.~Akesson,
  N.~Wehn, and K.~Goossens, ``{DRAMPower: Open-Source DRAM Power \& Energy
  Estimation Tool},'' \url{http://www.drampower.info}.

\bibitem{chang.thesis17}
K.~K. Chang, ``{Understanding and Improving the Latency of DRAM-Based Memory
  Systems},'' Ph.D. dissertation, Carnegie Mellon Univ., 2017.

\bibitem{Chang}
K.~K. Chang, D.~Lee, Z.~Chishti, A.~Alameldeen, C.~Wilkerson, Y.~Kim, and
  O.~Mutlu, ``{Improving DRAM Performance by Parallelizing Refreshes with
  Accesses},'' in \emph{{HPCA}}, 2014.

\bibitem{LISA}
K.~K. Chang, P.~J. Nair, D.~Lee, S.~Ghose, M.~K. Qureshi, and O.~Mutlu,
  ``{Low-Cost Inter-Linked Subarrays (LISA): Enabling Fast Inter-Subarray Data
  Movement in DRAM},'' in \emph{{HPCA}}, 2016.

\bibitem{chang.pomacs2017}
K.~K. Chang, A.~G. A.~G. Ya\u{g}l{\i}k\c{c}{\i}, S.~Ghose, A.~Agrawal,
  N.~Chatterjee, A.~Kashyap, D.~Lee, M.~O'Connor, H.~Hassan, and O.~Mutlu,
  ``{Understanding Reduced-Voltage Operation in Modern DRAM Devices:
  Experimental Characterization, Analysis, and Mechanisms},'' in
  \emph{SIGMETRICS}, 2017.

\bibitem{chang.sigmetrics2016}
K.~K. Chang, A.~Kashyap, H.~Hassan, S.~Ghose, K.~Hsieh, D.~Lee, T.~Li,
  G.~Pekhimenko, S.~Khan, and O.~Mutlu, ``{Understanding Latency Variation in
  Modern DRAM Chips: Experimental Characterization, Analysis, and
  Optimization},'' in \emph{SIGMETRICS}, 2016.

\bibitem{Chatterjee}
N.~Chatterjee, M.~O'Connor, D.~Lee, D.~R. Johnson, M.~Rhu, S.~W. Kecker, and
  W.~J. Dally, ``{Architecting an Energy-Efficient DRAM System for GPUs},'' in
  \emph{HPCA}, 2017.

\bibitem{Cacti}
K.~Chen, S.~Li, N.~Muralimanohar, J.~H. Ahn, J.~B. Brockman, and N.~P. Jouppi,
  ``{CACTI-3DD: Architecture-Level Modeling for 3D Die-Stacked DRAM Main
  Memory},'' in \emph{{DATE}}, 2012.

\bibitem{Chen}
X.~Chen, L.~Yang, R.~Dick, L.~Shang, and H.~Lekatsas, ``{A High-Performance
  Microprocessor Cache Compression Algorithm},'' \emph{{TVLSI}}, 2010.

\bibitem{Choi}
J.~Y. Choi, ``{LPDDR4: Evolution for New Mobile Worlds},'' in \emph{{MEMCON}},
  2013.

\bibitem{Cooper-Balis}
E.~Cooper-Balis and B.~Jacob, ``{Fine-Grained Activation for Power Reduction in
  DRAM},'' \emph{{IEEE Micro}}, 2010.

\bibitem{david.icac11}
H.~David, C.~Fallin, E.~Gorbatov, U.~R. Hanebutte, and O.~Mutlu, ``{Memory
  Power Management via Dynamic Voltage/Frequency Scaling},'' in \emph{ICAC},
  2011.

\bibitem{luz.dac2002}
V.~{De La Luz}, M.~Kandemir, and I.~Kolcu, ``{Automatic Data Migration for
  Reducing Energy Consumption in Multi-Bank Memory Systems},'' in \emph{{DAC}},
  2002.

\bibitem{delaluz.hpca01}
V.~{De La Luz}, M.~Kandemir, N.~Vijaykrishnan, A.~Sivasubramaniam, and M.~J.
  Irwin, ``{DRAM Energy Management Using Software and Hardware Directed Power
  Mode Control},'' in \emph{{HPCA}}, 2001.

\bibitem{delaluz.dac02b}
V.~{De La Luz}, A.~Sivasubramaniam, M.~Kandemir, N.~Vijaykrishnan, and M.~J.
  Irwin, ``{Scheduler Based DRAM Energy Management},'' in \emph{{DAC}}, 2002.

\bibitem{Deng1}
Q.~Deng, D.~Meisner, L.~Ramos, T.~F. Wenisch, and R.~Bianchini, ``{MemScale:
  Active Low-Power Modes for Main Memory},'' in \emph{{ASPLOS}}, 2011.

\bibitem{Deng2}
Q.~Deng, D.~Meisner, L.~Ramos, T.~F. Wenisch, and R.~Bianchini, ``{Active
  Low-Power Modes for Main Memory with MemScale},'' in \emph{{MICRO}}, 2012.

\bibitem{Diniz}
B.~Diniz, D.~Guedes, J.~W.~Meira, and R.~Bianchini, ``{Limiting the Power
  Consumption of Main Memory},'' in \emph{ISCA}, 2007.

\bibitem{dong.tcad12}
X.~Dong, C.~Xu, Y.~Xie, and N.~P. Jouppi, ``{NVSim: A Circuit-Level
  Performance, Energy, and Area Model for Emerging Nonvolatile Memory},''
  \emph{TCAD}, June 2012.

\bibitem{Dusser}
J.~Dusser, T.~Piquet, and A.~Seznec, ``{Zero-Content Augmented Caches},'' in
  \emph{{ICS}}, 2009.

\bibitem{Ekman}
M.~Ekman and P.~Stenstr{\"{o}}m, ``{A Robust Main-Memory Compression Scheme},''
  in \emph{{ISCA}}, 2005.

\bibitem{elmore.tr16}
R.~Elmore, K.~Gruchalla, C.~Phillips, A.~Purkayastha, and N.~Wunder, ``{An
  Analysis of Application Power and Schedule Composition in a High Performance
  Computing Environment},'' National Renewable Energy Laboratory, Tech Report
  NREL/TP-2C00-65392, 2016.

\bibitem{Fan}
X.~Fan, C.~S. Ellis, and A.~R. Lebeck, ``{Memory Controller Policies for DRAM
  Power Management},'' in \emph{{ISLPED}}, 2001.

\bibitem{ferdman.sigplan2012}
M.~Ferdman, A.~Adileh, O.~Kocberber, S.~Volos, M.~Alisafaee, D.~Jevdjic,
  C.~Kaynak, A.~D. Popescu, A.~Ailamaki, and B.~Falsafi, ``{Clearing the
  Clouds: A Study of Emerging Scale-Out Workloads on Modern Hardware},'' in
  \emph{ASPLOS}, 2012.

\bibitem{gauss.book1809}
C.~F. Gauss, \emph{{Theoria Motus Corporum Coelestium in Sectionibus Conicis
  Solem Ambientium}}.\hskip 1em plus 0.5em minus 0.4em\relax F. Perthes et I.
  H. Besser, 1809.

\bibitem{Hassan}
H.~Hassan, G.~Pekhimenko, N.~Vijaykumar, V.~Seshadri, D.~Lee, O.~Ergin, and
  O.~Mutlu, ``{ChargeCache: Reducing DRAM Latency by Exploiting Row Access
  Locality},'' in \emph{{HPCA}}, 2016.

\bibitem{hassan.hpca17}
H.~Hassan, N.~Vijaykumar, S.~Khan, S.~Ghose, K.~Chang, G.~Pekhimenko, D.~Lee,
  O.~Ergin, and O.~Mutlu, ``{SoftMC: A Flexible and Practical Open-Source
  Infrastructure for Enabling Experimental DRAM Studies},'' in \emph{{HPCA}},
  2017.

\bibitem{cacti7}
{Hewlett Packard Enterprise}, ``{CACTI 7.0},''
  \url{https://github.com/HewlettPackard/cacti}.

\bibitem{hollis.tcasii09}
T.~M. Hollis, ``{Data Bus Inversion in High-Speed Memory Applications},''
  \emph{TCAS II}, 2009.

\bibitem{holzle.book09}
U.~Holzle and L.~A. Barroso, \emph{{The Datacenter as a Computer: An
  Introduction to the Design of Warehouse-Scale Machines}}.\hskip 1em plus
  0.5em minus 0.4em\relax {Morgan \& Claypool}, 2009.

\bibitem{hong.iedm10}
S.~Hong, ``{Memory Technology Trend and Future Challenges},'' in \emph{IEDM},
  2010.

\bibitem{hwang.asplos12}
A.~Hwang, I.~Stefanovici, and B.~Schroeder, ``{Cosmic Rays Don't Strike Twice:
  Understanding the Nature of DRAM Errors and the Implications for System
  Design},'' in \emph{ASPLOS}, 2012.

\bibitem{inoue.ijsc1988}
M.~Inoue, T.~Yamada, H.~Kotani, H.~Yamauchi, A.~Fujiwara, J.~Matsushima,
  H.~Akamatsu, M.~Fukumoto, M.~Kubota, I.~Nakao, N.~Aoi, G.~Fuse, S.~Ogawa,
  S.~Odanaka, A.~Ueno, and H.~Yamamoto, ``{A 16-Mbit DRAM with a Relaxed
  Sense-Amplifier-Pitch Open-Bit-Line Architecture},'' \emph{JSSC}, 1988.

\bibitem{ddr3.jedec12}
{JEDEC Solid State Technology Assn.}, \emph{{JESD79-3F: DDR3 SDRAM Standard}},
  2012.

\bibitem{ddr3l.jedec13}
{JEDEC Solid State Technology Assn.}, \emph{{JESD79-3-1A.01: Addendum No.1 to
  JESD79-3 - 1.35V DDR3L-800, DDR3L-1066, DDR3L-1333, DDR3L-1600, and
  DDR3L-1866}}, 2013.

\bibitem{sodimm.jedec14}
{JEDEC Solid State Technology Assn.}, \emph{{JESD21C, Module 4.20.18: 204-Pin
  DDR3 SDRAM Unbuffered SO-DIMM Design Specification}}, 2014.

\bibitem{lpddr3.jedec15}
{JEDEC Solid State Technology Assn.}, \emph{{JESD209-3C: Low Power Double Data
  Rate 3 SDRAM (LPDDR3) Standard}}, 2015.

\bibitem{lpddr4.jedec17}
{JEDEC Solid State Technology Assn.}, \emph{{JESD209-4B: Low Power Double Data
  Rate 4 (LPDDR4) Standard}}, 2017.

\bibitem{Jung}
M.~Jung, D.~M. Mathew, {\'E}.~F. Zulian, C.~Weis, and N.~Wehn, ``{A New Bank
  Sensitive DRAMPower Model for Efficient Design Space Exploration},'' in
  \emph{{PATMOS}}, 2016.

\bibitem{Matthias}
M.~Jung, D.~M. Mathew, C.~C. Rheinl{\"{a}}nder, C.~Weis, and N.~Wehn, ``{A
  Platform to Analyze DDR3 DRAM's Power and Retention Time},'' \emph{{IEEE
  Design and Test}}, 2017.

\bibitem{kandemir.cases2004}
M.~Kandemir, O.~Ozturk, and M.~Karakoy, ``{Dynamic On-Chip Memory Management
  for Chip Multiprocessors},'' in \emph{{CASES}}, 2004.

\bibitem{kandemir.iccad2001}
M.~Kandemir, U.~Sezer, and V.~{De La Luz}, ``{Improving Memory Energy Using
  Access Pattern Classification},'' in \emph{{ICCAD}}, 2001.

\bibitem{kandemir.date2007}
M.~Kandemir, T.~Yemliha, S.~W. Son, and O.~Ozturk, ``{Memory Bank Aware Dynamic
  Loop Scheduling},'' in \emph{{DATE}}, 2007.

\bibitem{kaseridis.micro2011}
D.~Kaseridis, J.~Stuechelia, and L.~K. John, ``{Minimalist Open-Page: A DRAM
  Page-Mode Scheduling Policy for the Many-Core Era},'' in \emph{{MICRO}},
  2011.

\bibitem{keeth-dram}
B.~Keeth, R.~J. Baker, B.~Johnson, and F.~Lin, \emph{{DRAM Circuit Design:
  Fundamental and High-Speed Topics}}.\hskip 1em plus 0.5em minus 0.4em\relax
  Wiley-IEEE Press, 2007.

\bibitem{keysight34134a.manual}
{Keysight Technologies, Inc.}, \emph{{34134A AC/DC DMM Current Probe: User's
  Guide}},
  \url{https://literature.cdn.keysight.com/litweb/pdf/34134-90001.pdf}, 2009.

\bibitem{keysight34461a.manual}
{Keysight Technologies, Inc.}, \emph{{Keysight Truevolt Series Digital
  Multimeters: Operating and Service Guide}},
  \url{https://literature.cdn.keysight.com/litweb/pdf/34460-90901.pdf}, 2017.

\bibitem{Khan}
S.~Khan, D.~Lee, Y.~Kim, A.~R. Alameldeen, C.~Wilkerson, and O.~Mutlu, ``{The
  Efficacy of Error Mitigation Techniques for DRAM Retention Failures: A
  Comparative Experimental Study},'' in \emph{{SIGMETRICS}}, 2014.

\bibitem{PARBOR}
S.~Khan, D.~Lee, and O.~Mutlu, ``{PARBOR: An Efficient System-Level Technique
  to Detect Data Dependent Failures in DRAM},'' in \emph{{DSN}}, 2016.

\bibitem{Wilkerson}
S.~Khan, C.~Wilkerson, D.~Lee, A.~R. Alameldeen, and O.~Mutlu, ``{A Case for
  Memory Content-Based Detection and Mitigation of Data-Dependent Failures in
  DRAM},'' \emph{{CAL}}, 2016.

\bibitem{khan.micro17}
S.~Khan, C.~Wilkerson, Z.~Wang, A.~R. Alameldeen, D.~Lee, and O.~Mutlu,
  ``{Detecting and Mitigating Data-Dependent DRAM Failures by Exploiting
  Current Memory Content},'' in \emph{MICRO}, 2017.

\bibitem{kim.wwc2000}
H.~S. Kim, M.~Kandemir, N.~Vijaykrishnan, and M.~J. Irwin, ``{Characterization
  of Memory Energy Behavior},'' \emph{{WWC}}, 2000.

\bibitem{kim.hpca18}
J.~Kim, M.~Patel, H.~Hassan, and O.~Mutlu, ``{The {DRAM} Latency {PUF}: Quickly
  Evaluating Physical Unclonable Functions by Exploiting the
  Latency--Reliability Tradeoff in Modern {DRAM} Devices},'' in \emph{HPCA},
  2018.

\bibitem{kim.thesis15}
Y.~Kim, ``{Architectural Techniques to Enhance DRAM Scaling},'' Ph.D.
  dissertation, Carnegie Mellon Univ., 2015.

\bibitem{kim.micro10}
Y.~Kim, M.~Papamichael, O.~Mutlu, and M.~Harchol-Balter, ``{Thread Cluster
  Memory Scheduling: Exploiting Differences in Memory Access Behavior},'' in
  \emph{MICRO}, 2010.

\bibitem{SALP}
Y.~Kim, V.~Seshadri, D.~Lee, J.~Liu, and O.~Mutlu, ``{A Case for Exploiting
  Subarray Level Parallelism (SALP) in DRAM},'' \emph{{ISCA}}, 2012.

\bibitem{kim.cal15}
Y.~Kim, W.~Yang, and O.~Mutlu, ``{Ramulator: A Fast and Extensible DRAM
  Simulator},'' \emph{CAL}, 2015.

\bibitem{kim.isca14}
Y.~Kim, R.~Daly, J.~Kim, C.~Fallin, J.~H. Lee, D.~Lee, C.~Wilkerson, K.~Lai,
  and O.~Mutlu, ``{Flipping Bits in Memory Without Accessing Them: An
  Experimental Study of DRAM Disturbance Errors},'' in \emph{ISCA}, 2014.

\bibitem{Lebeck}
A.~R. Lebeck, X.~Fan, H.~Zeng, and C.~Ellis, ``{Power Aware Page Allocation},''
  in \emph{{ASPLOS}}, 2000.

\bibitem{lee.micro09}
C.~J. Lee, V.~Narasiman, O.~Mutlu, and Y.~N. Patt, ``{Improving Memory
  Bank-Level Parallelism in the Presence of Prefetching},'' in \emph{MICRO},
  2009.

\bibitem{lee.thesis16}
D.~Lee, ``{Reducing DRAM Energy at Low Cost by Exploiting Heterogeneity},''
  Ph.D. dissertation, Carnegie Mellon Univ., 2016.

\bibitem{lee.taco16}
D.~Lee, S.~Ghose, G.~Pekhimenko, S.~Khan, and O.~Mutlu, ``{Simultaneous
  Multi-Layer Access: Improving 3D-Stacked Memory Bandwidth at Low Cost},''
  \emph{ACM TACO}, 2016.

\bibitem{lee.sigmetrics17}
D.~Lee, S.~Khan, L.~Subramanian, S.~Ghose, R.~Ausavarungnirun, G.~Pekhimenko,
  V.~Seshadri, and O.~Mutlu, ``{Design-Induced Latency Variation in Modern DRAM
  Chips: Characterization, Analysis, and Latency Reduction Mechanisms},'' in
  \emph{{SIGMETRICS}}, 2017.

\bibitem{Tiered-Latency}
D.~Lee, Y.~Kim, V.~Seshadri, J.~Liu, L.~Subramanian, and O.~Mutlu,
  ``{Tiered-Latency DRAM: A Low Latency and Low Cost DRAM Architecture},'' in
  \emph{{HPCA}}, 2013.

\bibitem{lee.pact15}
D.~Lee, L.~Subramanian, R.~Ausavarungnirun, J.~Choi, and O.~Mutlu, ``{Decoupled
  Direct Memory Access: Isolating CPU and IO Traffic by Leveraging a
  Dual-Data-Port DRAM},'' in \emph{PACT}, 2015.

\bibitem{lee.hpca15}
D.~Lee, Y.~Kim, G.~Pekhimenko, S.~Khan, V.~Seshadri, K.~Chang, and O.~Mutlu,
  ``{Adaptive-Latency DRAM: Optimizing DRAM Timing for the Common-Case},'' in
  \emph{HPCA}, 2015.

\bibitem{Lefurgy}
C.~Lefurgy, K.~Rajamani, F.~Rawson, W.~Felter, M.~Kistler, and T.~Keller,
  ``{Energy Management for Commercial Servers},'' \emph{{Computer}}, 2003.

\bibitem{legendre.book1805}
A.-M. Legendre, \emph{{Nouvelles M\'{e}thodes pour la D\'{e}termination des
  Orbites des Com\`{e}tes}}.\hskip 1em plus 0.5em minus 0.4em\relax F. Didot,
  1805.

\bibitem{McPAT}
S.~Li, J.~H. Ahn, R.~D. Strong, J.~B. Brockman, D.~M. Tullsen, and N.~P.
  Jouppi., ``{McPAT: An Integrated Power, Area and Timing Modeling Framework
  for Multicore and Manycore Architectures.}'' in \emph{{MICRO}}, 2009.

\bibitem{Lin}
C.~H. Lin, D.~Y. Shen, Y.~J. Chen, C.~L. Yang, and M.~Wang, ``{SECRET:
  Selective Error Correction for Refresh Energy Reduction in DRAMs},'' in
  \emph{{ICCD}}, 2012.

\bibitem{Liu}
J.~Liu, B.~Jaiyen, Y.~Kim, C.~Wilkerson, and O.~Mutlu, ``{An Experimental Study
  of Data Retention Behavior in Modern DRAM Devices: Implications for Retention
  Time Profiling Mechanisms},'' in \emph{{ISCA}}, 2013.

\bibitem{liu.isca2012}
J.~Liu, B.~Jaiyen, R.~Veras, and O.~Mutlu, ``{RAIDR: Retention-Aware
  Intelligent DRAM Refresh},'' in \emph{{ISCA}}, 2012.

\bibitem{liu.sigplan2012}
S.~Liu, K.~Pattabiraman, T.~Moscibroda, and B.~G. Zorn, ``{Flikker: Saving DRAM
  Refresh-Power Through Critical Data Partitioning},'' in \emph{ASPLOS}, 2011.

\bibitem{pintool}
C.-K. Luk, R.~Cohn, R.~Muth, H.~Patil, A.~Klauser, G.~Lowney, S.~Wallace, V.~J.
  Reddi, and K.~Hazelwood, ``{Pin: Building Customized Program Analysis Tools
  with Dynamic Instrumentation},'' in \emph{PLDI}, 2004.

\bibitem{luo.hpca18}
Y.~Luo, S.~Ghose, Y.~Cai, E.~F. Haratsch, and O.~Mutlu, ``{{HeatWatch}:
  Improving 3D {NAND} Flash Memory Device Reliability by Exploiting
  Self-Recovery and Temperature Awareness},'' in \emph{HPCA}, 2018.

\bibitem{luo.sigmetrics18}
Y.~Luo, S.~Ghose, Y.~Cai, E.~F. Haratsch, and O.~Mutlu, ``{Improving 3D NAND
  Flash Memory Lifetime by Tolerating Early Retention Loss and Process
  Variation},'' in \emph{SIGMETRICS}, 2018.

\bibitem{Lyuh}
C.~Lyuh and T.~Kim, ``{Memory Access Scheduling and Binding Considering Energy
  Minimization in Multi-Bank Memory Systems},'' in \emph{{DAC}}, 2004.

\bibitem{malladi.isca12}
K.~T. Malladi, F.~A. Nothaft, K.~Periyathambi, B.~C. Lee, C.~Kozyrakis, and
  M.~Horowitz, ``{Towards Energy-Proportional Datacenter Memory with Mobile
  DRAM},'' in \emph{ISCA}, 2012.

\bibitem{malladi.micro12}
K.~T. Malladi, I.~Shaeffer, L.~Gopalakrishnan, D.~Lo, B.~C. Lee, and
  M.~Horowitz, ``{Rethinking DRAM Power Modes for Energy Proportionality},'' in
  \emph{MICRO}, 2012.

\bibitem{mathew.date18}
D.~M. Mathew, M.~Schultheis, C.~C. Rheinl{\"a}nder, C.~Sudarshan, C.~Weis,
  N.~Wehn, and M.~Jung, ``{An Analysis on Retention Error Behavior and Power
  Consumption of Recent DDR4 DRAMs},'' in \emph{DATE}, 2018.

\bibitem{Mathew}
D.~M. Mathew, \'{E}der F.~Zulian, S.~Kannoth, M.~Jung, C.~Weis, and N.~Wehn,
  ``{A Bank-Wise DRAM Power Model for System Simulations},'' \emph{{RAPIDO}},
  2017.

\bibitem{Meza}
J.~Meza, Q.~Wu, S.~Kumar, and O.~Mutlu, ``{Revisiting Memory Errors in
  Large-Scale Production Data Centers: Analysis and Modeling of New Trends from
  the Field},'' in \emph{{DSN}}, 2015.

\bibitem{jet5467a}
{MFactors}, ``{JET-5467A Product Page},''
  \url{http://www.mfactors.com/jet-5467a-ddr3-sodimm-extender-with-current-sensing/}.

\bibitem{micron.ddr3.design}
{Micron Technology, Inc.}, ``{DDR3 Point-to-Point Design Support},'' Technical
  Note TN-41-13, 2013.

\bibitem{micron.2015}
{Micron Technology, Inc.}, ``{Calculating Memory System Power for DDR3},''
  Technical Note TN-41-01, 2015.

\bibitem{micron.ddr4.design}
{Micron Technology, Inc.}, ``{DDR4 Point-to-Point Design Guide},'' Technical
  Note TN-40-40, 2018.

\bibitem{mukundan.isca13}
J.~Mukundan, H.~Hunter, K.~H. Kim, J.~Stuecheli, and J.~F. Martinez,
  ``{Understanding and Mitigating Refresh Overheads in High-Density DDR4 DRAM
  Systems},'' in \emph{ISCA}, 2013.

\bibitem{muller.iedm96}
K.~P. Muller, B.~Flietner, C.~L. Hwang, R.~L. Kleinhenz, T.~Nakao, R.~Ranade,
  Y.~Tsunashima, and T.~Mii, ``{Trench Storage Node Technology for Gigabit
  {DRAM} Generations},'' in \emph{IEDM}, 1996.

\bibitem{mutlu.date17}
O.~Mutlu, ``{The RowHammer Problem and Other Issues We May Face as Memory
  Becomes Denser},'' in \emph{DATE}, 2017.

\bibitem{mutlu.isca08}
O.~Mutlu and T.~Moscibroda, ``{Parallelism-Aware Batch Scheduling: Enhancing
  Both Performance and Fairness of Shared {DRAM} Systems},'' in \emph{ISCA},
  2008.

\bibitem{mutlu.imw2013}
O.~Mutlu, ``{Memory Scaling: A Systems Architecture Perspective},'' in
  \emph{IMW}, 2013.

\bibitem{Ohsawa}
T.~Ohsawa, K.~Kai, and K.~Murakami, ``{Optimizing the DRAM Refresh Count for
  Merged DRAM/Logic LSIs},'' in \emph{{ISLPED}}, 1998.

\bibitem{ozturk.isqed06}
O.~Ozturk and M.~Kandemir, ``{Data Replication in Banked DRAMs for Reducing
  Energy Consumption},'' in \emph{{ISQED}}, 2006.

\bibitem{patel.dac11}
A.~Patel, F.~Afram, S.~Chen, and K.~Ghose, ``{MARSSx86: A Full System Simulator
  for x86 CPUs},'' in \emph{DAC}, {2011}.

\bibitem{Patel}
M.~Patel, J.~Kim, and O.~Mutlu, ``{The Reach Profiler (REAPER): Enabling the
  Mitigation of DRAM Retention Failures via Profiling at Aggressive
  Conditions},'' in \emph{{ISCA}}, 2017.

\bibitem{paul.isca15}
I.~Paul, W.~Huang, M.~Arora, and S.~Yalamanchili, ``{Harmonia: Balancing
  Compute and Memory Power in High-Performance GPUs},'' in \emph{ISCA}, 2015.

\bibitem{pearson.prsl1895}
K.~Pearson, ``{Notes on Regression and Inheritance in the Case of Two
  Parents},'' \emph{Proc. Royal Soc.\ London}, 1895.

\bibitem{pekhimenko.hpca16}
G.~Pekhimenko, E.~Bolotin, N.~Vijaykumar, O.~Mutlu, T.~C. Mowry, and S.~W.
  Keckler, ``{A Case for Toggle-Aware Compression for {GPU} Systems},'' in
  \emph{HPCA}, 2016.

\bibitem{pekhimenko.hpca15}
G.~Pekhimenko, T.~Huberty, R.~Cai, O.~Mutlu, P.~B. Gibbons, M.~A. Kozuch, and
  T.~C. Mowry, ``{Exploiting Compressed Block Size as an Indicator of Future
  Reuse},'' in \emph{HPCA}, 2015.

\bibitem{pekhimenko.micro13}
G.~Pekhimenko, V.~Seshadri, Y.~Kim, H.~Xin, O.~Mutlu, M.~A. Kozuch, P.~B.
  Gibbons, and T.~C. Mowry, ``{Linearly Compressed Pages: A Low-Complexity,
  Low-Latency Main Memory Compression Framework},'' in \emph{MICRO}, 2013.

\bibitem{pekhimenko.pact12}
G.~Pekhimenko, V.~Seshadri, O.~Mutlu, M.~A. Kozuch, P.~B. Gibbons, and T.~C.
  Mowry, ``{Base-Delta-Immediate Compression: Practical Data Compression for
  On-Chip Caches},'' in \emph{{PACT}}, 2012.

\bibitem{Qureshi}
M.~K. Qureshi, D.~H. Kim, S.~Khan, P.~J. Nair, and O.~Mutlu, ``{AVATAR: A
  Variable-Retention-Time (VRT) Aware Refresh for DRAM Systems},'' in
  \emph{{DSN}}, 2015.

\bibitem{rambus.rdram.website}
{Rambus, Inc.}, ``{RDRAM Memory Architecture},''
  \url{https://www.rambus.com/memory-and-interfaces/rdram-memory-architecture/}.

\bibitem{rixner.isca00}
S.~Rixner, W.~J. Dally, U.~J. Kapasi, P.~Mattson, and J.~D. Owens, ``{Memory
  Access Scheduling},'' in \emph{ISCA}, 2000.

\bibitem{rosenfeld-cal2011}
P.~Rosenfeld, E.~Cooper-Balis, and B.~Jacob, ``{DRAMSim2: A Cycle Accurate
  Memory System Simulator},'' \emph{CAL}, 2011.

\bibitem{voltron.github}
{SAFARI Research Group}, ``{Characterization Results of Modern DRAM Devices
  Under Reduced-Voltage Operation --- GitHub Repository},''
  \url{https://github.com/CMU-SAFARI/DRAM-Voltage-Study}.

\bibitem{ramulator.github}
{SAFARI Research Group}, ``{Ramulator: A DRAM Simulator --- GitHub
  Repository},'' \url{https://github.com/CMU-SAFARI/ramulator}.

\bibitem{softmc.repo}
{SAFARI Research Group}, ``{SoftMC --- GitHub Repository},''
  \url{https://github.com/CMU-SAFARI/SoftMC}.

\bibitem{vampire.github}
{SAFARI Research Group}, ``{VAMPIRE --- GitHub Repository},''
  \url{https://github.com/CMU-SAFARI/VAMPIRE}.

\bibitem{Schroeder}
B.~Schroeder, E.~Pinheiro, and W.~Webe, ``{DRAM Errors in the Wild: A
  Large-Scale Field Study},'' in \emph{{SIGMETRICS}}, 2009.

\bibitem{seshadri.micro17}
V.~Seshadri, D.~Lee, T.~Mullins, H.~Hassan, A.~Boroumand, J.~Kim, M.~A. Kozuch,
  O.~Mutlu, P.~B. Gibbons, and T.~C. Mowry, ``{Ambit: In-Memory Accelerator for
  Bulk Bitwise Operations Using Commodity DRAM Technology},'' in \emph{MICRO},
  2017.

\bibitem{seshadri.thesis16}
V.~Seshadri, ``{Simple DRAM and Virtual Memory Abstractions to Enable Highly
  Efficient Memory Systems},'' Ph.D. dissertation, Carnegie Mellon University,
  2016.

\bibitem{seshadri.micro13}
V.~Seshadri, Y.~Kim, C.~Fallin, D.~Lee, R.~Ausavarungnirun, G.~Pekhimenko,
  Y.~Luo, O.~Mutlu, P.~B. Gibbons, M.~A. Kozuch, and T.~C. Mowry, ``{RowClone:
  Fast and Energy-Efficient In-{DRAM} Bulk Data Copy and Initialization},'' in
  \emph{MICRO}, 2013.

\bibitem{seshadri.bookchapter17}
V.~Seshadri and O.~Mutlu, ``{Simple Operations in Memory to Reduce Data
  Movement},'' in \emph{Advances in Computers, Volume 106}, 2017.

\bibitem{Shin}
W.~Shin, J.~Yang, J.~Choi, and L.-S. Kim, ``{NUAT: A Non-Uniform Access Time
  Memory Controller},'' in \emph{{HPCA}}, 2014.

\bibitem{Sridharan}
V.~Sridharan, N.~DeBardeleben, S.~Blanchard, K.~B. Ferreira, J.~Stearley,
  J.~Shalf, and S.~Gurumurthi, ``{Memory Errors in Modern Systems: The Good,
  the Bad, and the Ugly},'' in \emph{{ASPLOS}}, 2015.

\bibitem{Sridharan2}
V.~Sridharan and D.~Liberty, ``{A Study of DRAM Failures in the Field},'' in
  \emph{{SC}}, 2012.

\bibitem{stan.tvlsi95}
M.~R. Stan and W.~P. Burleson, ``{Bus-Invert Coding for Low-Power I/O},''
  \emph{TVLSI}, 1995.

\bibitem{stan.glsvlsi95}
M.~R. Stan and W.~P. Burleson, ``{Coding a Terminated Bus for Low Power},'' in
  \emph{GLSVLSI}, 1995.

\bibitem{henning.can06}
{Standard Performance Evaluation Corp.}, ``{SPEC CPU2006 Benchmarks},''
  http://www.spec.org/cpu2006.

\bibitem{stuecheli.isca2010}
J.~Stuecheli, D.~Kaseridis, D.~Daly, H.~C. Hunter, and L.~K. John, ``{The
  Virtual Write Queue: Coordinating DRAM and Last-Level Cache Policies},'' in
  \emph{{ISCA}}, 2010.

\bibitem{stuecheli.micro2011}
J.~Stuecheli, D.~Kaseridis, D.~Daly, H.~C. Hunter, and L.~K. John,
  ``{Coordinating DRAM and Last-Level-Cache Policies with the Virtual Write
  Queue},'' \emph{{IEEE Micro}}, 2011.

\bibitem{sudan.asplos2010}
K.~Sudan, N.~Chatterjee, D.~Nellans, M.~Awasthi, R.~Balasubramonian, and
  A.~Davis, ``{Micro-Pages: Increasing DRAM Efficiency with Locality-Aware Data
  Placement},'' in \emph{{ASPLOS}}, 2010.

\bibitem{udipi.hipc09}
A.~N. Udipi, N.~Muralimanohar, and R.~Balasubramonian, ``{Non-Uniform Power
  Access in Large Caches with Low-Swing Wires},'' in \emph{HiPC}, 2009.

\bibitem{Udipi}
A.~N. Udipi, N.~Muralimanohar, N.~Chatterjee, R.~Balasubramonian, A.~Davis, and
  N.~P. Jouppi, ``{Rethinking DRAM Design and Organization for
  Energy-Constrained Multi-Cores},'' in \emph{{ISCA}}, 2010.

\bibitem{Venkatesan}
R.~Venkatesan, S.~Herr, and E.~Rotenberg, ``{Retention-Aware Placement in DRAM
  (RAPID): Software Methods for Quasi-Non-Volatile DRAM},'' in \emph{{HPCA}},
  2006.

\bibitem{vijaykumar.isca15}
N.~Vijaykumar, G.~Pekhimenko, A.~Jog, A.~Bhowmick, R.~Ausavarungnirun, C.~Das,
  M.~T. Kandemir, T.~C. Mowry, and O.~Mutlu, ``{A Case for Core-Assisted
  Bottleneck Acceleration in GPUs: Enabling Flexible Data Compression with
  Assist Warps},'' in \emph{ISCA}, 2015.

\bibitem{vogelsang.micro10}
T.~Vogelsang, ``{Understanding the Energy Consumption of Dynamic Random Access
  Memories},'' in \emph{MICRO}, 2010.

\bibitem{wang2014bigdatabench}
L.~Wang, J.~Zhan, C.~Luo, Y.~Zhu, Q.~Yang, Y.~He, W.~Gao, Z.~Jia, Y.~Shi,
  S.~Zhang, C.~Zheng, G.~Lu, K.~Zhan, X.~Li, and B.~Qiu, ``{BigDataBench: A Big
  Data Benchmark Suite From Internet Services},'' in \emph{HPCA}, 2014.

\bibitem{Ware}
F.~A. Ware and C.~Hampel, ``{Improving Power and Data Efficiency with Threaded
  Memory Modules},'' in \emph{{ICCD}}, 2006.

\bibitem{ware.power7}
M.~Ware, K.~Rajamani, M.~Floyd, B.~Brock, J.~C. Rubio, F.~Rawson, and J.~B.
  Carter, ``{Architecting for Power Management: The IBM POWER7 Approach},'' in
  \emph{{HPCA}}, 2010.

\bibitem{Wilson}
P.~R. Wilson, S.~F. Kaplan, and Y.~Smaragdakis, ``{The Case for Compressed
  Caching in Virtual Memory Systems},'' in \emph{{USENIX ATC}}, 1999.

\bibitem{virtex6.website}
{Xilinx, Inc.}, ``{Virtex-6 FPGA Family},''
  \url{https://www.xilinx.com/products/silicon-devices/fpga/virtex-6.html}.

\bibitem{ml605.manual}
{Xilinx, Inc.}, ``{ML605 Hardware User Guide},''
  \url{https://www.xilinx.com/support/documentation/boards_and_kits/ug534.pdf},
  2012.

\bibitem{xilinx_ar_36719}
{Xilinx, Inc.}, ``{{MIG} 7 Series and Virtex-6 {DDR}2/{DDR}3 Solution Center -
  Design Assistant - Memory Controller Efficiency and Possible Improvements},''
  \url{https://www.xilinx.com/support/answers/36719.html}, 2017.

\bibitem{Yang}
J.~Yang, Y.~Zhang, and R.~Gupta, ``{Frequent Value Compression in Data
  Caches},'' in \emph{{MICRO}}, 2000.

\bibitem{yoon.isca12}
D.~H. Yoon, J.~Chang, N.~Muralimanohar, and P.~Ranganathan, ``{BOOM: Enabling
  Mobile Memory Based Low-Power Server DIMMs},'' in \emph{ISCA}, 2012.

\bibitem{Yoon}
D.~H. Yoon, M.~K. Jeong, and M.~Erez, ``{Adaptive Granularity Memory Systems: A
  Tradeoff Between Storage Efficiency and Throughput},'' in \emph{{ISCA}},
  2011.

\bibitem{Zhang}
T.~Zhang, K.~Chen, C.~Xu, G.~Sun, T.~Wang, and Y.~Xie, ``{Half-DRAM: A
  High-Bandwidth and Low-Power DRAM Architecture from the Rethinking of
  Fine-Grained Activation},'' in \emph{{ISCA}}, 2014.

\bibitem{Zheng}
H.~Zheng, J.~Lin, Z.~Zhang, E.~Gorbatov, H.~David, and Z.~Zhu, ``{Mini-Rank:
  Adaptive DRAM Architecture for Improving Memory Power Efficiency},'' in
  \emph{{MICRO}}, 2008.

\bibitem{zuravleff.patent97}
W.~Zuravleff and T.~Robinson, ``{Controller for a Synchronous DRAM That
  Maximizes Throughput by Allowing Memory Requests and Commands to Be Issued
  Out of Order},'' U.S. Patent No. 5,630,096, 1997.

\end{thebibliography}

\newpage

\appendix

\section*{Appendix: Full DRAM Energy Model Parameters}

Our new DRAM power model, \ch{VAMPIRE} (see Section~\ref{sec:model}), depends on accurately
capturing the current consumed by read and write operations when operations are
interleaved across different columns and banks, as we discuss in 
Section~\ref{sec:datadep:toggle}.  
In Section~\ref{sec:datadep:models}, we introduce a linear model 
\ch{(Equation~\ref{eq:model})} that captures
the energy consumed by each read or write operation, as a function of
(1)~the number of ones in the cache line,
(2)~the number of bits \ch{that are toggled}, and
(3)~whether the operations are interleaved across different banks and/or
columns.
Because different types of interleaving make use of different switching
circuitry (see Figure~\ref{fig:toggle}), we require \ch{a separate set of model 
parameters to use in Equation~\ref{eq:model}} for each
type of operation interleaving.  Table~\ref{tbl:fullmodel} provides the
model parameters (in \si{\milli\ampere}) for each type of interleaving.  All of these parameters
are derived using linear least-squares regression on data collected from
real DRAM modules.

\begin{table}[h]
  \centering
  \small
  \caption{Parameters to quantify current consumption ($I_{total}$) due to data dependency, for the model 
  \chI{$I_{total} = I_{zero} + \Delta I_{one} N_{ones} + \Delta I_{toggle} N_{toggles}$}, where $N_{ones}$ is the number of ones in the cache line, $N_{toggles}$ is the number of bits that are \ch{toggled}, \ch{and $I_{zero}$, \chII{$\Delta I_{one}$, and $\Delta I_{toggle}$} are the parameters}.}%
  \label{tbl:fullmodel}%
  \vspace{-7pt}%
    \begin{tabular}{c||c|c|c||c|c|c}
        \hline
        \multicolumn{7}{c}{\bf No Interleaving (Same Bank \& Column)} \\
        \hhline{=======}
        \multirow{2}{*}{\bf Vendor} & \multicolumn{3}{c||}{\bf Read} & \multicolumn{3}{c}{\bf Write} \\
        \cline{2-7}
        & $I_{zero}$ (\si{\milli\ampere}) & $\Delta I_{one}$ (\si{\milli\ampere}) & $\Delta I_{toggle}$ (\si{\milli\ampere}) & $I_{zero}$ (\si{\milli\ampere}) & $\Delta I_{one}$ (\si{\milli\ampere}) & $\Delta I_{toggle}$ (\si{\milli\ampere}) \\
        \hhline{=#===#===}
        A & 250.88 & 0.449 & 0.0000 & 489.61 & -0.217 & 0.0000 \\ \hline
        B & 226.69 & 0.164 & 0.0000 & 447.95 & -0.191 & 0.0000 \\ \hline
        C & 222.11 & 0.134 & 0.0000 & 343.41 & -0.000 & 0.0000 \\
        \hline 
        \multicolumn{7}{c}{} \\
        \hline
        \multicolumn{7}{c}{{\bf Column Interleaving Only} (same as Table~\ref{tbl:model} in Section~\ref{sec:datadep:models})} \\
        \hhline{=======}
        \multirow{2}{*}{\bf Vendor} & \multicolumn{3}{c||}{\bf Read} & \multicolumn{3}{c}{\bf Write} \\
        \cline{2-7}
        & $I_{zero}$ (\si{\milli\ampere}) & $\Delta I_{one}$ (\si{\milli\ampere}) & $\Delta I_{toggle}$ (\si{\milli\ampere}) & $I_{zero}$ (\si{\milli\ampere}) & $\Delta I_{one}$ (\si{\milli\ampere}) & $\Delta I_{toggle}$ (\si{\milli\ampere}) \\
        \hhline{=#===#===}
        A & 246.44 & 0.433 & 0.0515 & 531.18 & -0.246 & 0.0461 \\ \hline
        B & 217.42 & 0.157 & 0.0947 & 466.84 & -0.215 & 0.0166 \\ \hline
        C & 234.42 & 0.154 & 0.0856 & 368.29 & -0.116 & 0.0229 \\
        \hline 
        \multicolumn{7}{c}{} \\
        \hline
        \multicolumn{7}{c}{\bf Bank Interleaving Only} \\
        \hhline{=======}
        \multirow{2}{*}{\bf Vendor} & \multicolumn{3}{c||}{\bf Read} & \multicolumn{3}{c}{\bf Write} \\
        \cline{2-7}
        & $I_{zero}$ (\si{\milli\ampere}) & $\Delta I_{one}$ (\si{\milli\ampere}) & $\Delta I_{toggle}$ (\si{\milli\ampere}) & $I_{zero}$ (\si{\milli\ampere}) & $\Delta I_{one}$ (\si{\milli\ampere}) & $\Delta I_{toggle}$ (\si{\milli\ampere}) \\
        \hhline{=#===#===}
        A & 287.24 & 0.244 & 0.0200 & 534.93 & -0.249 & 0.0225 \\ \hline
        B & 228.14 & 0.159 & 0.0364 & 419.99 & -0.179 & 0.0078 \\ \hline
        C & 289.99 & 0.034 & 0.0455 & 304.33 & -0.054 & 0.0455 \\
        \hline
        \multicolumn{7}{c}{} \\
        \hline
        \multicolumn{7}{c}{\bf Bank and Column Interleaving} \\
        \hhline{=======}
        \multirow{2}{*}{\bf Vendor} & \multicolumn{3}{c||}{\bf Read} & \multicolumn{3}{c}{\bf Write} \\
        \cline{2-7}
        & $I_{zero}$ (\si{\milli\ampere}) & $\Delta I_{one}$ (\si{\milli\ampere}) & $\Delta I_{toggle}$ (\si{\milli\ampere}) & $I_{zero}$ (\si{\milli\ampere}) & $\Delta I_{one}$ (\si{\milli\ampere}) & $\Delta I_{toggle}$ (\si{\milli\ampere}) \\
        \hhline{=#===#===}
        A & 277.13 & 0.267 & 0.0200 & 537.58 & -0.249 & 0.0225 \\ \hline
        B & 223.61 & 0.152 & 0.0364 & 420.43 & -0.179 & 0.0078 \\ \hline
        C & 266.51 & 0.099 & 0.0090 & 323.22 & -0.072 & 0.0090 \\
        \hline
    \end{tabular}%
\end{table}

\end{document}